\documentclass[a4paper, oneside, 12pt,table]{book}
  \pdfoutput=1
\usepackage[export]{adjustbox}
\usepackage{pdfpages}
\usepackage{perpage} %the perpage package
\MakePerPage{footnote} %the perpage package command
\usepackage[utf8]{inputenc}
\usepackage[T1]{fontenc}
\usepackage{tablefootnote}
\usepackage[english,french]{babel}
\usepackage{xcolor} 
\usepackage[top=2cm, bottom=2cm, left=2.5cm, right=2cm]{geometry}
\usepackage{setspace}
\usepackage{multirow}
\definecolor{gainsboro}{rgb}{0.86, 0.86, 0.86}
\usepackage{hyperref}
\usepackage{hyperref}
\hypersetup{
    colorlinks=true,
    linkcolor=black,
    filecolor=black,      
    urlcolor=black,
}
  \usepackage{color,soul}
\setul{0.5ex}{0.3ex}
\definecolor{Green}{rgb}{0,0.5,0}
\setulcolor{Green}

\newcommand{\subsubsubsection}[1]{\paragraph{#1}\mbox{}}
\usepackage{comment}

\usepackage{booktabs}% http://ctan.org/pkg/booktabs
\usepackage{amsfonts}
\usepackage{subfig}
\usepackage{cite}
\usepackage{apacite}
\usepackage[toc,page]{appendix}
\usepackage{chngcntr}% to change table counting system.
\usepackage{hyperref}%to change the hyperrefernces style.
\usepackage{graphicx}%for working with figures.
\usepackage{caption}
\usepackage{longtable}
\usepackage{subfig}%used for subfigures, many figures in one line
\usepackage{subfloat}
\usepackage{array}%used to personalize tabular in order fo change width and other properties
\usepackage{tabulary}
\usepackage{makecell}

\usepackage{wrapfig}
\usepackage{amsmath}

\usepackage{amssymb}
\usepackage{mathrsfs}
\usepackage{enumitem}
\usepackage{listings}
\usepackage{pdflscape}
\usepackage{tikz}
\usepackage{remreset}
\usetikzlibrary{positioning,calc}
\usepackage{verbatim}%for multiline comments
\usepackage{fancyhdr}
\usepackage[explicit]{titlesec}
\usepackage[Sonny]{fncychap}
\usepackage{etoolbox}
\usepackage{hhline}
\usepackage{refcount}
\usepackage{float}
\usepackage[autostyle]{csquotes}

\usepackage{filecontents}
\usepackage[nottoc,numbib]{tocbibind}
\usepackage[ruled,vlined,linesnumbered,french,onelanguage]{algorithm2e}
\SetKwRepeat{Do}{faire}{tant que}

\SetKwInput{kwParam}{Données}
\SetKwInput{kwSortie}{Sorties}

\ChNameVar{\rmfamily\large\bfseries}
\ChNumVar{\rmfamily\large\bfseries}
\ChTitleVar{\rmfamily\normalsize\mdseries}
\numberwithin{equation}{chapter}
\usepackage{amssymb}% http://ctan.org/pkg/amssymb
\usepackage{pifont}% http://ctan.org/pkg/pifont
\newcolumntype{C}[1]{>{\centering}m{#1}}
\usepackage{adjustbox}

\hypersetup{
	colorlinks=false,
    linkcolor=black,    
	urlcolor=black,
	citecolor=black,
}
\counterwithin{table}{chapter}% specify table counting style. Section number.Table number.
\counterwithin{figure}{chapter}
\newenvironment{compitemize} {\begin{itemize}[noitemsep,topsep=0pt,parsep=0pt,partopsep=0pt, leftmargin=2.4mm]} {\end{itemize}}
\titleformat{\paragraph}[runin]
{\normalfont\normalsize\bfseries}{}{0pt}{\theparagraph)\hspace*{0.66em}#1\\}
\renewcommand\theparagraph{\Alph{paragraph}}
\makeatletter
\renewcommand{\@makechapterhead}[1]{%
	\vspace*{7\p@}%
	{\parindent \z@ \raggedleft \normalfont
		\ifnum \c@secnumdepth >\m@ne
		\Large \bfseries \@chapapp\space \thechapter
		\vspace{18pt}%
		\hrule
		\par\nobreak
		\vskip 12\p@
		\fi
		\interlinepenalty\@M
		\normalsize \normalfont #1\par
		\vskip 12\p@
		\hrule
		\vskip 30\p@
}}
\@removefromreset{chapter}{part}
\makeatother

\counterwithout{table}{chapter}

\makeatletter
     \renewcommand*\l@figure{\@dottedtocline{1}{1em}{3.2em}}
\makeatother
\makeatletter
     \renewcommand*\l@table{\@dottedtocline{1}{1em}{3.2em}}
\makeatother
\begin{document}
%page de garde	

\includepdf[pages=-]{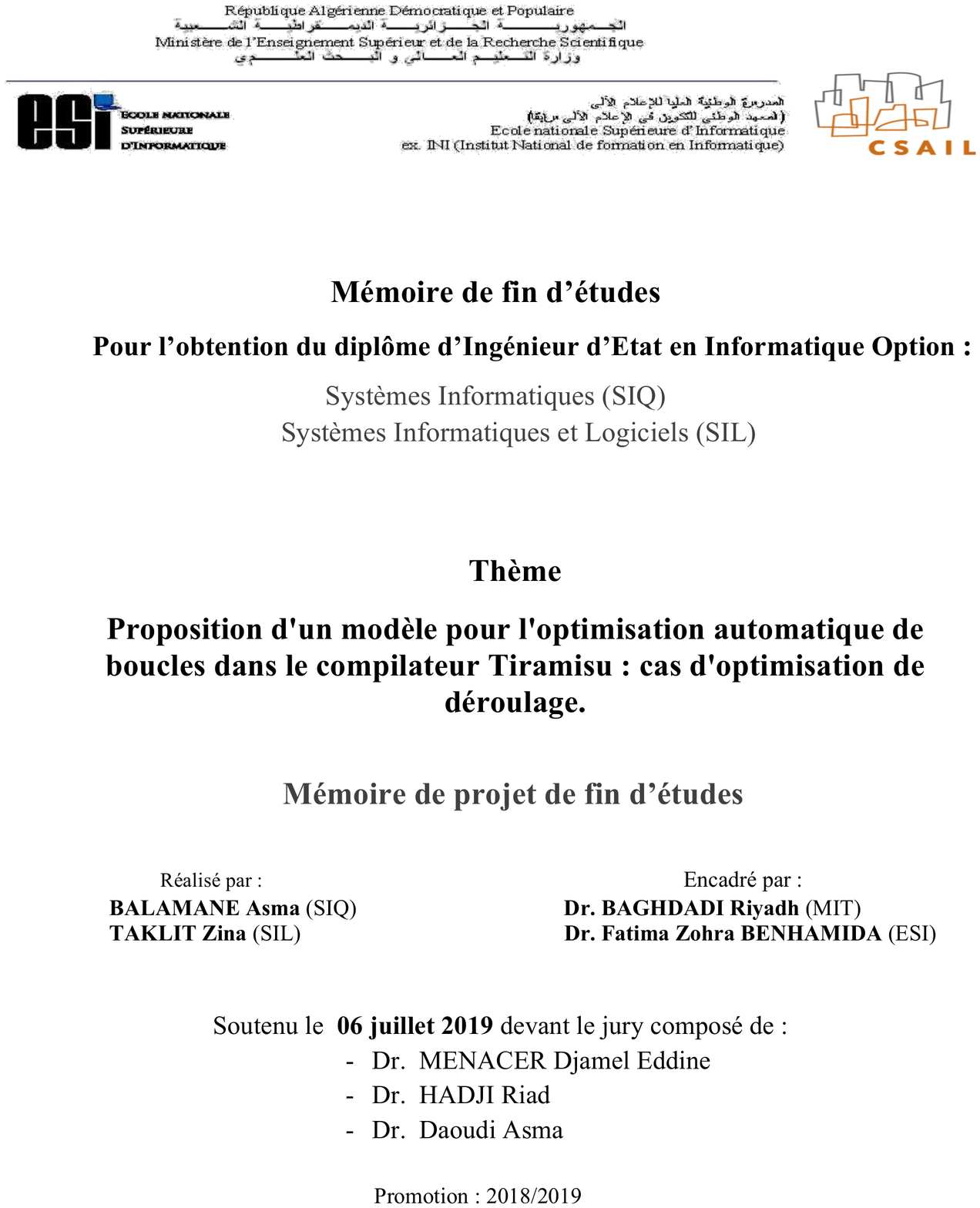}
%	\newpage
\pagestyle{plain}
\setcounter{tocdepth}{3}
\frontmatter
\pagenumbering{Roman}
\clearpage
\thispagestyle{empty}
\mbox{}\vfill
\begin{onehalfspace}
\begin{center}
\vspace{-2cm}
    \Huge{ \textbf{Dédicaces}}
 
\addcontentsline{toc}{chapter}{\textsc{Dédicaces}}
\vspace{2cm}
%	\begin{flushright}
	\large{
	\textit{Tous les mots ne sauraient exprimer ma gratitude, et mon profond amour}\\
	\textit{Merci pour ta tendresse, ton attention, ta patience et tes encouragements} \\
	\textit{Merci pour tout ma bien aimée MAMAN}\\ \vspace{0.5cm}
	\textit{Ma chère grand-mère, mon adorable sœur Imane et ma chère tante Saida}\\
	\textit{Mon cher frère Islam}\\
	\textit{Aucune dédicace ne peut exprimer la profondeur des sentiments d’amour et d’attachement que j’éprouve à votre égard} \\\vspace{0.5cm}
		\textit{ Mon aimable amie Selma}\\ 
		\textit{Merci d'être toujours à mes côtés jusqu'à la fin }\\
    	\textit{ Mes chères amies Yasmine, Chama et Amel}\\ 
    	\textit{Ensembles nous avons passé les plus agréables moments}\\ \vspace{0.5cm}
    	\textit{ Mes amies d'enfance Zineb et Lidya}\\
    	  	\textit{ Mon amie et binôme Zina }\\
    	\textit{Merci... }\\ \vspace{0.5cm}

	             \textbf{\textit{-Asma- }}
		}
	
	\end{center}

\end{onehalfspace}

\vfill
\thispagestyle{empty}
\mbox{}\vfill
\clearpage 
\vspace{-2 cm}
\begin{center}
\begin{onehalfspace}
\begin{center}
  %\Huge{ \textbf{Dédicaces}}
  \end{center}
 \vspace{5cm}
\par \textit{Je dédie ce travail à ma petite famille, qui m'a toujours soutenu et encouragé pour réaliser mes rêves} \\ \textit{particulièrement à ma mère qui me soulage toujours, me remonte le moral, m'encourage pour aller plus loin dans ma vie et prie pour moi jour et nuit.}\\ \vspace{0.5cm}
\par \textit{Je dédie aussi ce travail a mes amis qui m'ont consolidé psychiquement et m'ont aidé à l'accomplir. Plus particulièrement, Miled, Selma, Amel et Fahima qui ont veillé pour m'aider, m'encourager et  me donner des précieux conseils.}\\
\textit{À Amina également qui m'a fort orienté pour implémenter le modèle du projet.}\\ \vspace{0.5cm}
\par \textit{À Mon binôme Asma. Ensembles nous avons travaillé main à main pour réaliser et finaliser ce travail.}\\\vspace{0.5cm}
\par \textit{Aux membres du club GDG et Code\&Share avec lesquels j'ai passé de très bons moments cette année}.\\ \vspace{0.5cm}
\par \textit{Je ne peux trouver les mots justes et sincères pour vous exprimer mon affection et ma gratitude. Vous êtes pour moi les plus agréables personnes...\\ Merci d'être toujours à mes cotés.} \vspace{0.5cm}
 \par \textbf{\textit{-Zina- }}

\end{onehalfspace}
\end{center}
\vfill
\thispagestyle{empty}
\mbox{}\vfill
\clearpage 
\thispagestyle{plain}
\vspace{5cm}

 \begin{center}
  %\Huge{ \textbf{Dédicaces}}
  \end{center}
\begin{center}
    \Huge{ \textbf{Remerciment}}

 \end{center}

\addcontentsline{toc}{chapter}{\textsc{Remerciment}}
\begin{onehalfspace}

\begin{center}
 \vspace{2cm}
%%%%%%%%%%%%%%%%%%%%%%%%%%%%%%Remerciment%%%%%%%%%%%%%%%%%%%%%%%%%%%%%%%%%%%%%%%%

\textit{Nous remercions Allah de nous avoir donné la patience et le courage pour accomplir ce travail.} \vspace{0.5cm}

  \textit{Nous tenons à remercier notre promoteur Dr. Riyadh BAGHDADI pour son suivi continu, sa patience, sa disponibilité et surtout ses judicieux conseils, qui ont contribué à alimenter notre réflexion et nous orienter pour prendre les meilleurs choix.} \vspace{0.5cm}

\textit{Nous remercions également Dr. Fatima Zohra BENHAMIDA, notre encadrante à l’École nationale Supérieure de l’Informatique, pour son accueil et ses conseils qui nous ont permis d'améliorer considérablement ce mémoire.}\\\vspace{0.5cm}

\textit{Nous remercions les membres de jury pour l’intérêt qu’ils portent à ce travail et d’avoir accepté de l’examiner et de le juger.}\\

 \textit{Nous tenons à remercier également Mme. AIT ALI YAHIA Dahbia, pour tous ses efforts afin d’assurer le bon déroulement des stages de fin d’étude, et pour sa compréhension et sa gentillesse.}\\ \vspace{0.5cm}
 \textit{Pour terminer nous témoignons nos reconnaissances et notre profonde gratitude envers toutes les personnes qui ont contribué de loin ou de près à la réalisation de ce travail.}
\end{center}
\end{onehalfspace}

%%%%%%%%%%%%%%%%%%%%%%%% Résumé  %%%%%%%%%%%%%%%%%%%%%%%%%%%%%%%%%%%%%
\vfill
\clearpage 
\section*{\Huge{Résumé}} \vspace{0.5in}
\addcontentsline{toc}{chapter}{Résumé}
\vspace*{-2em}
\begin{onehalfspace}
\par Les architectures des ordinateurs deviennent de plus en plus complexes. Ceci nécessite des efforts pour développer des techniques d'amélioration de programmes permettant une exploitation efficace des ressources matérielles. De ce fait, des transformations sont appliquées sur divers niveaux d’abstraction de code, à savoir le haut niveau, où la représentation du code est proche du langage haut niveau, et le bas niveau, où la représentation est proche du code machine. Ces transformations s'appellent les optimisations de code. 
\par L’optimisation des programmes requiert une expertise profonde. D'une part, elle représente une tâche fastidieuse, car elle nécessite plusieurs tests pour trouver les meilleures combinaisons d’optimisations à appliquer avec leurs bons paramètres. D’une autre part, cette tâche est critique, car elle risque de dégrader les performances du programme au lieu de les améliorer. L’automatisation de l’optimisation de programmes permet de faciliter cette tâche et d’obtenir de bons résultats.
\par Notre projet de fin d'étude consiste à concevoir et développer un modèle pour l’optimisation automatique de boucles dans Tiramisu. Ce dernier est un nouveau langage pour créer des codes de haute performance. Il permet de séparer entre le programme et ses optimisations. Nous avons pris comme cas d’étude l’optimisation de déroulage (\textit{loop unrooling}). Notre contribution vise à automatiser le choix du meilleur facteur de l’optimisation de déroulage de boucles pour un programme écrit en Tiramisu. La solution proposée se base sur les réseaux de neurones profonds.

\par Pour évaluer notre solution, nous avons d'abord comparé le modèle proposé avec les autres algorithmes du \textit{machine learning} (K plus proches voisin et les arbres de décision) utilisés dans les travaux précédents. Notre modèle présente une précision compétitive à ces deux méthodes. Une seconde évaluation est effectuée sur un ensemble de benchmarks. Nous avons comparé entre les facteurs prédits par le modèle et ceux trouvés exhaustivement. Le modèle a pu prédire correctement le meilleur facteur de déroulage dans $4/15$ des instances. Nous avons également évalué l'accélération en temps d'exécution des benchmarks sans application de déroulage et avec l'application du déroulage dont le facteur est prédit par notre modèle. Notre méthode a pu améliorer le temps d'exécution dans $5/15$ des instances. Ce résultat confirme que les réseaux de neurones profonds peuvent être utilisés pour résoudre le problème de choix du facteur de déroulage. Le modèle apprend et donne une précision supérieure à la prédiction aléatoire. \\ \\

\textbf{\large{Mots clés}} : \\ 
\par Optimisation automatique de boucles, sélection de paramètres d’optimisation, déroulage de boucle, Tiramisu, approche d’optimisation automatique.
\end{onehalfspace}
\vfill
%\newpage
\selectlanguage{english}
\section*{\Huge{Abstract}} \vspace{0.5in}
\addcontentsline{toc}{chapter}{Abstract}
\vspace*{-2em}
\begin{onehalfspace}
\par Computer architectures become more and more complex. It requires more effort to develop techniques that improve programs’ performance and allow to exploit material resources efficiently. As a result, many transformations are applied on various levels of code abstraction. The first level is the high level, where the representation is close to the high level language. The second one is the low level, where the presentation is close to the machine code. Those transformations are called code optimizations. 
\par Optimizing programs requires deep expertise. On one hand, it is a tedious task, because it requires a lot of tests to find out the best combination of optimizations to apply with their best factors. On the other hand, this task is critical, because it may degrade the program’ performance instead of improving it. The automatization of this task can deal with this problem and permit to obtain good results.
\par Our end of study project consists on proposing a novel approach based on neural networks to automatically optimize loops in Tiramisu. Tiramisu is a new language to create a code of high performance. It allows to separate between the algorithm and its optimizations. We have chosen loop unrolling as a study case. Our contribution aims to automate the choice of the best loop unrolling factor for a program written in Tiramisu.

\par Initially, we have compared our model with other machine learning algorithms ( KNN and decesion treee) from previous work. Our model was competitive to those two methods.  A second evaluation has been performed on a set of five benchmarks. We compared between the factor predicted by our model and the best one found exhaustively. Our model has been able to predict correctly the best unrolling factor in $4/15$ of the instances. We have also evaluated the acceleration in the execution time of the benchmarks without applying loop unrolling optimization and with the application of loop unrolling with predicted unrolling factor. Our method could improve execution time in $5/15$ of the instances. This result confirms that deep neural networks can be used to solve the problem of choosing the best unrolling factor. The model has been able to  learn and it gives greater precision than random prediction. \\ \\ 

\par \textbf{\large{Key words}} : \\
\par Automatic optimization of loops, selection of the best factor of optimizations, loop unrolling, Tiramisu, automatic optimization approaches.
\end{onehalfspace}

%\phantomsection
%\section*{} \vspace{0.5in}
\addcontentsline{toc}{chapter}{Résumé en arabe}
%\includepdf[pages=-]{resumear.pdf}
\includepdf[pages=-]{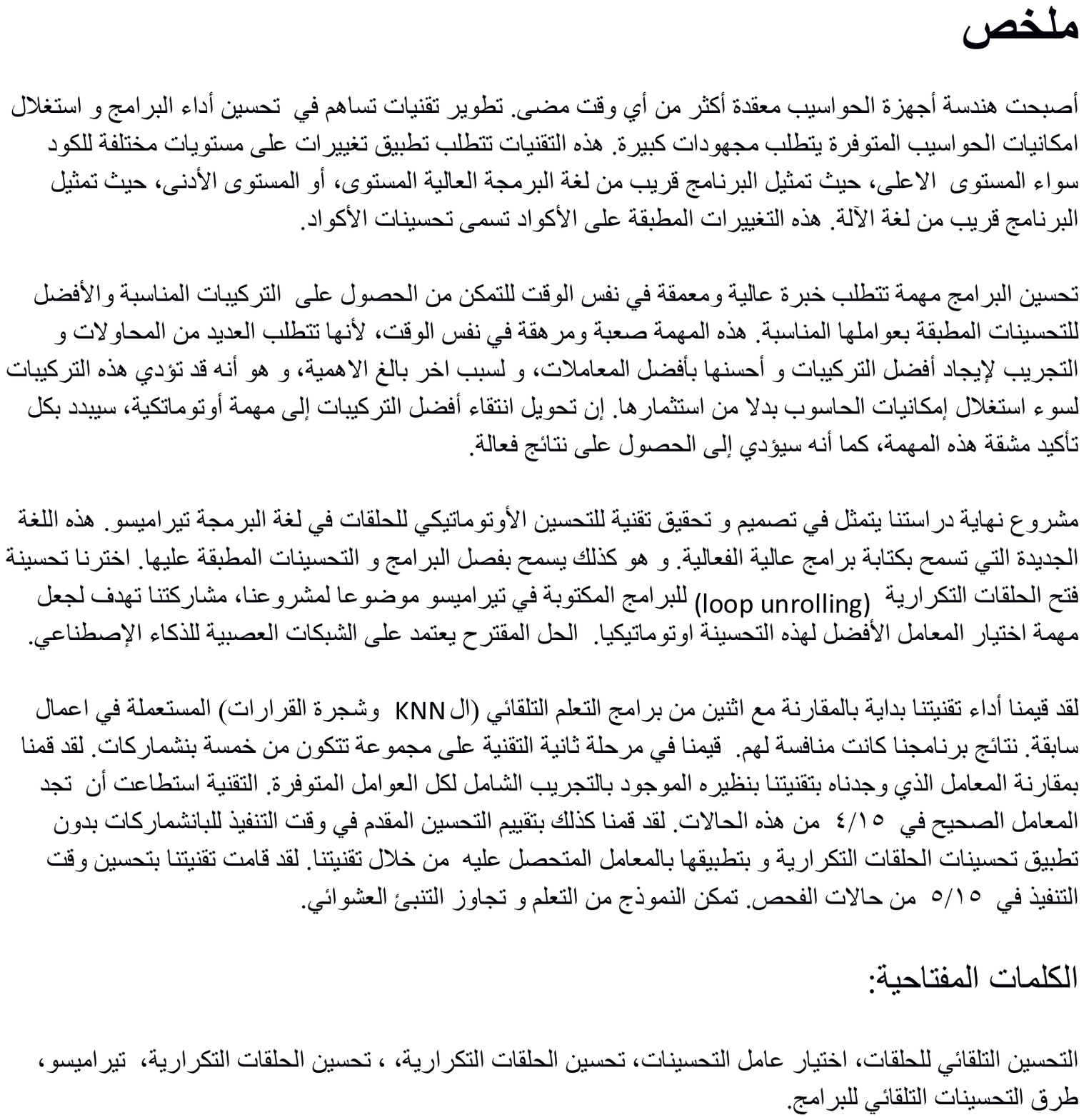}
\cleardoublepage
% arabic abstract%
%\newpage
\selectlanguage{french}
\clearpage
\cleardoublepage
%\chapter{Remerciements}
%\chapter{Résumé}
%\chapter{Abstract}
%\section*{\Huge{Table de matière}} \vspace{0.5in}
\begingroup
\color{black}
\tableofcontents
\endgroup

\clearpage %\cleardoublepage %for openright
\phantomsection

\begingroup
\color{black}
\listoffigures
\endgroup

\clearpage %\cleardoublepage %for openright
\phantomsection

\begingroup
\color{black}
\listoftables
\endgroup

\clearpage
\phantomsection
\chapter*{Glossaire}

\addcontentsline{toc}{chapter}{Liste des abréviations}
\renewcommand{\arraystretch}{1.4}
\begin{tabular}{ll}
  \textbf{AG} & Algorithme Génétique  \\
  \textbf{ANN} & \textit{Artificial Neural Network} \\
  \textbf{AST} & Arbre Syntaxique Abstrait \\
  \textbf{CNN} & \textit{Convolutional Neural Network} \\
  \textbf{CPU} & \textit{Central Processing Unit} \\
  \textbf{CSV} & \textit{Comma-Separated Values} \\
  \textbf{DAG} & \textit{Directed Acyclic Graph} \\
  \textbf{FPGA} &  \textit{Field-Programmable Gate Array} \\
  \textbf{GPU} & \textit{Graphics Processing Unit} \\
   \textbf{HPC} & \textit{High Performance Computing} \\
   \textbf{LSD} & Langage Spécifique au domaine \\
     \textbf{ML} & \textit{Machine Learning} \\
  \textbf{MLP} & \textit{Multi-Layer Perceptron} \\
  \textbf{NLP} & \textit{Natural language processing} \\
  \textbf{RI} & Représentation Intermédiaire \\
  \textbf{ReLU} & \textit{Rectified Linear Unit} \\
  \textbf{RMSprop} & \textit{Root Mean Square prop} \\
   \textbf{RNA} & Réseaux de Neurones Artificiels \\
 \textbf{RNNs} & \textit{Recurrent Neural Network} \\
  \textbf{SGD} & \textit{Stochastic Gradient Descent} \\
  \textbf{TSS} & \textit{Tile Size Selection} \\
\end{tabular}
\begin{comment}
\begin{itemize}
	\setlength{\itemsep}{1pt}
	\item[] \textbf{AG}  Algorithme Génétique.
	\item[] \textbf{ANN}  \textit{Artificial Neural Network}.
	\item[] \textbf{AST}  Arbre Syntaxique Abstrait.
	\item[] \textbf{CNN} \textit{Convolutional Neural Network}
	\item[] \textbf{CPU} \textit{Central Processing Unit}
	\item[] \textbf{CSV} \textit{Comma-Separated Values}
	\item[] \textbf{DAG}  \textit{Directed Acyclic Graph}.
	\item[] \textbf{FPGA}  \textit{Field-Programmable Gate Array}
	\item[] \textbf{GPU} \textit{Graphics Processing Unit}
	\item[] \textbf{HPC}  \textit{High Performance Computing}.
	\item[] \textbf{LSD}  Langage Spécifique au domaine.
	\item[] \textbf{ML}  \textit{Machine Learning}.
	\item[] \textbf{MLP}  \textit{Multi-Layer Perceptron}.
	\item[] \textbf{RI}  Représentation Intermédiaire.
	\item[] \textbf{ReLU}  \textit{Rectified Linear Unit}.
	\item[] \textbf{RNNs}  \textit{Recurrent Neural Network}.
	\item[] \textbf{SGD}  \textit{Stochastic Gradient Descent}.
	\item[] \textbf{TSS}  \textit{Tile Size Selection}.

\end{itemize}
\end{comment}
\renewcommand{\arraystretch}{1} 
\cleardoublepage
\mainmatter
\cleardoublepage
\thispagestyle{plain}
\vspace{0.5cm}
%%%%%%%%%%%%%%%%%%%%%%%%%%%%Introduction Général %%%%%%%%%%%%%%%%%%%%%
\section*{\Huge{Introduction générale}} \vspace{0.5in}
\vspace*{-2em}

\addcontentsline{toc}{chapter}{\textsc{Introduction générale}}
\begin{onehalfspace}

\par Afin d’atteindre des niveaux de performances avancés, les architectures des ordinateurs deviennent de plus en plus complexes.  Dans certains domaines, les applications requièrent une exploitation efficace des ressources matérielles. Les développeurs doivent introduire des transformations (optimisations) sur divers niveaux  d’abstraction de codes. Ces optimisations  visent à améliorer certaines métriques notamment l’utilisation de la mémoire et le temps d’exécution. 

\par L’optimisation des programmes n’est pas une tâche simple. Elle requiert une expertise profonde afin de donner les bonnes optimisations avec leurs meilleurs facteurs à appliquer. Il s’agit d’une tâche qui est d’une part fastidieuse, car elle nécessite plusieurs tests pour trouver les meilleures combinaisons d’optimisations. D’une autre part, c'est une tâche critique, car elle risque de dégrader les performances du programme au lieu de les améliorer. En effet, L'optimisation dépend de plusieurs contraintes, à savoir les caractéristiques matérielles de la machine d’exécution, les dépendances entre les instructions du programme et les interactions entre les optimisations de codes à appliquer. Ces interdépendances agissent significativement sur l’effet des optimisations.
\par L’automatisation de l’optimisation des programmes permet de faciliter cette tâche et d’obtenir de bons résultats compétitifs aux optimisations soigneusement établies par des experts. Plusieurs approches ont été proposées afin d’améliorer l’automatisation de l'optimisation de codes donnant naissance à des nouvelles techniques intégrées aux compilateurs. Il existe plusieurs problèmes ouverts dans le domaine d’optimisation automatique : la sélection des optimisations bénéfiques à appliquer, l'estimation des bons paramètres des optimisations et la définition de l'ordre d'application des optimisations pour avoir les meilleures performances.	
\par L’optimisation du code touche principalement deux niveaux, à savoir l'optimisation haut niveau, où la représentation du code est proche du langage haut niveau, et l'optimisation bas niveau où la représentation est proche du code machine. L’optimisation haut niveau du code s’avère primordiale avant toute optimisation bas niveau notamment dans certains langages dit spécifiques au domaine (LSD). Ces langages fournissent des fonctionnalités de programmation spéciales qui ne sont pas facilement offertes par les autres langages de programmation à objectifs généraux \cite{TouatiAdvancedBackendOptim}.
\par Tiramisu est un nouveau langage et compilateur qui permet de générer des codes très rapides et de cibler différentes architectures matérielles (multicore, GPU, FPGA et systèmes distribués). Il a été  lancé en 2018 par l'équipe COMMIT de CSAIL\footnote{Computer Science and Artificial Intelligence Laboratory.} du MIT, elle travaille principalement sur le développement de nouveaux compilateurs qui facilitent l’écriture des codes optimisés. Tiramisu offre la particularité  de séparer l’algorithme et les optimisations à appliquer sur le code dans une partie appelée \textit{Schedule}. Cette partie contient des commandes d'optimisation introduites par le programmeur manuellement, c'est-à-dire que le programmeur doit choisir parmi les optimisations, celles qui rendent son code plus rapide. Or, l'optimisation manuelle est compliquée, le programmeur risque de ne pas choisir les meilleures combinaisons d'optimisations.

\par Dans le cadre de notre projet de fin d’étude, nous visons à contribuer dans l’optimisation automatique de boucles dans Tiramisu. Nous prenons comme cas d’étude l’optimisation de déroulage (\textit{loop unrooling}). Le problème de sélection du meilleur facteur de déroulage représente le problème principal ciblé par notre projet de fin d’étude.\\ 
Concrètement, les objectifs de notre contribution sont :
 \begin{itemize}[label=\textendash]

\item Concevoir un modèle pour automatiser le choix du meilleur facteur de l’optimisation de déroulage de boucles pour un programme écrit en Tiramisu (un programme déjà optimisé ou non optimisé), et ce, pour des architectures matérielles à base de CPU.
\item Explorer une nouvelle approche de sélection automatique du facteur de déroulage. La solution proposée doit présenter une précision qui dépasse la sélection aléatoire (\textit{random selection}) et qui est compétitive aux travaux précédents. 

\end{itemize}

\par Ce document est organisé comme suit : dans la première partie, nous commençons d’abord par une étude théorique répartie en trois chapitres. Dans le chapitre \ref{ch:chapterOne} nous introduisons l’optimisation du code en donnant une explication sur le fonctionnement des optimisations haut niveau du code les plus utilisées. Le chapitre \ref{ch:chapterTwo} présente les approches et les techniques adoptées par les compilateurs pour optimiser automatiquement les programmes. Dans le chapitre \ref{ch:chapterThree}, nous exposons le compilateur Tiramisu ciblé par notre projet. Dans la deuxième partie, nous détaillons notre contribution tout au long de trois chapitre. Dans le chapitre \ref{ch:conception}, nous expliquons les phases de conception du système proposé en détaillant notre modèle de prédiction du meilleur facteur de déroulage. Dans le chapitre \ref{ch:realisation}, nous expliquons les étapes de réalisation du système tout en justifiant les différents choix technologiques considérés. \\
Pour clore ce mémoire, nous exposons dans le chapitre \ref{ch:tests}, les résultats de comparaison du modèle avec deux autres algorithmes de \textit{machine learning} (K plus proches voisin et les arbres de décision) utilisés dans les travaux précédents pour le problème de sélection du facteur de déroulage. Nous exposons également dans le dernier chapitre l'évaluation effectuée sur des benchmarks implémentés sur Tiramisu.

\end{onehalfspace}
%%%%%%%%%%%%%%%%%%%%%%%%%%%%%%%%%%%%%%%%%%%%%%%%%%%%%%%%%%%%%%%%%%%%%%%

\part*{\textsc{Etat de l'art}}
\addcontentsline{toc}{part}{Etat de l'art}
\pagestyle{fancy}
\renewcommand{\chaptermark}[1]{ \markboth{#1}{} }
\fancyhf{}
\rhead{\textbf{Chapitre \thechapter: \leftmark}}
%\fancyhead[L]{}
\renewcommand{\footrulewidth}{0.4pt}
\cfoot{\thepage}
\renewcommand{\thechapter}{\Roman{chapter}}
\setcounter{chapter}{0}
\chapter{\textsc{Généralités sur les optimisations de code}}
\label{ch:chapterOne}
\begin{onehalfspace}
\section*{Introduction} \vspace{0.3in}
\addcontentsline{toc}{chapter}{\textsc{Introduction}}
\vspace*{-2em}
\begin{onehalfspace}
 \par Dans certains domaines d'application, il ne suffit pas d'écrire des codes qui répondent aux spécifications, mais aussi des codes qui exploitent efficacement les ressources matérielles avec un temps d'exécution minimal \cite{Knijnenburg2002}.
\par Les optimisations des programmes sont des transformations appliquées sur la structure du code dans le but d'améliorer ses performances notamment le temps d'exécution. Cependant, dans certains cas, l'application des optimisations peut les dégrader. En effet, déterminer les transformations susceptibles d'améliorer les performances dépend de plusieurs facteurs parfois incontrôlables par le programmeur, à savoir la gestion des caractéristiques de l'architecture de la machine . D'autre part, la connaissance de l'impact d'application de chaque optimisation à part et celui de sa combinaison avec d'autres est très compliquée ce qui fait de l'optimisation du code une tâche assez complexe et nécessite une expertise profonde.
\par Dans ce chapitre, nous allons exposer les optimisations de codes utilisées dans le compilateur Tiramisu, expliquer leurs objectifs et déterminer les paramètres à considérer pour les appliquer. Enfin, nous allons clôturer le chapitre par une explication des difficultés rencontrées afin de choisir les bonnes combinaisons d'optimisations à appliquer.
\end{onehalfspace}
\end{onehalfspace}
\section{Niveaux d'optimisation du code}
\par Un compilateur est un programme qui traite les instructions écrites dans un langage de programmation donné pour les traduire en langage machine, utilisé par le processeur d'un ordinateur. Les algorithmes peuvent être exprimés sous divers langages de programmation, les codes seront ensuite transformés en une représentation intermédiaire ou un code machine. De ce fait, un code peut avoir une représentation haut niveau qui est proche aux langages de programmation haut niveau, et une représentation bas niveau proche aux instructions du processeur.
\par les optimisations de code représentent l'ensemble des techniques utilisées afin d'améliorer les performances d'un programme, à savoir le temps d'exécution, l'utilisation de la mémoire ou encore la consommation de ressources. L'optimisation des programmes s'avère primordiale dans certains domaines d'application nécessitant un calcul immense et donc une grande consommation des ressources. La théorie de la compilation définit des frontières entre l'optimisation haut niveau et l'optimisation bas niveau, cette classification est liée aux niveaux de la représentation de code, à savoir la représentation haut niveau et la représentation bas niveau.\\
\par \textbf{- Optimisation bas niveau (\textit{backend optimization}):} il s'agit de l’ensemble des transformations appliquées sur le code sous sa représentation finale proche aux instructions destinées aux processeurs comme les instructions assembleur et le code à trois adresses. Les métriques optimisées à ce niveau d’abstraction sont généralement liées à l’architecture du processeur : le nombre des instructions générées, la taille du code, l’ordonnancement des instructions, l’allocation et l’affectation des registres, l’optimisation du cache, l’optimisation du mode d’adressage, etc. \\
\par \textbf{- Optimisation haut niveau (\textit{front-end optimization}):} elle représente l’ensemble des transformations appliquées sur le code écrit dans le langage haut niveau directement ou sous sa représentation intermédiaire proche au langage haut niveau, cette représentation contient des structures syntaxiques sophistiquées telles que les boucles et les structures de contrôles ainsi que les structures de données complexes. Parmi ces optimisations se trouvent les transformations des nids de boucles, l’analyse de dépendance de données et de procédures, la parallélisation des instructions et des blocs  et l’analyse des alias. 
L’optimisation haut niveau de code est généralement effectuée par le programmeur, qui gère les entités abstraites et tient compte du fonctionnement du programme en général pour optimiser la conception du système. L’analyse et l’optimisation à ce niveau d’abstraction du programme permettent d’améliorer des métriques indépendamment des architectures d’exécution \cite{TouatiAdvancedBackendOptim}. \\
\par Ce travail s'intéresse à l’optimisation haut niveau des codes qui représente la phase basique de l’optimisation de codes.
Dans des langages spécifiques aux domaines (LSD) comme Halide  \cite{Kelly2013Halide}, Tiramisu \cite{Baghdadi2018TiramisuV3}, PolyMage \cite{PolyMage} qui sont utilisés dans le domaine de la programmation de hautes performances, l’optimisation haut niveau est primordiale avant toute optimisation bas niveau. En effet, les LSD fournissent des fonctionnalités spéciales de programmation optimisée  qui ne sont pas fournies par les autres langages de programmation comme C, Java, etc.

\section{Optimisations de boucles}
\begin{onehalfspace}
\par Les outils de profilage\footnote{\textit{Code profiling} consiste à analyser l'exécution d'un programme pour connaître son comportement à l'exécution.} de code ainsi que les démonstrations basées sur la théorie de la complexité des algorithmes ont affirmé que les programmes passent plus de 90\% du temps d'exécution au niveau des boucles, il est donc plus intéressant de focaliser sur l'optimisation des nids de boucles. De plus, ils utilisent voracement les ressources matérielles, tel que la mémoire cache et les unités de calcul \cite{Yaroub2013BclnestOptm}.

\subsection{Nids de boucles}
\label{nidsDeBoucles}
\begin{figure}[ht]
	\begin{center}
		 \frame{\includegraphics[scale=0.8]{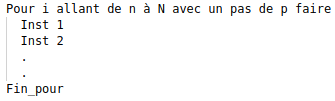}} 
	\end{center}
	\caption{Structure générale d'une boucle.}
	\label{fig:boucle_links}
\end{figure}

Une "boucle" est considérée comme étant une structure itérative d’un programme fini. La structure algorithmique est schématisée dans la figure \ref{fig:boucle_links} \cite{Ciorba2008ParallelLops}.

 \par
\begin{figure}[ht]
	\begin{center}
		 \frame{\includegraphics[scale=0.8]{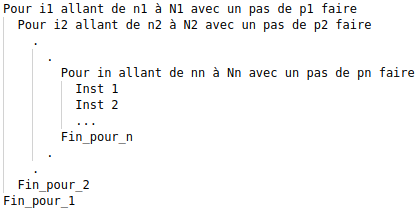}} 
	\end{center}
	\caption{Structure générale d'un nid de boucle.}
	\label{fig:niddeboucle_links}
\end{figure}

 Un nid de boucles est une imbrication d’un nombre fini n (n>1) de boucles ${(B1,B2 ..Bn)}$ (voir figure \ref{fig:niddeboucle_links}). Il est dit "parfait" si toutes les instructions appartiennent à la boucle la plus interne. 

 \par Une grande partie du temps d'exécution est consommée au niveau des boucles. De ce fait, il est très utile d'optimiser ces parties de codes, ceci permet d'améliorer efficacement les performances globales des programmes dans nombreuses applications.

 \par Différentes techniques sont  utilisées pour augmenter les  performances de boucles.  Cependant, les  optimisations ne sont pas  toujours valides et certaines  peuvent détériorer les  performances du  programme ou même  fausser ses résultats.  L'analyse de dépendance est  une des théories qui  permettent de décider la  possibilité d'appliquer une  optimisation sur un  programme sans avoir changé sa logique. Nous allons  aborder ici les optimisations  de boucles les plus  importantes et les plus  utilisées, à savoir la  parallélisation, la  vectorisation, la fusion de  boucles, l'interversion de boucles, le déroulage, tuilage, etc.
 %%%%%%%%%%%%%Les optimisations de boucles %%%%%%%%%%%%%%%%%%%%%%%
 \subsection{Parallélisation (\textit{Parallelization})}
 \par
\begin{figure}[ht]
	\begin{center}
		\includegraphics[scale=0.71]{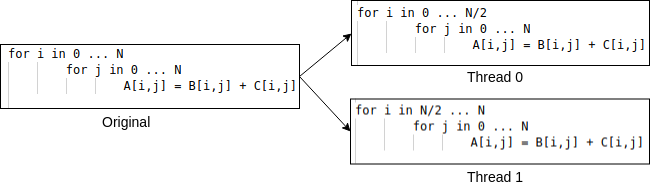}
	\end{center}
	\caption{Parallélisation des itérations d’une boucle imbriquée sur deux \textit{threads}.}
	\label{fig:exmp_parallel_links}
\end{figure}

 Le parallélisme de boucle est un type très courant de parallélisme du code. Il a comme principal objectif de détecter et d'exploiter le parallélisme dans des programmes séquentiels. Cette transformation consiste à affecter chaque itération indépendante à des \textit{threads} parallèles puis lancer l'exécution de ces derniers de manière asynchrone (voir figure \ref{fig:exmp_parallel_links}). Cela nécessite de trouver la  bonne transformation de  boucle qui permet de lancer le parallélisme.
\par Pour effectuer cette transformation d'une manière efficace, il faut prendre en considération les paramètres de la machine. En effet, plus que la machine dispose d'unités fonctionnelles plus la parallélisation est meilleure, de ce fait, la machine doit avoir au moins deux unités de traitement pour pouvoir exécuter les instructions en parallèle.

\par Il est intéressant de noter que la gestion des \textit{threads} peut engendrer un coût  supplémentaire. Donc, pour que l'optimisation soit vraiment utile, il faut que le temps perdu dans la gestion des \textit{threads} soit strictement inférieur au temps gagné par la parallélisation. D'autre part, pour que la parallélisation soit valide, il faut s'assurer qu'elle n'inverse pas le sens de dépendance\footnote{Il y a une dépendance de données entre deux instructions, S1 et S2, lorsque elles tentent d’accéder à la même localité mémoire M, et au moins l’une de ces deux instructions tente de modifier M.} entre les itérations \cite{Allen2002Optim}.

 \subsection{Vectorisation (\textit{Vectorization})}
\par La vectorisation est le processus permettant de transformer un programme qui opère sur des données élémentaires à un programme qui opère sur des vecteurs de données chacun porte sur deux éléments ou plus \cite{Byunghyun2010DataTransformations}. 

\par
\begin{figure}[ht]
	\begin{center}
		\includegraphics[scale=0.9]{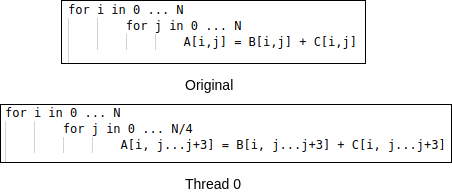} 
	\end{center}
	\caption{Vectorisation de la boucle y avec un facteur de quatre.}
	\label{fig:exmp_vect_links}
\end{figure}

 Dans une boucle, cette transformation consiste à regrouper un ensemble de $X$ éléments qui subissent le même traitement à travers les itérations pour les rediriger vers des registres vectoriels afin d'effectuer un traitement commun \cite{Allen2002Optim}. La figure \ref{fig:exmp_vect_links}  montre un exemple de vectorisation de boucles.

\par L'objectif de cette transformation est d'utiliser efficacement les ressources matérielles notamment les registres. La vectorisation est une technique très utile pour exploiter le parallélisme au niveau de données. La taille des registres vectoriels affecte le choix du facteur $k$ dans la transformation de vectorisation, car plus la taille des registres est grande plus il est possible d'appliquer la vectorisation pour gagner en temps d'exécution \cite{Byunghyun2010DataTransformations}.

\subsection{Interversion de boucles (\textit{Loop reordering})}
\par
\begin{figure}[ht]
	\begin{center}
		\includegraphics[scale=0.78]{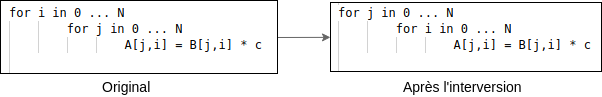} 
	\end{center}
	\caption{Interversion de deux boucles.}
	\label{fig:exmp_reorder_links}
\end{figure}

  Cette transformation permet  de changer l'ordre  d'exécution de niveaux de boucle dans une boucles imbriquée i.e la boucle la plus interne devient la plus externe et vice versa (voir la figure \ref{fig:exmp_reorder_links}). L'interversion de boucles permet d'exploiter le parallélisme et d'améliorer la localité de données \cite{David1994Compilertransformations} et par conséquent, d'utiliser le cache efficacement et d'améliorer l'accès aux données.

\par Cette transformation permet d'excellentes améliorations, mais elle n'est pas toujours légale. Pour effectuer cette transformation de manière à préserver le sens original du programme, il faut s'assurer qu'elle soit valide à appliquer, ceci est fait en analysant la dépendance entre les instructions des boucles à intervertir. 
 
\subsection{Fusion de boucles (\textit{Loop fusion})}
\par
\begin{figure}[ht]
	\begin{center}
		\includegraphics[scale=0.85]{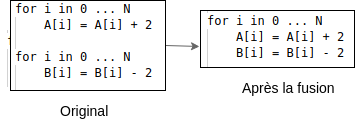} 
	\end{center}
	\caption{Fusion de deux boucles adjacentes en une seule boucle.}
	\label{fig:loop_fusion}
\end{figure}

 Cette transformation consiste à fusionner deux boucles adjacentes en une seule boucle (voir la figure \ref{fig:loop_fusion}) afin de réduire la surcharge de la boucle et d'améliorer les performances d'exécution.
\par Il est important de noter que cette transformation n'améliore pas toujours le temps d'exécution, car cela dépend des contraintes architecturales de la machine. Par exemple, l'architecture de la mémoire peut offrir de meilleures performances si les deux tableaux sont initialisés dans des boucles séparées, plutôt que de les initialiser simultanément dans une seule boucle. D'autre part, cette transformation n'est pas toujours légale, elle doit être soumise aux conditions de validité en utilisant l'analyse de dépendance.

\subsection{Découpage en bandes (\textit{Loop splitting})}
\par
\begin{figure}[ht]
	\begin{center}
		 \includegraphics[scale=0.88]{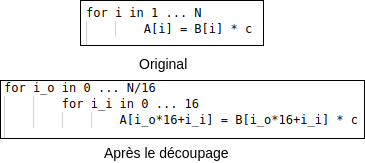}
	\end{center}
	\caption{Découpage en bandes d'une boucle avec un facteur de 16.}
	\label{fig:exmp_split_links}
\end{figure}

 Le découpage en bandes est l'une des transformations de  boucles les plus puissantes  qui consiste à découper la dimension par un facteur $k$ pour créer de nouvelles dimensions en divisant la variable d’itération en sous variables internes et externes. Ce facteur est connu sous le nom de "facteur de découpage". Il transforme la boucle ayant par exemple l'indice i et la dimension N en une boucle imbriquée : la boucle la plus interne avec la dimension k et la boucle externe avec la dimension $N/k$ qui est le résultat de la division de l'étendue de $N$ sur le facteur de découpage k (voir la figure \ref{fig:exmp_split_links}). Après le découpage, les références à l'index d'origine deviennent : ($externe \times facteur + interne$)
\cite{Jonathan2012Decouplingalgorithms}.

\par Cette transformation est utilisée pour les nids de boucles qui manipulent les données multidimensionnelles (tableaux à 2 dimensions ou à 3 dimensions). L’objectif principal de cette transformation est de séparer les itérations indépendantes de la boucle principales, ce qui permet de les exécuter en parallèle. 

\par L'optimisation de découpage en bandes ne change pas l'ordre d'évaluation des instructions. Cette transformation ne fait rien en elle-même, mais cela ouvre d'autres possibilités d'optimisations comme le déroulage lorsqu'il est combiné avec d'autres transformations. La vectorisation et le déroulage l'utilisent souvent, ils sont précédés par une division de la dimension de boucles par le facteur de vectorisation ou de déroulage. Ensuite, la vectorisation ou le de déroulage de la boucle interne est effectué.

\subsection{Déroulage de boucle (\textit{Loop unrolling})}
\label{loopUnrollingSection}

\par
\begin{figure}[ht]
	\begin{center}
	\includegraphics[scale=0.90]{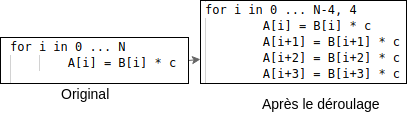}
	\end{center}
	\caption{Déroulage d’une boucle avec un facteur de déroulage de quatre.}
	\label{fig:loop_unrol}
\end{figure}

 L'optimisation de déroulage (\textit{loop unrolling}) est une transformation qui consiste à répliquer les instructions de la boucle la plus interne une ou plusieurs fois, selon le facteur de déroulage k donné. Ce qui permet de réduire le nombre d'itérations. Par exemple, si $k=4$ alors le nombre d'itérations (initialement égal à $N$) est réduit à $N/4$ (voir la figure \ref{fig:loop_unrol}). Le déroulage est appliqué lorsque plusieurs itérations de boucles rechargent les mêmes valeurs.
\par La transformation de déroulage permet la réutilisation des registres, si les instructions déroulées sont indépendantes entre elles, il est possible de les lancer en parallèle à condition que la machine dispose d'un nombre suffisant de registres pour appliquer la parallélisation. Un autre avantage de cette transformation est qu'elle peut être appliquée sur n'importe quelle boucle sans restriction sur le langage source. Le déroulage améliore les performances presque dans tous les cas où il est appliqué d'une manière significative \cite{David1994Compilertransformations}.

\par Il est recommandé d'appliquer cette transformation sur la boucle la plus interne, car le fait de l'appliquer sur la boucle externe risque d'augmenter la surcharge du contrôle de la boucle au lieu de la diminuer (à cause de la réplication de la boucle interne). Le déroulage de la boucle interne dans le cas des boucles partageant des données qui se chevauchent réduit le nombre total de chargements (\textit{load}). Il permet aussi de réduire le nombre de tests d'instruction de branchements : $N$ conditions d'arrêt seront testées $(N/k)$ fois, les pénalités de branchements seront réduites d'un facteur de $k$ dans les architectures à base de pipeline. 

\par Le facteur de déroulage $k$ doit être bien choisi en prenant en considération les contraintes architecturales de la machine : si $k$ est grand, le chargement de toutes les instructions dans le cache à la fois ne sera pas possible, ainsi le taux de défauts de cache augmente et de même le temps d'exécution. L'optimisation de déroulage peut être intégrée automatiquement dans les compilateurs grâce à sa simplicité  \cite{David1994Compilertransformations}. Pour plus de détails sur cette optimisation voir l'annexe A.

\subsection{Tuilage de boucles (\textit{Loop tiling})} 

\par
\begin{figure}[ht]
	\begin{center}
		\includegraphics[scale=0.78]{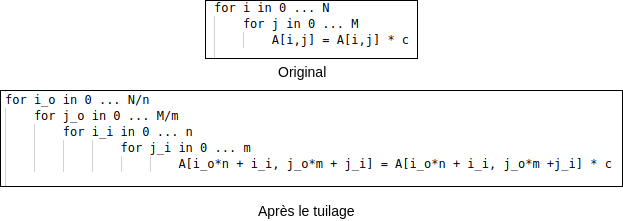} 
	\end{center}
	\caption{Un tuilage avec facteur : $\displaystyle n\times m$. Après le tuilage, A est accessible en blocs de taille $\displaystyle n\times m$.}
	\label{fig:exmp_tile_links}
\end{figure}

 Le tuilage ou encore le \textit{tiling} ou le \textit{blocking} est une combinaison de découpage en bandes et d'interversion de boucles. Tout d'abord, nous découpons $i$ et $j$ avec un facteur de $n$ et $m$ respectivement ($\displaystyle n\times m$ est appelé le facteur de tuilage). Ensuite, nous réordonnons les variables pour exprimer le tuilage (voir la figure \ref{fig:exmp_tile_links}). Le tuilage est une transformation qui permet de découper la matrice en petits blocs rectangulaires : il itère à l'extérieur sur les blocs, et à l'intérieur, il itère sur les instructions de chaque bloc \cite{Baghdadi2018TiramisuV3}.

\par Le tuilage est la généralisation multidimensionnelle de découpage en bandes. Le tuilage est principalement utilisé pour améliorer la réutilisation du cache en divisant l'espace d'itération\footnote{L'ensemble de toutes les instances d'exécution d'une instruction.} en blocs \cite{David1994Compilertransformations}. Le tuilage permet aussi de maximiser la localité de données (localité spatiale et temporelle). Il permet de créer de nouvelles boucles internes de sorte que les données accessibles dans les boucles internes s'insèrent dans le cache. Le tuilage peut augmenter le taux de parallélisme, car si les données des blocs ne dépendent pas l'une des autres, il est possible de les paralléliser. Cette transformation s'avère indispensable pour atteindre des hautes performances dans les calculs de multiplication sur les matrices denses ou encore dans l'algèbre linéaire. Elle constitue aussi un facteur clé pour améliorer les performances des réseaux de neurones convolutifs\footnote{Les \textit{CNN} estun type de réseaux de neurones connus par l'opération convolution. Ils s’adaptent bien pour
des données organisées en grille telles que les images.}.

\par Le tuilage est une transformation complexe. Le facteur de tuilage doit être choisi soigneusement. D'une part, il doit être assez petit pour s'assurer que le bloc peut être chargé entièrement dans le cache et pour que la parallélisation soit bénéfique. D'une autre part, il doit être assez grand pour bien exploiter l'espace du cache. Il n'est pas toujours possible d'appliquer cette transformation comme elle se base sur la transformation de l'interversion de boucles qui n'est pas toujours légale. Elle doit être soumise au processus de validation en utilisant l'analyse de dépendance avant son application. Pour plus de détails sur cette optimisation voir l'annexe A.

\subsection{Torsion de boucles (\textit{Loop skewing})} 
\par la torsion de boucle ou \textit{Loop skewing} est une transformation qui redéfinit l'espace d'itération pour permettre d'exploiter la parallélisation par la suite. Elle est principalement utile en combinaison avec la transformation d'interversion de boucles \cite{David1994Compilertransformations}.
\par
\begin{figure}[ht]
	\begin{center}
		\includegraphics[scale=0.95]{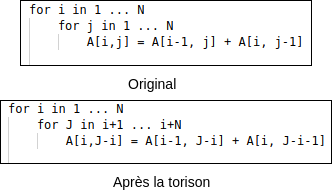} 
	\end{center}
	\caption[Exemple de \textit{loop skewing}, J sert à une nouvelle variable pour remplacer j avec $J=j-1$.]{Exemple de \textit{loop skewing}, J sert à une nouvelle variable pour remplacer j avec $J=j-1$~\protect\cite{David1994Compilertransformations}.}
	\label{fig:exmp_skew_links}
\end{figure}

 Cette optimisation est effectuée en ajoutant l'index de la boucle externe multipliée par le facteur d'inclinaison $k$, aux bornes de la variable de la boucle interne, puis soustraire la même valeur pour chaque utilisation de la variable de la boucle interne à l'intérieur de la boucle (voir la figure \ref{fig:exmp_skew_links}). Elle change les bornes de la boucle et change aussi l'utilisation des index correspondants pour garder le même sens que le programme original. Cette transformation ne change pas le sens du programme et elle est toujours légale\cite{David1994Compilertransformations}.

\par Les boucles qui résultent de la transformation d'inclinaison sont extrêmement trapézoïdales\footnote{Trapézoïdal est un adjectif qui désigne quelque chose en forme de trapèze, c'est-à-dire avec quatre côtés dont deux sont parallèles.} ce qui produit des bornes complexes. Les pénalités résultant d'une telle boucle irrégulière sont plus sévères sur les machines vectorielles mais moins graves sur les multi-processeurs asynchrones. Cependant, même sur ces architectures, l'inclinaison de boucle doit être appliquée avec précaution pour éviter des déséquilibres de charges significatifs \cite{Allen2002Optim}.

\subsection{Calcul redondant} 
\par Cette transformation permet de calculer la donnée à chaque fois que c'est nécessaire au lieu d'aller la charger à partir de la mémoire. En effet, la donnée doit être recalculée d'une manière redondante à chaque fois qu'elle est utilisée même si le résultat existe déjà dans la mémoire. Cette optimisation est très utile lorsque le coût  arithmétique (le temps pris) de calcul est inférieur par rapport au coût  de son chargement à partir de la mémoire \cite{Menaceri2008HalideAuto}.

\subsection{Réorganisation de calcul} 
\par Cette transformation permet de réordonner les instructions du programme \cite{David1994Compilertransformations}. Elle consiste à rapprocher les instructions consommatrices (celles qui utilisent la donnée produite par d'autres instructions) des productrices (celle qui produit ces données). Cette transformation assure que la donnée consommée se trouve toujours dans le cache de données, ce qui permet d'améliorer la localité des données par conséquent \cite{Menaceri2008HalideAuto}.
\section {Difficultés du choix des bonnes optimisations} 
\par Plus les architectures des processeurs deviennent complexes, plus le nombre de transformations possibles à appliquer augmente et donc le processus de sélection des bonnes optimisations devient très compliqué aussi \cite{David1994Compilertransformations}.

\par Ainsi, la structure de l'architecture matérielle est un facteur déterminant pour le choix des optimisations à effectuer, car il faut maximiser l’utilisation des ressources, à savoir les processeurs, les unités fonctionnelles et les registres, etc. D’autre part, il faut minimiser le nombre d’opérations effectuées, les défauts de cache et la taille mémoire requise.  
\par
\begin{figure}[ht]
	\begin{center}
		\includegraphics[scale=0.9]{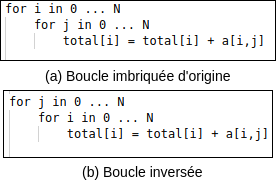} 
	\end{center}
	\caption[Exemple de l'interversion de boucle pour montrer la difficulté de choix de la meilleure optimisation.]{Exemple de l'interversion de boucle pour montrer la difficulté de choix de la meilleure optimisation~\protect\cite{David1994Compilertransformations}.}
	\label{fig:exmp_diff_links}
\end{figure}

 Parfois, ces objectifs ne peuvent pas être réunis. Considérons la figure \ref{fig:exmp_diff_links}(a), supposons que la machine stocke les tableaux colonne par colonne, l’accès à chaque élément de la matrice "a" nécessite de charger entièrement la colonne contenant cet élément à chaque itération de la boucle interne. Par contre, en inversant l’ordre des niveaux de la boucle, les éléments seront accessibles séquentiellement colonne par colonne ce qui correspond à l’ordre de stockage défini par la machine. Ceci permet de réduire la distance entre deux accès successifs de la mémoire et donc, de minimiser les défauts de cache. D'autre part, la version (a) du programme exprimé dans la figure \ref{fig:exmp_diff_links} permet au tableau "total[i]" d'être chargé dans le registre une seule fois pour chaque ligne de la matrice "a". Mais, la version optimisée (b) augmente le nombre des opérations de chargement/stockage de $"total[i]"$ (de $2n$ à $2n^2$) car l'élément $"total[i]"$ est chargé dans le registre pour chaque itération de la boucle interne de la matrice $"a"$ \cite{David1994Compilertransformations}.
\par Dans la Figure \ref{fig:exmp_diff_links} (a), la boucle interne accède aux éléments du tableau a avec un stride\footnote{Le nombre de locations mémoires entre deux éléments successifs dans un tableau.}-n. En inversant l'ordre des boucles, l'accès devient en stride-1, comme le montre la figure \ref{fig:exmp_diff_links} (b).

\par Une autre difficulté peut se poser aussi, la transformation peut être illégale, donc il faut d'abord étudier la validité d'appliquer cette transformation, une tâche qui n'est pas simple.
\par Nous avons vu que le choix d'une seule optimisation est une tâche assez complexe. La difficulté se multiplie énormément lorsqu'on veut avoir une bonne combinaison d'optimisations. \\ Considérons la multiplication généralisée de matrice (gemm), qui calcule $C = \alpha A.B + \beta C$ (avec $A$,$B$ et $C$ sont des matrice, $\alpha$ et $\beta$ sont des scalaires). Elle  représente un bloc de construction de nombreux algorithmes, notamment les RNCs (Réseau de Neurones Convolutifs). Les implémentations hautement optimisées nécessitent la fusion des boucles de multiplication et d'addition, ainsi que l'application du tuilage de boucles à deux niveaux, de la vectorisation, du déroulage de la boucle, etc. \cite{Baghdadi2018TiramisuV3}.
\par Générer un code de telle complexité est très difficile. En effet, pour la plupart des programmeurs, il est difficile d'obtenir ce niveau de performance, car l'effort requis pour explorer l'espace d'implémentations possibles est insoluble lors du codage manuel de transformations de code complexe. L'optimisation du code nécessite des connaissances très approfondies dans le domaine ainsi que de l'expertise. À titre d'exemple chez Google, il y a plus de 150 ingénieurs qui écrivent des algorithmes en langage Halide\footnote{Halide \cite{Kelly2013Halide} est un nouveau compilateur et langage spécifique au domaine de traitement d'images, la principale particularité de ce language est la séparation entre les optimisations (\textit{schedule}) et l’algorithme.} non optimisés alors qu'il y a uniquement deux personnes qui disposent de l'expertise nécessaire pour les optimiser. \cite{Baghdadi2018TiramisuV3}.

\par Par conséquent, le choix des optimisations à appliquer sur un programme est une tâche très fastidieuse qui nécessite la connaissance de l'architecture matérielle, le recensement des avantages et des inconvénients de chaque optimisation. Elle nécessite également des tests sur l'effet de combiner les optimisations et s'assurer que la logique de l'algorithme reste correcte après l'application des optimisations. 
\section*{Conclusion}
\addcontentsline{toc}{chapter}{\textsc{Conclusion}}
\par  Dans ce chapitre, nous avons abordé les principales optimisations de boucles qui permettent d'améliorer les performances et de réduire le temps d'exécution. Nous avons exposé principalement les optimisations de boucles les plus importantes dans l'optimisation de code.
\par Il est important de noter qu'il ne faut pas appliquer les optimisations sans avoir effectué une étude au préalable, car elles peuvent détériorer les performances. Il faut effectivement prendre plusieurs contraintes en considération comme l'impact de l'utilisation de chaque optimisation à part et l'effet de les combiner. Il faut aussi prendre en considération les propriétés de l'architecture d'exécution (nombres d'unité de traitements, la taille des registres, etc). Quelques optimisations n'ont aucun effet sur le programme mais s'avère très utiles comme prétraitement pour appliquer d'autres. Ainsi, il se trouve que l'optimisation du programme n'est pas une tâche simple. C'est une tâche très complexe et fastidieuse pour les programmeurs. Elle nécessite beaucoup de temps pour trouver la meilleure combinaison d'optimisations qui utilise efficacement les ressources matérielles et améliore davantage le temps d'exécution. C'est pour cette raison, des compilateurs qui permettent d'optimiser le code d'une manière automatique sont apparus. Dans le chapitre suivant,  nous allons aborder les approches et les techniques utilisées par les compilateurs pour optimiser le code d'une manière automatique.
\end{onehalfspace}

%=====================================================================
% chapitre 2 

\chapter{\textsc{Optimisation automatique dans les compilateurs}}
\label{ch:chapterTwo}
\begin{onehalfspace}
\setcounter{footnote}{0}
\section*{Introduction}
\addcontentsline{toc}{chapter}{\textsc{Introduction}}
 \par Dans le chapitre précèdent, nous avons exposé les différentes optimisations de codes qui permettent d'améliorer ses performances et d'exploiter efficacement les caractéristiques qu'offrent les architectures matérielles. La tâche de sélection de bonnes optimisations à appliquer est très complexe, elle nécessite une compréhension profonde des optimisations, de leurs effets sur le code ainsi que leurs interactions. Ce qui rend cette tâche très difficile voire infaisable manuellement pour des programmes assez complexes. Des techniques d'optimisation automatique ont été intégrées dans les compilateurs afin d'alléger cette tâche et de trouver les meilleures combinaisons d’optimisations à appliquer sur les programmes. Cependant, l'automatisation mène à résoudre un problème NP-difficile où l’espace de recherche est immense.
 \par Dans ce chapitre, nous allons exposer les différentes approches d'optimisation automatique en comparant les avantages et les défis de chacune d'elle. Nous allons également cité quelques techniques de pointes basées sur ces approches. 
 
\section{Motivation pour adopter l'optimisation automatique}
\par
\begin{figure}[ht]
	\begin{center}
		\includegraphics[scale=0.54]{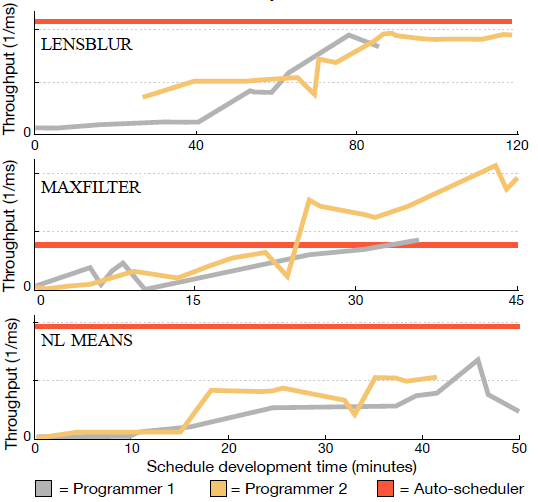} 
	\end{center}
	\caption[Comparaison entre l’optimisation automatique et manuelle effectuée sur trois benchmarks.]{Comparaison entre l’optimisation automatique et manuelle effectuée sur trois benchmarks~\protect\cite{Mullapudi2016HalidAutoScheduler}.}
	\label{fig:motivationExemple}
\end{figure}

 Pour bien illustrer l’effort nécessaire lors de l'optimisation manuelle par un développeur expert comparativement à l’optimisation automatique, \cite{Mullapudi2016HalidAutoScheduler} ont recruté deux développeurs professionnels dans le langage Halide \cite{Kelly2013Halide} pour mettre en défi l’Auto-scheduler de Halide. Les experts ont sélectionné trois benchmarks qu’ils n’ont jamais optimisés auparavant\footnote{LENSBlur , NLMeans et MAXFilter.} et ils ont implémenté les algorithmes originaux de ces benchmarks sur Halide. Chaque expert a défini des optimisations qu'il a jugées optimales pour chaque benchmark. Tout au long du processus d'optimisation manuelle, les experts ont pris des mesures de performances de leurs optimisations actuelles. Ensuite, ils ont comparé les performances de leur code optimisé manuellement à celui optimisé automatiquement par l’Auto-scheduler de Halide\footnote{Le test est effectué sur une machine de 4 cœurs d’Intel E5-2690 proposée par les deux experts.}. \\ 
Les résultats de la comparaison sont illustrés par la figure \ref{fig:motivationExemple}. L’axe des $x$ dans chaque graphe indique le temps de développement (en minutes) des optimisations manuelles. L’axe des $y$ montre les performances des benchmarks\footnote{Mesuré en tant que débit en pixels.}. La ligne horizontale correspond aux performances des optimisations générées par l’Auto-scheduler (en secondes). Les lignes jaune et grise correspondent chacune au progrès de chaque programmeur. 
%%%% Add l'analyse de la figure 
\par  Pour deux benchmarks parmi trois, même après presque une heure de travail, l'optimisation manuelle donne des résultats moins performants que ceux obtenus par l'optimisation automatique (générés en quelques secondes). L'optimisation automatique de codes est compétitive et parfois dépasse l'optimisation manuelle en termes de performances.   
\section{Défis de l’optimisation automatique du code}
\par Le défi majeur du choix des bonnes combinaisons d'optimisations réside essentiellement dans leurs dépendances aux langages de programmation, à la logique des algorithmes et aux architectures matérielles. De plus, les optimisations peuvent dégrader les performances du code dans certains cas. Comprendre le comportement de ces optimisations, leurs effets sur le code source et leurs interactions est un problème de modélisation très complexe qui donne naissance à des thématiques de recherche de pointe. En effet, il existe plusieurs problèmes ouverts dans le domaine d’optimisation des compilateurs : la sélection des optimisations bénéfiques à appliquer, l'estimation des bons paramètres des optimisations (le facteur de tuilage, facteur de déroulage de boucles, etc.) et la définition de l'ordre d'application des optimisations pour avoir les meilleures performances.

\subsection{Problème de sélection des bonnes optimisations}
\par Le problème est défini en se limitant à la sélection des optimisations bénéfiques pour un programme donné tout en ignorant l’ordre d'application de ces optimisations. Les chercheurs ont montré que l’interdépendance et l'interaction entre l'activation et la désactivation des optimisations dans une combinaison peuvent altérer considérablement les performances du programme en cours d'exécution, même en ignorant leur ordre d’application \cite{Ashouri2018Survey}.\\

\begin{figure}[ht]
	\begin{center}
		\includegraphics[scale=0.5]{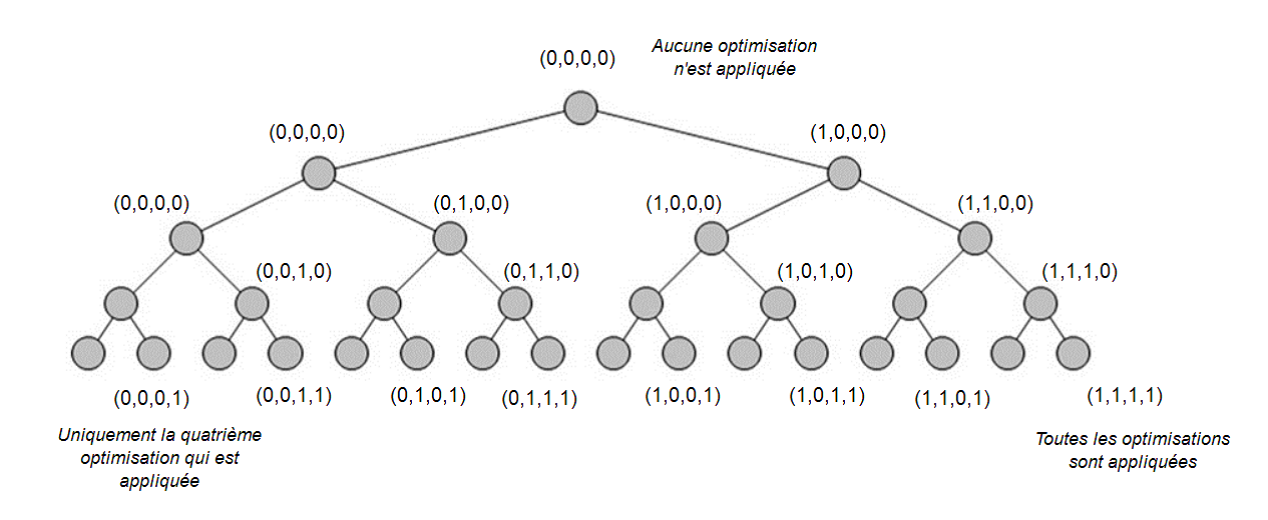} 
	\end{center}
	\caption[Exemple d’espace de recherche d’une combinaison de quatre optimisations.]{Exemple d’espace de recherche d’une combinaison de quatre optimisations~\protect\cite{Menaceri2008HalideAuto}.}
	\label{fig:exmp_explorationSpace}
\end{figure}

\textbf{L’espace de recherche} : l’ensemble des optimisations peut être modélisé sous forme d'un vecteur booléen (soit $\Omega = \{0,1\}^n$). Un élément d’optimisation $(o_i)$ peut avoir deux valeurs : $(o_i =1)$ si l’optimisation est activée, sinon ($o_i =0$) quand elle est désactivée. L’espace de toutes les combinaisons d’optimisations est $2^n$ (voir la figure \ref{fig:exmp_explorationSpace}).

\par L’espace d’optimisation du problème de sélection est un espace exponentiel. Si les facteurs que peut prendre chaque optimisation sont pris en considération, la version étendue de l’espace d’optimisation dépasse le choix binaire (entre activer/désactiver l'optimisation) à une variante à choix multiples pour chaque optimisation. De plus, certaines optimisations peuvent prendre plusieurs paramètres. Pour simplifier, si nous considérons $m$ comme étant le nombre total des options d’optimisations, l’espace devient $ \{0,1,…,m\}^n $.
\subsection{Problème du choix de l'ordre des optimisations (\textit{phase-ordering problem})}
\par L'ordre dans lequel les optimisations sont appliquées influence sur l'accélération apportée : le choix d’un ordre d’optimisations prédéfini peut dégrader certaines optimisations qui auraient pu être plus efficaces si un autre ordre a été choisi. 

\par \textbf{L’espace de recherche} : l’espace des choix d’ordre possibles des optimisations est factoriel ($n!$  permutations) tel que $n$ est le nombre d’optimisations à appliquer. Si de plus, le cas d'une taille variable d’une séquence d’optimisation est pris en considération\footnote{Possible de considérer cinq optimisations ou se restreindre à trois ou encore deux, de plus, une optimisation peut être réappliquée plusieurs fois, donc le nombre des optimisations dans la séquence est variable.}, le cardinal de l'espace devient alors  $\sum_{i=0}^l n^i$, où $n$ représente le nombre d’optimisations  possibles et $l$ représente la taille maximale que peut prendre une séquence d’optimisations. Même avec des valeurs raisonnables de $ n $ et de $ l $, l’espace de recherche d’optimisations formé est immense. Par exemple, en supposant que $ n=10$  et  $l=10$, l'espace de recherche d’optimisations est formé de plus de 11 billions séquences d’optimisations différentes. Le problème de trouver le bon ordre d’optimisations est non déterministe étant donné que la taille d’une séquence d’optimisation est non bornée. 
\par Le problème du choix du bon ordre d'optimisations reste un problème ouvert dans le domaine d'optimisation des compilateurs. L'incapacité des chercheurs à résoudre complètement le problème les conduit à concentrer leurs efforts sur le problème de la sélection de l'ensemble des bonnes optimisations sans prendre en considération le problème d'ordre \cite{Ashouri2018Survey}.

\section{Approches d'optimisation automatique du code}
\label{sec:Approche_optim_automatic}
\par
\begin{figure}[ht]
	\begin{center}
		\includegraphics[scale=0.4]{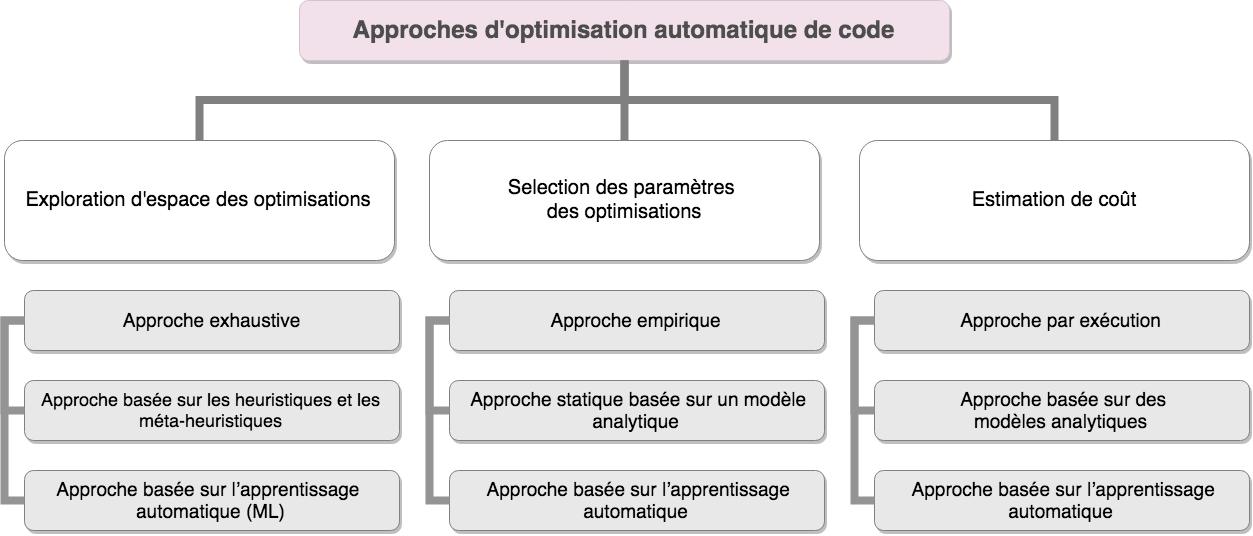} 
	\end{center}
	\caption{Approches d'optimisation automatique de code.}
	\label{fig:Approches}
\end{figure}

 L'optimisation automatique du code traite de la génération automatique des codes optimisés en utilisant différents scénarios et architectures. Elle vise à choisir différents facteurs de code qui influencent en maximisant ou en minimisant une fonction objectif \cite{Ashouri2018Survey}. 
Différentes approches ont été proposées, chaque approche traite une problématique d'optimisation de code pour répondre à la fonction objectif en favorisant certaines contraintes par rapport à d'autres. Cependant, elles reposent toutes sur deux principales notions : l'espace des optimisations à appliquer et le coût d'application des optimisations. Nous avons donné une classification des approches d'optimisation de code en se basant sur la problématique  traitée par chaque classe. La figure \ref{fig:Approches} donne une vision globale sur les approches d'optimisation automatique du code.
\par L'espace des optimisations à appliquer représente les différentes optimisations de code connues, chacune vise à améliorer certaines caractéristiques de performance définies au niveau de la fonction objectif, certaines optimisations nécessitent des paramètres à définir. Cet espace de transformations crée de nombreuses versions du programme. L'exploration de cet espace des optimisations et leurs paramètres peut être coûteuse ce qui mène à définir des approches dédiées pour une exploration plus efficace. D'autre part, le coût d'application d'un sous-ensemble d'optimisations $S$, représenté sous forme d'une instance mesurée de la fonction objectif, permet de décider l'efficacité de $S$ et aide à orienter l'exploration.
\subsection{Approches d'exploration d'espace d'optimisations}
\par L'espace d'optimisations est constitué de différentes optimisations de boucles pouvant améliorer les performances d'un programme si une "bonne combinaison" a été choisie.\\ L'espace est très souvent grand ce qui impose de définir des approches d'exploration permettant de trouver la solution en un temps raisonnable. \\ Les approches exposées sont : (1) l'approche exhaustive, (2) l'approche basée sur les heuristiques et les métaheuristiques et (3) l'approche basée sur l'apprentissage automatique. 
\subsubsection{Approche exhaustive}
\par Cette méthode consiste à explorer tout l'espace de recherche en essayant toutes les optimisations possibles afin de trouver la meilleure combinaison. Par exemple, ATLAS une bibliothèque spécialisée dans l'optimisation automatique des programmes de multiplication de matrices. Elle utilise la méthode exhaustive pour trouver certains paramètres d'optimisation pré-élaborés ce qui permet d'améliorer la multiplication de matrices sur l'architecture d'exécution cible. ATLAS explore exhaustivement des plages de valeurs des paramètres d'optimisations pour différentes tailles de matrices. Ensuite, cette bibliothèque sauvegarde les meilleurs modèles d'implémentation du programme pour les différentes tailles de  matrices afin de les utiliser ultérieurement pendant l'exécution \cite{Atlasref}. 
\par Cette approche permet donc d'essayer toutes les possibilités pour choisir la meilleure avec précision. Cependant, elle requière un temps de traitement énorme exponentiellement lié à la dimension de l'espace d'exploration. Cette approche est utilisée alors pour des espaces restreints qui ont été déjà réduits grâce à un certain mécanise de présélection.  
\subsubsection{Approches d'heuristiques et de métaheuristiques}
Plusieurs algorithmes basés sur des heuristiques\footnote{Les heuristiques sont des règles empiriques simples basées sur l’expérience. Elles ne donnent pas forcément la solution la plus optimale mais plutôt des solutions assez proches de l'optimale en un temps raisonnable. Elles sont définies pour un type de problème d'optimisation.} et des métaheuristiques\footnote{Des stratégies d’exploration d’espace de recherche (qui peut être très grand). Elles permettent d’explorer cet espace d’une manière efficace pour se rapprocher de la solution optimale à des problèmes d'optimisation plus généraux.} ont été proposées pour résoudre les problèmes d’optimisation. Dans le cas où l’espace de recherche est vaste, la recherche exhaustive, les méthodes itératives ou les méthodes heuristiques simples ne sont pas assez pratiques, alors que les métaheuristiques permettent souvent de trouver de bonnes solutions avec moins d’effort de calcul. L’algorithme génétique (AG) est l’une des métaheuristiques les plus utilisées dans les problèmes d'optimisation automatique du code. Les méthodes d’optimisation automatique basées sur l'algorithme génétique génèrent d’abord une population aléatoire d’individus. Ensuite, les différents opérateurs de l’algorithme génétique (mutation, croisement, sélection, etc.) sont appliqués sur ces individus pendant $n$ itérations. La sortie de cet algorithme est la meilleure combinaison d’optimisations explorées.
  \par\cite{ Cooper1999AG} ont utilisé l'algorithme génétique pour la sélection des optimisations avec la compilation itérative. \cite{ iterativeCompilationFactors} ont aussi proposé une technique itérative où ils ont utilisé l’algorithme génétique pour la sélection du facteur de tuilage et de déroulage des boucles et ils ont pu montrer que leur méthode peut fonctionner sur plusieurs différentes architectures. \cite{Jonathan2012Decouplingalgorithms} ont aussi utilisé l’algorithme génétique pour permettre l’optimisation automatique des programmes écrits dans le langage Halide. 
\par Cependant, les méthodes approchées peuvent dans certains cas donner des solutions qui ne sont pas proches de la solution optimale. Ceci est dû souvent au mauvais choix des hyperparamètres caractérisant les heuristiques et les métaheuristiques notamment. D'autre part, le choix de la méthode la plus appropriée au problème est souvent difficile.

\subsubsection{Approche d'apprentissage automatique}
\label{MLExploration}
\par L'apprentissage automatique est utilisé pour concevoir des modèles de résolution aux principaux problèmes d’optimisation de code : le choix de la meilleure optimisation à appliquer, l'estimation des paramètres pour les optimisations sélectionnées et le choix de l’ordre d’application des optimisations. Les méthodes basées sur cette approche consistent à construire des modèles de prédiction utilisant différentes classes d'algorithmes d'apprentissage automatique. Elles prennent en entrée les caractéristiques du code à optimiser et donnent en sortie la prédiction associée \cite{ Park2011IterativeCompilation}.
\subsubsubsection{Caractéristiques du programme (\textit{program features})} \label{featuresSec}
Pour construire le modèle d’apprentissage, le concepteur doit décider les caractéristiques les plus représentatives du programme et qui aident le modèle à apprendre mieux pour donner des bonnes estimations. Les caractéristiques sont représentées sous forme d'une structure (vecteurs, graphes) identifiant distinctement les programmes : plus les caractéristiques sont précises et détaillées plus elles sont représentatives. La représentation sous forme de graphe permet de mettre en relief les dépendances entre les instructions. Par exemple \cite{KosekiGraphe} ont utilisé les graphes comme structure de représentation pour trouver les bons facteurs de déroulage à appliquer sur les boucles. 
Or, la construction d'une structure volumineuse des caractéristiques est inefficace et peut ralentir les processus ML. Plusieurs projets basés sur l’apprentissage automatique utilisent des techniques d'extraction des caractéristiques des programmes : l'analyse statique des caractéristiques, l'analyse dynamique et l'analyse hybride \cite{Ashouri2018Survey}. 
\par \textbf{a) Extraction statique} : il s'agit d'extraire les caractéristiques d'un programme à partir de son code source seulement, sans prendre en considération les fonctionnalités du programme en cours de son exécution. Ces caractéristiques peuvent être simples comme le nom de la fonction où plus complexes telles que le nombre des opérations de chargement/stockage en mémoire dans les boucles. Il existe de nombreux extracteurs de caractéristiques de code source utilisés. Par exemple, le framework Milepost GCC \cite{milepost} est utilisé en tant qu’un plugin pour le compilateur GCC pour extraire les  caractéristiques du code source \cite{Ashouri2018Survey}.
\par
\begin{figure}[ht]
   \centering
   \subfloat[Ensemble des caractéristiques]{{\includegraphics[width=6cm]{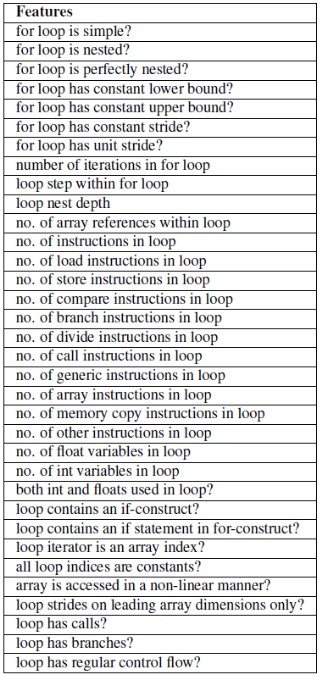} }}
    \qquad
    \subfloat[Ensemble des optimisations]{{\includegraphics[width=6cm]{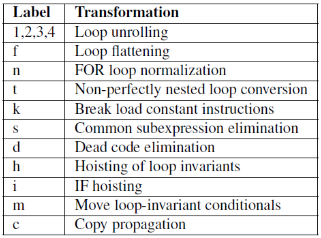} }}

    \caption{Ensemble des optimisations et des caractéristiques utilisées 
    dans la technique de~\protect\cite{Agakov2006}.}
 \label{fig:Agakov}
\end{figure}

 \cite{Agakov2006} focalisent sur l'optimisation de boucles, ils proposent 33 caractéristiques statiques pour décrire les boucles à optimiser (voir figure \ref{fig:Agakov} (a)), dans le but de sélectionner les cinq meilleures optimisations parmi un ensemble d'optimisations proposé (voir figure \ref{fig:Agakov} (b)). \par Cependant, la caractérisation statique ne décrit pas suffisamment le programme, certains détails relatifs à l'exécution du programme ne sont pas pris en considération alors qu'ils influencent sur le choix des optimisations. L'utilisation de la caractérisation statique génère des modèles qui ne sont pas très précis \cite{GrapheModeling}. 

\par \textbf{b) Extraction dynamique} : cette analyse consiste à collecter des informations sur le programme en cours d’exécution ce qui permet  de caractériser son comportement dynamique sur l’architecture d’exécution. Elle est aussi utilisée pour déterminer comment plusieurs ressources du système sont utilisées. Les machines actuelles offrent des registres pour extraire les caractéristiques des programmes au cours d’exécution : le comportement du cache (nombre de défauts de cache, nombre de succès de cache), le nombre d’erreurs de prédiction de branchement, etc. Pour pouvoir aboutir à cette analyse, il faut exécuter le programme plusieurs fois pour extraire les caractéristiques nécessaires ce qui est gourmand en termes de temps d'exécution. D'autre part, la nature des registres qui diffèrent d’une machine à une autre rend cette analyse non portable entre les plateformes d'exécution. De ce fait, des chercheurs ont proposé d’autres façons de collection dynamique qui peut être portable sur plusieurs plateformes à condition qu'elles disposent de la même structure de jeu d’instructions. Cette nouvelle façon de caractérisation s’appelle l’instrumentation et elle est réalisée à l’aide des outils d’analyse dynamique de programmes
\cite{Ashouri2018Survey}.
 \par \cite{Cavazos2007PerformanceCounters} ont proposé un modèle basé sur ML (un modèle basé sur l'apprentissage automatique hors ligne\footnote{Il se fait en temps de compilation.}) qui peut être utilisé pour prédire la bonne séquence d’optimisations. Leur méthode utilise cette analyse pour construire le vecteur de caractéristiques des programmes en entrée du modèle.

\par \textbf{c) Extraction hybride} : la caractérisation hybride consiste à combiner les deux techniques précédentes ce qui permet d’extraire plus d’informations sur le programme au cours d’exécution. Elle s’avère très intéressante car elle prend en considération les différents niveaux de caractérisation. \cite{GrapheModeling} ont utilisé la caractérisation hybride pour construire un modèle de prédiction qui aide à choisir efficacement une bonne combinaison d’optimisations. \cite{ Ashouri2018Survey} ont aussi utilisé le framework  MilePost \cite{milepost}  pour extraire les caractéristiques statiques et MICA\footnote{MICA est un outil pour l'extraction des caractéristiques dynamiques des programmes indépendantes de la machine comme  la moyenne d’accès aux registres par instruction, la prédiction de branchement.} pour extraire les caractéristiques dynamiques. 
\par Cependant, il n’est pas évident de connaître les caractéristiques à retenir et qui aident le plus pour choisir des bonnes optimisations. Parfois, des algorithmes d’apprentissage automatique (tel que l'analyse en composantes principales ) sont utilisés pour réduire la dimension des caractéristiques et se restreindre aux caractéristiques fondamentales pour la construction du modèle \cite{Ashouri2018Survey}.
 \subsubsubsection{Algorithmes d’apprentissage automatique}
 Plusieurs algorithmes d’apprentissage automatique sont utilisés pour concevoir les modèles de prédiction des bonnes combinaisons d’optimisations. Ils sont classés en trois grandes catégories : apprentissage supervisé, apprentissage non supervisé et autres méthodes (y compris l’apprentissage par renforcement, techniques basées sur les graphes et méthodes statiques).  Dans cette section nous allons exposer brièvement l’utilisation de certains algorithmes dde l'apprentissage automatique.
\par  \textbf{a) Apprentissage supervisé}
\par L’apprentissage supervisé permet l’apprentissage d’une fonction à partir des données étiquetées de l’ensemble d’apprentissage : le modèle reçoit un ensemble d’exemples étiquetés en tant que données d’apprentissage pour apprendre à déterminer les étiquettes de classe pour les nouvelles instances non vues \textit{(unseen points)}. Les modèles linéaires et les SVMs\footnote{SVM (Support Vector Machine ou Machine à vecteurs de support) appartient à la catégorie des classificateurs linéaires (qui utilisent une séparation linéaire des données).} ainsi que les arbres de décision et les forêts aléatoires sont les algorithmes les plus adoptés pour les problèmes de classification et de régression modélisant l'optimisation automatique du code.
\begin{itemize}[label=\textendash]
\item \textbf{Modèles linéaires et SVMs} : les modèles linéaires, à savoir la régression linéaire, l'algorithme des K-plus proches voisins, etc. représentent les méthodes d’apprentissage supervisé les plus populaires. Ils sont généralement très stables i.e les sorties n'enregistrent pas des fluctuations importantes par rapport aux modifications mineures apportées à l'ensemble d'apprentissage. D'autre part, les SVMs utilisent une fonction noyau pour assurer une transformation non linéaire des données vers un espace intermédiaire \textit{(feature space)}. Ceci permet d’appliquer une classification linéaire qui sépare les points vers les différentes classes. Ils sont utilisés dans la construction des modèles pour estimer si une optimisation est bénéfique pour le programme, dans ce cas la sortie appartient à la classe "vrai". Dans le cas contraire la sortie sera "faux" \cite{Ashouri2018Survey}.
	
\item \textbf{Arbres de décision et forêts aléatoires (\textit{Decision Trees and Random Forests})} : 
 \label{arbre_decision}
il s’agit de trouver un partitionnement des individus  à représenter sous la forme d’un arbre de décision. L’objectif est de produire des groupes d’individus les plus homogènes possibles du point de vue de la variable à prédire qui prend une valeur binaire. 
\par Les forêts aléatoires d'arbres décisionnels désignent une famille de méthodes de classification, composée de différents algorithmes d’induction d’ensemble d’arbres de décision. La méthode  commence par construire de nombreux arbres de décision au moment de l’apprentissage et fournit en sortie la classe correspondante dans le cas de la classification ou encore, la moyenne des classes dans le cas de la régression. Les forêts aléatoires viennent pour corriger le problème de surapprentissage des arbres de décisions. \cite{Monsifrot2002} ont utilisé des arbres de décision pour suivre le comportement de l’optimisation de déroulage afin de décider si l'application de l'optimisation de déroulage de boucles est bénéfique sur une architecture donnée\footnote{Sur les machines UlraSPARC et IA-64.}.
\\

\item \textbf{Réseaux de neurones artificiels} : les réseaux de neurones artificiels (RNA), ou \textit{Artificial Neural Network} en anglais, sont des réseaux constituent d'un ensemble de couches dont chacune est constitue de nombreuses unités élémentaires appelées les neurones. Chaque couche calcule une sortie sur la base des informations qu'elle reçoit. Les réseaux de neurones artificiels constituent, entre autre, une alternative intéressante aux statistiques traditionnelles pour le traitement des données. Les réseaux de neurones sont une façon de construire des modèles paramétriques, c'est-à-dire pour lesquels la fonction objectif est explicite. Contrairement à d'autres algorithmes paramétriques comme la régression linéaire, ils permettent de construire facilement des modèles très complexes et non linéaires.

\par Les réseaux de neurones sont souvent utilisés pour la construction de modèles de prédiction des optimisations de code. \cite{Kulkarni2012phaseOrderingANN} ont proposé une technique qui utilise la neuro-évolution (NEAT)\footnote{NEAT est un algorithme génétique pour la génération de réseaux de neurones artificiels (ANNs). Il représentent des modèles puissants pour l’apprentissage des problèmes complexes, car ils sont capables de changer la topologie du réseau et les paramètres de pondération pour trouver la fonction de coût (\textit{fitness}) la plus équilibrée.} pour construire un réseau de neurones artificiel capable de prédire un ordre des optimisations bénéfique pour une partie de code en cours d'optimisation. \\

\end{itemize}

\textbf{b) Apprentissage non supervisé}
\par L’apprentissage non supervisé regroupe l'ensemble des algorithmes d’apprentissage automatique permettant  d'identifier des groupes d’objets ou d’individus similaires (\textit{clusters}) à partir d’un ensemble de données sans en connaître au préalable la structure (à partir de données non étiquetées) ce qui les distinguent des algorithmes d'apprentissage supervisé. \\
Les méthodes de partitionnement de données (\textit{clustering}) représentent la classe des algorithmes les plus utilisés pour l’apprentissage non supervisé, elles permettent de construire des classes automatiquement en se basant sur un critère de similarité entre les données. Le \textit{clustering} est utilisé pour regrouper les nids de boucles indépendantes afin de préparer le programme à la phase d'optimisation. Il est également utilisé pour réduire l'espace d'optimisation. Par exemple, \cite{ Martins2016Clustering} proposent une technique basée sur \textit{le clustering}, elle consiste à  regrouper les fonctions d'un programme pour réduire l'espace d'exploration des combinaisons d’optimisations, pour chaque groupe, l'espace contient les optimisations précédemment suggérées pour les fonctions qu'il inclut. \\

\par Comme toutes les approches déjà exposées, les modèles basés sur le ML présentent des défis relatifs. D'une part, à leurs implémentations car la précision du modèle implique sa complexité. D'une autre part, des défis relatifs à la complexité de la phase d'apprentissage (\textit{training}) des données, à savoir le type et la masse importante des données et les problèmes du mauvais apprentissage (le surapprentissage\footnote{Le modèle s’adapte bien aux données de traitement (\textit{Training Set}) et se généralisera mal pour d'autres données non déjà vues.} ou encore le sous-apprentissage\footnote{Le modèle s’adapte mal aux données de traitement (\textit{Training Set}) et n’arrive même pas à capturer ses corrélations : le coût d’erreur en phase d’apprentissage reste grand.} du modèle).

\subsubsection {Comparaison entre les approches d'exploration d'espace des optimisations}

\par Dans le tableau \ref{tab:comparaisonApproches}, nous comparons entre les différentes approches en se basant sur leurs degrés de précision et le temps de calcule pris. Cette comparaison vise principalement à mettre en relief les avantages et les inconvénients de chaque approche. 

\begin{table}[h!]
  \centering
  \caption{Tableau comparatif entre les approches d’exploration d'espace d'optimisations.}
  \begin{adjustbox}{max width=\textwidth}
\begin{tabular}{|p{7cm}|p{7cm}|p{7cm}|}
\hline
\textbf{Critère} /\textbf{Approche} & \textbf{Degré de précision} & \textbf{Temps de calcul }\\ \hline
 Exhaustive & Précision la plus élevée & Exponentiel \\ \hline
Basée sur les heuristiques et les métaheuristiques & Précision moyenne comparée à l'exhaustive et dépend de la complexité du modèle & Le temps est réduit comparativement à l'approche exhaustive et dépend de la taille du problème \\ \hline
Basé sur le ML & Précision moyenne comparée à l'exhaustive et dépend de la complexité de l'algorithme et les hyperparamètres du modèle & assez réduit  \\ \hline
\end{tabular}%
\end{adjustbox}
\label{tab:comparaisonApproches}
\end{table}

\par L’approche exhaustive parcourt tout l’espace de recherche pour avoir la solution. Elle est simple, mais vu qu’elle prend beaucoup de temps, d’autres approches sont apparues. Les méthodes rapprochées (basées sur les heuristiques et les métaheuristiques) permettent de réduire l’espace de recherche par rapport à l’approche exhaustive. Cependant, elles sont lentes et ne garantissent pas de trouver la solution la plus optimale. L’approche basée sur l’apprentissage automatique vient pour permettre la génération des modèles d’une manière automatique grâce à l'apprentissage sur un grand ensemble de programmes pour assurer davantage de précision. 
\par Il est à noter que l'hybridation entre les approches permet de bénéficier à la fois de plusieurs avantages des approches \cite{Agakov2006}. Très souvent, il s'agit d'utiliser une approche pour résoudre une partie du problème comme la sélection ou l'ordre des optimisations, combinée avec une autre approche pour résoudre d'autres parties du problème comme l'estimation des bons paramètres pour les optimisations sélectionnées. 

\subsection{Approches de sélection des bons paramètres des optimisations }
\par
\begin{figure}[ht]
	\begin{center}
		\includegraphics[scale=0.90]{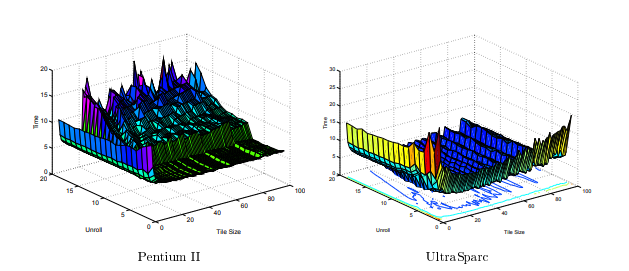} 
	\end{center}
	\caption[Variation du temps d'exécution en fonction des paramètres de tuilage et du déroulement de boucles.]{Variation du temps d'exécution en fonction des paramètres de tuilage et du déroulement de boucles~\protect\cite{iterativeCompilationFactors}.}
	\label{fig:ParamsFig}
\end{figure}

 Certaines optimisations de boucles nécessitent un ou plusieurs paramètres qui influencent sur l'efficacité de l'optimisation. La définition de ces paramètres dépend de différents critères, à savoir la hiérarchie mémoire, la complexité du cache, la taille et le nombre des registres vectoriels, etc. D'autre part, l'interdépendance et l'interaction entre les optimisations peuvent être renforcées à cause des paramètres choisis, la figure \ref{fig:ParamsFig}\footnote{Pour la multiplication de matrices $M \times M$ sur une architecture Pentium II et UltraSPARC.} montre qu’un léger écart par rapport aux "bons" paramètres des optimisations de tuilage de boucles et de déroulement peut entraîner une augmentation considérable du temps d’exécution voire même un ralentissement par rapport au programme initial \cite{iterativeCompilationFactors}. 

\par L'espace des paramètres des optimisations peut être intégré dans l'espace des optimisations, la technique d'optimisation automatique effectue alors une exploration générale de l'espace résultant, cette approche prend en considération l'interaction entre les paramètres des différentes optimisations. Cependant, l'espace résultant est considérablement grand. La technique risque d'être complexe et lente. De ce fait, les recherches de pointe se sont orientées vers une approche "séparatrice" entre l'espace des optimisations et l'espace de leurs paramètres, traitant ainsi séparément le problème de sélection/ordre des optimisations et le problème d'estimation des bons paramètres pour les optimisations sélectionnées. 
\par Des modèles ont été conçus afin de prédire le meilleur paramètre à utiliser pour une ou plusieurs optimisations en entrée, Les approches adoptées sont principalement : (1) l'approche empirique (estimation par exécution), (2) l'approche statique basée sur un modèle analytique et (3) l'approche basée sur l'apprentissage automatique. 
\subsubsection{Approche empirique (estimation par exécution)}
\par Dans cette approche, une exploration de l'espace des paramètres est effectuée et pour chaque solution une version du programme est générée, ensuite exécutée. Le paramètre permettant d'enregistrer le meilleur temps d'exécution est sélectionné. La bibliothèque d'optimisation automatique des programmes d'algèbre linéaire (ATLAS) effectue une recherche empirique afin d'obtenir le meilleur paramètre dans des plages de valeurs pour des optimisations pré-élaborées.
\par \cite{iterativeCompilationFactors} ont utilisé la compilation itérative afin de choisir le meilleur facteur de tuilage et de déroulage de boucles. Cependant, la taille de l'espace de recherche impose des limites sur cette approche. Par exemple, il est impossible de l'utiliser pour trouver le paramètre de tuilage sur des espaces rectangulaires car l'espace de recherche pour le tuilage de boucle rectangulaire est exponentiellement plus grand que celui de tuilage carré \cite{Yuki}. 

\subsubsection{Approche statique basée sur un modèle analytique}
\par Le modèle analytique est conçu en se basant sur l'architecture matérielle et de certains détails sur le comportement d'un ensemble de programmes. Ce modèle statique préconçu donne directement en sortie le paramètre d'une optimisation appliquée sur un programme donné pour une combinaison (architecture, compilateur) spécifique \cite{Yuki}. 

\par Les modèles visent principalement à améliorer l'utilisation de la mémoire cache et les registres. Par exemple, \cite{SarkarTileAnalytic} ont présenté un algorithme à temps constant qui estime le facteur de tuilage le plus optimal pour les nids de boucles. Ils ont formulé une fonction du coût sous forme d'une équation quadratique qui dépend de la taille du problème et de la taille du cache. Ils ont pu essayer tous les candidats pour les minima locaux de cette fonction en un temps constant. 
\par L'inconvénient majeur de cette approche est la nature statique des modèles analytiques. Les modèles analytiques sont développés sur la base d'une analyse détaillée du programme et de l'architecture. Lorsqu'un nouveau facteur de l'architecture est ajouté, les modèles doivent être mis à jour suite à une analyse exhaustive de ce nouveau facteur. Ceci est très coûteux, car cela nécessite des connaissances spécialisées et éventuellement change  toute la  conception du modèle \cite{Yuki}.

\subsubsection{Approche basée sur l'apprentissage automatique}
\par Dans cette approche, un modèle est conçu grâce à des algorithmes d'apprentissage automatique. Cette approche permet de mettre en place plusieurs techniques de pointe d'estimation des paramètres d'optimisations. 

\par Des modèles basés sur la classification utilisant l'algorithme des K-plus proches voisins (KNN) et les SVMs ont été proposés comme outils d'estimation du meilleur facteur de déroulage de boucles, le modèle considère un sous ensemble des paramètres après une restriction en se basant sur les caractéristiques du programme en entrée. \cite{Yuki} ont utilisé l'apprentissage supervisé, à savoir un réseau de neurones artificiel pour prédire le meilleur facteur de l'optimisation de tuilage. La technique prend en entrée un vecteur de caractéristiques du programme et donne en sortie l'estimation du meilleur paramètre du tuilage de boucles (voir la section \ref{TssYuki}). \cite{Xiaoming2008Matrix} ont proposé un système de classification basé sur le machine learning pour optimiser la multiplication de matrices. Leur système permet de déterminer le nombre de niveaux de tuilage et la taille de tuilage à chaque niveau selon la plateforme ciblée.

\subsection{Approches d'estimation de coût}
\par  L'estimation du niveau de performance atteinte, suite à l'application des optimisations, est un facteur décisif pour mesurer l'efficacité de ces optimisations sur le programme. Le coût  d'application d'une combinaison d'optimisations est relatif aux objectifs\footnote{Autres objectifs tels que la taille du code ou la consommation d'énergie.} visés notamment le temps d'exécution du programme optimisé qui représente le principal objectif définissant le coût  de l'application des optimisations. De ce fait, tout au long de cette rubrique, le coût  des optimisations fera référence exclusivement au temps d'exécution du programme optimisé.  

Pour estimer le coût, plusieurs approches ont été proposées, il n'existe pas de meilleure approche puisque chacune adopte un compromis entre certaines contraintes, à savoir la précision, le temps d'exécution et la complexité de l'approche d'estimation \cite{mendis2018ithemal}. 
\subsubsection{Estimation par exécution }
\par
\begin{figure}[ht]
	\begin{center}
		\includegraphics[scale=0.55]{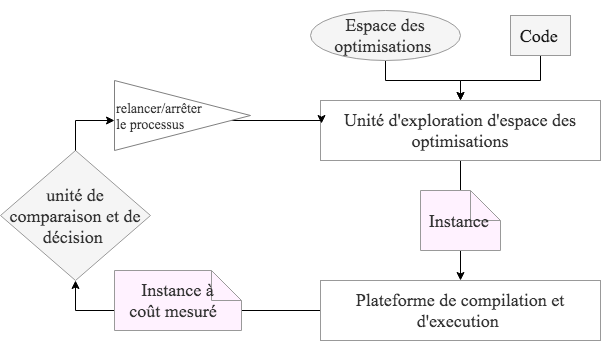} 
	\end{center}
	\caption{Principe général de l'approche d'estimation par exécution.}
	\label{fig:CoutExec}
\end{figure}

 Cette approche permet d'estimer le coût  en exécutant réellement le programme. Elle est souvent utilisée dans des domaines exigeant une précision assez élevée. Le principe de cette approche est simple : chaque instance de programme optimisé générée\footnote{Code avec ses optimisations.} par l'unité d'exploration d'espace des optimisations, sera compilée et exécutée pour mesurer réellement son temps d'exécution qui sera renvoyé vers une unité de comparaison et de décision afin de classer cette instance par rapport aux autres. Ce processus est répété autant de fois qu'une nouvelle instance est générée. Le critère d'arrêt peut être après un nombre prédéfini d'itérations ou après avoir atteint un coût  d'optimisation prédéfini \cite{Knijnenburg2002}. La figure \ref{fig:CoutExec} illustre le principe de cette approche. 

\par En théorie, cette approche peut être utilisée pour n’importe quel ensemble d’optimisations de compilateur. Cependant, il est recommandé de l'utiliser pour des espaces d'exploration qui ont été déjà réduits. Plusieurs techniques d'optimisation automatique adoptent cette approche telles que les techniques de compilation itérative\footnote{Les transformations successives d'optimisation sont appliquées sur un programme, leurs coûts sont déterminés par l'exécution réelle du code résultant. Plusieurs versions différentes du programme sont générées et exécutées, la version la plus optimale en termes de temps d'exécution est sélectionnée.} \cite{Knijnenburg2002}. Elle est aussi très utilisée pour estimer le coût  pour des parcours partiels d'espace de paramètres pour certaines optimisations, notamment les facteurs de tuilage de boucles et du déroulage de boucles \cite{iterativeCompilationFactors}. Elle est également utile dans les cas où l'architecture visée change, car cette stratégie n'impose pas de connaissances sur les plateformes d'exécution.
\par Cependant, cette approche impose des restrictions parfois bloquantes dans la pratique, les espaces de recherche peuvent être extrêmement grands, d'ailleurs c'est le cas le plus fréquent, ce qui impose un prétraitement de sélection pour restreindre l'espace de recherche. De ce fait, l'efficacité de cette approche dépend étroitement de la technique d'exploration de l'espace des optimisations. 

\subsubsection{Estimation basée sur des méthodes analytiques }
\par Cette approche est basée sur la conception d'un modèle analytique qui peut être un algorithme ou une fonction mathématique capable de prédire automatiquement le coup d'application d'une optimisation sans avoir besoin de l'exécuter.

\par
\begin{figure}[ht]
	\begin{center}
		\includegraphics[scale=0.5]{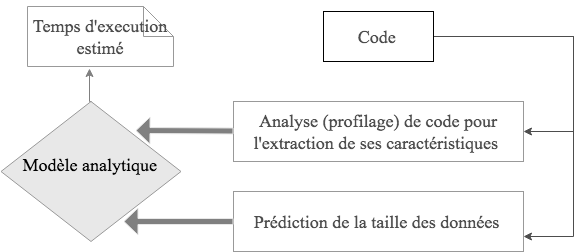}
	\end{center}
	\caption{Principe général de l'approche d'estimation du coût  utilisant le modèle analytique.}
	\label{fig:CoutAnalytique}
\end{figure}

 Le principe général d'un prédicteur analytique consiste à construire une fonction prenant en entrée un tuple $(F_P, T)$, où $F$ est le vecteur caractérisant le programme (soit $P$)  à optimiser, $T$ représente une des combinaisons d'optimisations possible à appliquer sur $P$. La sortie du modèle est le coût  prévu que la séquence T devrait apporter suite à son application sur le programme $P$  \cite{Ashouri2018Survey}. La figure \ref{fig:CoutAnalytique} résume le fonctionnement général de cette approche. 

\par Le modèle est synthétisé par des expressions mesurant le coût  en se basant sur les caractéristiques statiques ou dynamiques du code en entrée (voir section \ref{MLExploration} (\ref{featuresSec})) ainsi que la taille des données qui influence considérablement le modèle. Le temps d'exécution total est décomposé en général comme suit :
    $$T_{total} = T_{CPU} + T_{MEM} + T_{COMM} + T_{I/O}$$   
Où $T_{CPU}$ est le temps de calcul passé par le processeur lui-même, $T_{MEM}$ est le temps nécessaire pour accéder aux hiérarchies de la mémoire, $T_{COMM}$ est le temps de communication interprocessus (les threads) et $T_{I/O}$ est le temps écoulé pour effectuer les $E/S$. Chaque terme de cette équation constitue une formule mathématique exprimée en fonction de valeurs d’entrée du programme, de caractéristiques du programme et éventuellement de quelques paramètres caractérisant la machine cible en se basant, très souvent, sur la documentation du fournisseur \cite{Cascaval2000AnalyticTimeEstimation}.

\par Chaque sous-système de la formule principale est traité en détail. En effet, pour estimer $T_{CPU}$, le sous-système compte le nombre d'opérations exécutées dans les unités fonctionnelles du CPU. Les opérations ensuite sont regroupées en classes (blocs) dont l'estimation du coût  d'exécution est définie selon les modèles proposés des architectures d'exécution. Le terme $T_{MEM}$ est estimé en comptant le nombre des accès pour chaque niveau de la hiérarchie mémoire dont la pénalité est calculée selon différents modèles. Ces derniers prennent en considération les caractéristiques de chaque niveau de la hiérarchie mémoire et se basent essentiellement sur le microbenchmarking\footnote{Des programmes \textit{benchmarks} utilisés pour estimer les performances du compilateur et de la plateforme d'exécution.} \cite{Smith1995MemoryTimeMesuring}.
\par Par ailleurs, les architectures modernes ont des organisations internes très complexes. Les modèles de machine simplifiés, qui prennent en compte des abstractions des architectures réelles, fournissent des estimations de performances souvent très approximatives. En effet, plusieurs contraintes s'opposent à l'approche analytique :
 \begin{itemize}[label=\textendash]
\item \textbf{Extension des micro-opérations} : chaque instruction est étendue aux micro-opérations dans le processeur. Ainsi, les pipelines, les dépendances et la gestion des ressources se produisent au niveau des micro-opérations, ce qui représente un niveau plus granulaire que les instructions considérées dans les modèles analytiques. 

\item \textbf{Exécution dans le désordre et les architectures superscalaires} : les processeurs qui adoptent une exécution dans le désordre\footnote{En anglais \textit{out of order execution} consiste à réorganiser l'ordre dans lequel les instructions vont s'exécuter dans le processeur. Ces instructions ne sont alors pas forcément exécutées dans l'ordre dans lequel elles apparaissent dans le programme.} exploitent la structure de dépendance des données d'un bloc de base pour mapper son ordre  d'exécution optimisant le parallélisme des micro-opérations au niveau des instructions. Les unités d'exécution superscalaires\footnote{Un processeur est dit superscalaire s'il est capable d'exécuter plusieurs instructions simultanément parmi une suite d'instructions. Pour cela, il comporte plusieurs unités de calcul et il est capable de détecter l'absence de dépendances entre instructions.} permettent l'exécution simultanée de plusieurs instructions d'une même opération. Le problème d'estimation du coût  qui en résulte est donc non linéaire.
\item\textbf{Caractéristiques non spécifiées} : la documentation vague de certaines caractéristiques micro-architecturales pose un défi supplémentaire. Par exemple, lorsque l'exécution dans le désordre est spécifiée, la taille du tampon de réorganisation peut ne pas l'être. Ce qui mène à considérer des estimations vagues pour les modèles proposés \cite{mendis2018ithemal}.
\end{itemize}
\subsubsection{Estimation basée sur l'apprentissage automatique }
Cette approche permet de prédire le coût  des optimisations sans exécution du programme optimisé. En général, l'approche prend en entrée un tuple $(F, T)$ où $F$ est une structure caractérisant le programme (voir section \ref{MLExploration} (\ref{featuresSec})) et $T$ est l'une des séquences possibles d'optimisations à appliquer, la sortie ( soit $f_e$) est le coût  du tuple $(F, T)$ estimé grâce à un ou plusieurs algorithmes de l'apprentissage automatique.
\begin{figure}[ht]
	\begin{center}
		\includegraphics[scale=0.45]{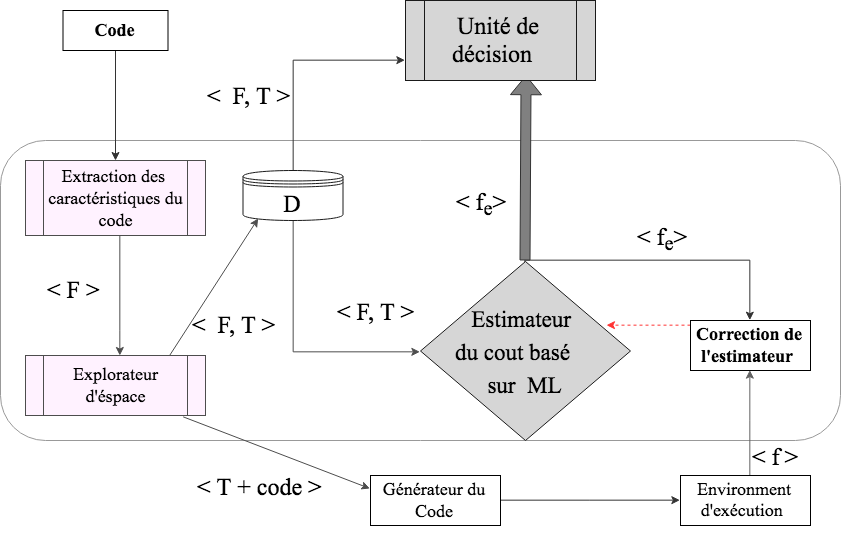}
	\end{center}
	\caption{Principe général de l'approche d'estimation basée sur l'apprentissage automatique.}
	\label{fig:CoutML}
\end{figure}
\par Pour concevoir l'estimateur, plusieurs algorithmes du machine learning ont été utilisés. Nous exposons une approche basée sur \textbf{les réseaux de neurones artificiels} qui ont donné de bon résultats d'estimation dans plusieurs techniques de pointe. 

\par Les modèles basés sur les réseaux de neurones exploitent les avantages qu'ils offrent. En effet, les réseaux de neurones permettent de prédire le temps d'exécution d'un programme en entrée dont les caractéristiques sont modélisées sous forme d'une structure (vecteur, graphe, etc.). La phase d'apprentissage (\textit{training}) permet au réseau d'apprendre en corrigeant ses différents coefficients.  L'ensemble de données d'apprentissage (\textit{Data\-Set}) est composé des couples de (caractéristiques de $P_i$, temps d'exécution de $T_i$) avec $P_i$ un programme généré aléatoirement (pour éviter le surapprentissage  ou encore le sous-apprentissage du modèle). 
\par
\begin{figure}[ht]
	\begin{center}
		\includegraphics[scale=0.45]{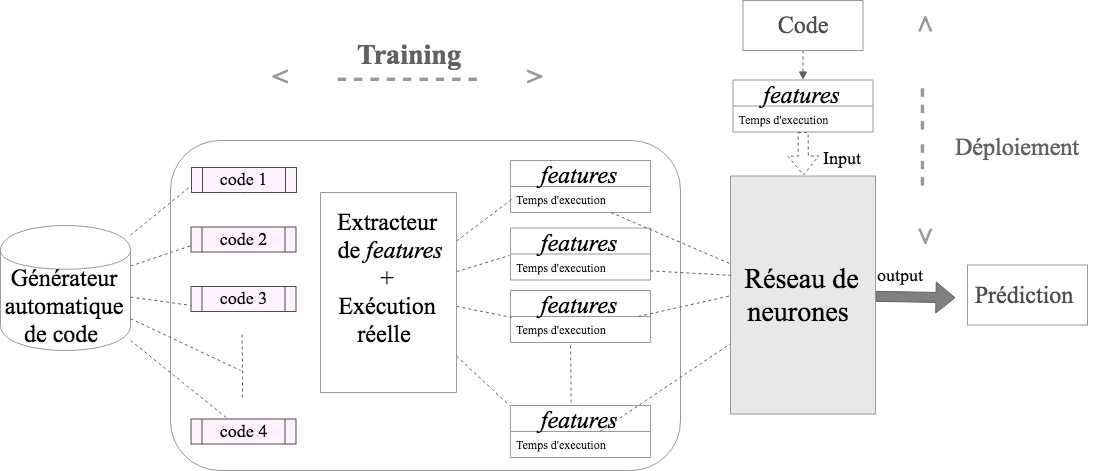}
	\end{center}
	\caption{Vue d'ensemble sur le modèle d'estimation du coût  basé sur les réseaux de neurones.}
	\label{fig:RNN}
\end{figure}

 Le passage d'une couche du réseau à une autre se fait grâce à des fonctions dont le choix influence sur les performances du modèle, la profondeur et le nombre des nœuds dans chaque couche présentent aussi des hyperparamètres critiques, ces deniers peuvent être réglés grâce à des tests de performance du modèle effectué pour chaque hyper-paramètre. La figure \ref{fig:RNN} résume les principales composantes du modèle.

\par \cite{Rahmani2010TSS}, dans leur technique de sélection du meilleur paramètre de l'optimisation du tuilage de boucles (TSS), ont utilisé les réseaux de neurones pour estimer le temps d’exécution des programmes. Le modèle de prédiction est constitué de trois couches. Il reçoit en entrée les caractéristiques du programme ainsi que les trois facteurs de tuilage et donne en sortie le temps d’exécution prédit du programme pour chaque facteur. Cette estimation permet donc de sélectionner le meilleur facteur de tuilage.

\par D'autres techniques de pointe basées sur l'utilisation des réseaux de neurones utilisent les réseaux de neurones récurrents (\textit{RNNs}) pour prédire le débit d’un ensemble d’instructions de bloc de base (voir la section  \ref{ithemal}). Le bloc d'instruction est représenté par un graphe orienté acyclique (DAG\footnote{\textit{Directed Acyclic Graph} un graphe orienté qui ne possède pas de circuit. Un tel graphe peut être vu comme une hiérarchie.}). Ce mécanisme de représentation est inspiré des techniques de traitement de langage naturel.
\par Les modèles basés sur l'apprentissage automatique ont pu atteindre des performances remarquables. Cependant, ces modèles sont très sensibles aux hyperparamètres qu'ils utilisent, les ensembles de données d'apprentissage (problème de sous- apprentissage et surapprentissage) ainsi que la définition des caractéristiques représentatives du programme ( \textit{Features engineering}). La conception de ces modèles doit être méticuleusement étudiée au niveau de chaque phase.  
\subsubsection{Comparaison entre les approches d'estimation de coût}
\par Afin de comparer les trois approches d'estimation du coût, il faut préciser les critères à considérer. Il s'agit de la précision, le temps d'exécution et la complexité de l'estimateur.
 \begin{itemize}[label=\textendash]

\item  \textbf{La précision} : l'estimateur du coût doit être précis. Tout le processus du choix des optimisations dépend de cette estimation. En effet, à une échelle d'ordre aussi petite, l'erreur doit être minimale car si l'estimateur présente un taux d'erreur important, la prédiction des bonnes optimisations sera erronée et la méthode d'automatisation sera par la suite fausse.  
\item \textbf{Temps d'exécution de l'estimateur} : L'estimateur doit être rapide. En effet, un estimateur fait partie de tout un processus d'automatisation. L'exploration d'espace de recherche, qui est assez grand, risque de prendre un temps important. Si de plus, l'exécution de l'estimateur prend un temps considérable, la méthode d'automatisation sera trop lente et inutile.
\begin{comment}
\item \textbf{Portabilité} : L'estimateur doit être portable, le changement de l'architecture d'exécution du programme ne doit pas influencer la précision de l'estimation. Pour adapter manuellement l'estimateur aux différentes architectures, le modèle risque de perdre son efficacité d'autant plus que plusieurs caractéristiques changent d'une architecture à une autre. 
\end{comment}

\item \textbf{Complexité de l'estimateur} : l'estimateur ne doit pas être trop complexe, c'est-à-dire que l'estimateur ne doit pas dépendre de plusieurs prétraitements et paramètres qui risquent d'influencer sa précision. D'autre part, la complexité d'un estimateur implique l'intervention de plusieurs contributeurs et nécessite très souvent des outils plus complexes.
\end{itemize}

\par  Pour concevoir une méthode d'optimisation automatique, le choix de l'approche d'estimation du coût se fait selon les contraintes favorisées. La taille de l'espace des optimisations à appliquer et leurs paramètres peuvent orienter le concepteur. Ceci impose des limitations sur la précision de l'approche. De ce fait, plusieurs travaux ont adopté une phase de présélection pour réduire l'espace de recherche \cite{OptimSpace}. Le tableau \ref{tab:CoutCompar} résume la comparaison entre les trois approches d'estimation de coût.

\begin{table}[h!]
  \centering
  \caption{Comparaison entre les trois approches d'estimation du coût.}
  \begin{adjustbox}{max width=\textwidth}
\begin{tabular}{|p{4.3cm}|p{4cm}|p{4cm}|p{4cm}|}
  \hline 
  \textbf{Approche / critère} & \textbf{La précision}  & \textbf{Temps d'exécution } & \textbf{complexité } \\ \hline \hline
 Par exécution & Très élevée et présente un référentiel pour les autres approches  & Très élevé dans le cas de grands espaces d'optimisation & Simple à concevoir   \\ 
 \hline
  \mbox{Basée sur des méthodes} analytique & Moins élevée et dépend de la complexité de la méthode et la  définition des modèles d'architectures &  Très réduit & Plus le modèle est précis plus il est complexe \\ 
   \hline
 Basée sur l'apprentissage automatique & Elevée mais dépend de la complexité du modèle  &  réduit & Complexe(dépendance aux hyperparamètres, de données d'apprentissage, etc.) \\
  \hline
\end{tabular}%
\end{adjustbox}
\label{tab:CoutCompar}
  \end{table}
  
\section{Exemples de techniques d'optimisation automatique}
\par Plusieurs techniques d'optimisation automatique ont été intégrées dans les compilateurs. Elles ont permis d'apporter des améliorations remarquables sur leurs fonctionnements et ce, sur différents niveaux, à savoir l'accélération de l'exploration d'espace des optimisations, la sélection des meilleurs paramètres des optimisations et l'estimation du coût des programmes optimisés. Dans cette section, nous allons exposer trois techniques. Chacune propose une solution pour automatiser un de ces trois niveaux.   

\subsection{Autoscheduler de Halide basé sur l’approche analytique } \label{auto-schedulerHlide}
\par
\begin{figure}[ht]
	\begin{center}
		\includegraphics[scale=1]{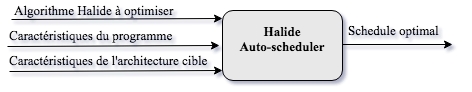}
	\end{center}
	\caption{Les entrées et la sortie de l’Auto-Schedule de Halide.}
	\label{fig:AutoSchedule}
\end{figure}

 L’auto-scheduler \cite{Mullapudi2016HalidAutoScheduler} est une technique de génération automatique des \textit{Schedules}\footnote{Équivalent à "Planning" en français, ce terme représente l'ensemble des optimisations à appliquer sur le code Halide.}. Elle permet de résoudre le problème du temps perdu lors de la compilation et d’exécution de chaque \textit{schedule} candidat pour un espace de recherche immense. Il s'agit alors d'une estimation des performances de chaque \textit{schedule}. En entrée, l’auto-scheduler reçoit le programme à optimiser, des informations complémentaires sur le programme tel que l’étendue des boucles de chaque fonction et la taille estimée des images ainsi que l’architecture de la machine (taille du registre vectoriel, coût du chargement mémoire) (voir la figure \ref{fig:AutoSchedule}). 

\par Le processus de l’Auto-scheduler passe par quatre phases. D’abord, la phase de pré-calcul de chaque fonction, à savoir l'estimation de son coût  arithmétique. Ensuite, l’Auto-scheduler forme des groupes de fonctions\footnote{Une fonction Halide est équivalente à une boucle imbriquée dont la profondeur est supérieure ou égale au nombre de variables manipulées par cette fonction.} liées par des relations producteur-consommateur\footnote{Le résultat de calcul de la fonction productrice est utilisé dans le calcul de la fonction consommatrice.}, chaque groupe est optimisé indépendamment afin d’augmenter la localité et maximiser la réutilisation de données dans chaque groupe. Il s'agit d'appliquer des transformations de tuilage sur chaque fonction et de définir les points optimaux pour fusionner les fonctions producteur-consommateur. 
Enfin, il faut définir l’ordonnancement des groupes pour assurer une meilleure localité. Puis, dérouler les fonctions, les vectoriser et enfin paralléliser les boucles externes des fonctions dans chaque groupe. L'algorithme suivant  résume les phases du modèle de l’auto-scheduler \cite{Mullapudi2016HalidAutoScheduler}.

\begin{algorithm}[h!]
\SetAlgoLined
\KwResult{Schedule du programme P}
G = \{\}\;
\For {(fonction $f_i$ dans $P$)}{ 
Tuiler $f_i$ et la mettre dans un groupe singleton $g_i$\;
 $G = G \cup g_i$\;
}
 \While{(Il existe des groupements potentiels avec un gain dans $G$)}{
Soit $(g1, g2)$ le groupement potentiel avec le plus petit coût  (plus grand gain)\;
coût \_fusion $=$ le coût  de la fusion de $g1$ et $g2$ \;
coût \_Non\_fusion $=$ le coût  lors de l’absence de fusion entre $g1$ et $g2$ \;
\eIf{ (coût \_fusion $>$ coût \_Non\_fusion) }{
Ne pas fusionner $g1$ et $g2$\;}{
 Fusionner$g1$ et $g2$\;}
 Mettre à jour $G$ avec $g1$ et $g2$ dans le même groupe\;
\For{ (groupe $g_i$ dans $G$) }{
 Soit $f_i$ la fonction de sortie de $g_i$\;
Choisir la meilleure interversion de boucle pour $f_i$ : celle qui améliore la réutilisation des données dans $f_i$\;
Dérouler et vectoriser deux niveaux de boucles internes de petite étendue de $f_i$\;
 Paralléliser le niveau de boucle externe de $f_i$\;
 }
}
\caption{Pseudo algorithme de l’Auto-Scheduler}
\end{algorithm}
\begin{figure}[H]
	\begin{center}
		\includegraphics[scale=0.55]{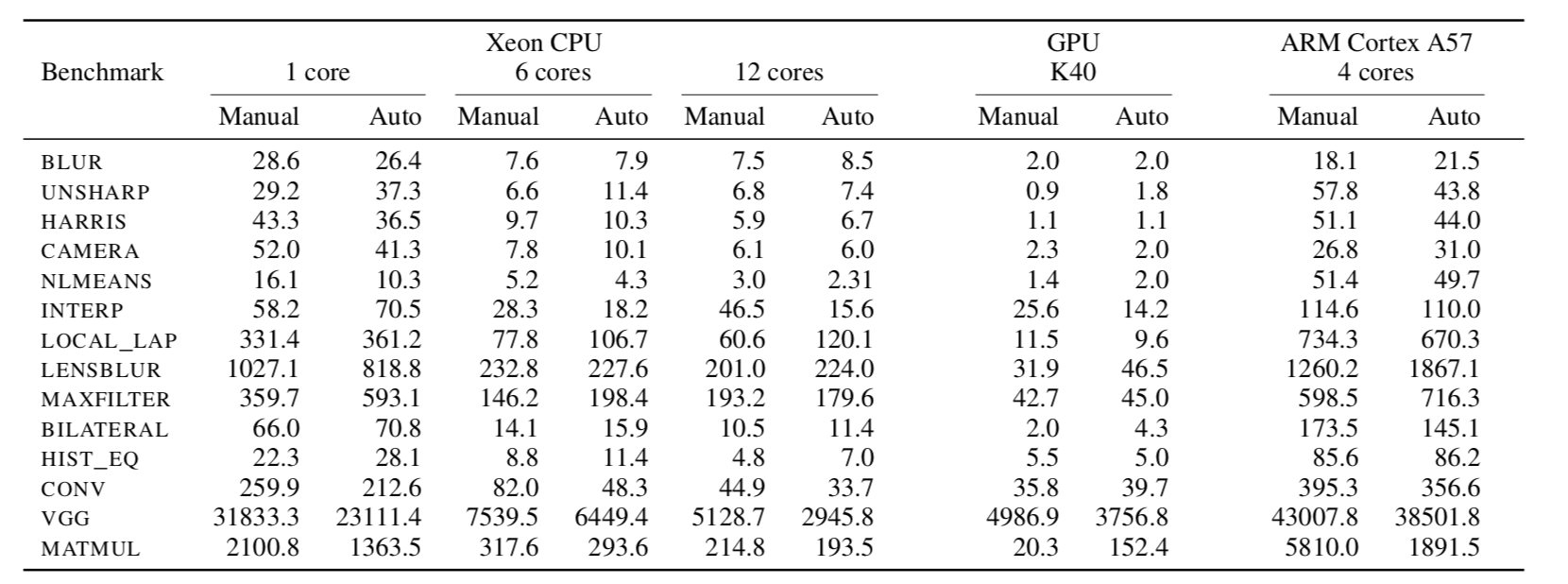}
	\end{center}
	\caption[Comparaison entre les temps d'exécution de quelques benchmarks optimisés manuellement et par Auto-Schedule de Halide.]{Comparaison entre les temps d'exécution de quelques benchmarks optimisés manuellement et par Auto-Schedule de Halide~\protect\cite{Mullapudi2016HalidAutoScheduler}.}
	\label{fig:autoscheduleResutls}
\end{figure}

\par L’Auto-Scheduler de Halide a été testé par un ensemble de 14 benchmarks différents sur la même architecture d’exécution\footnote{Intel Xeon E5-2620 v3 CPU.}. Dans 8 parmi 14 des benchmarks, la version optimisée à la main a donné des performances meilleures que celle optimisée par l’Auto-Scheduler, mais les écarts entre les performances sont réduits, l’Auto-Scheduler reste très compétitif à l'optimisation manuelle (voir la figure \ref{fig:autoscheduleResutls}).

\subsection{Modèle automatique pour le problème de \textit{Tile Size Selection} (TSS)} 
\label{TssYuki}
\par Une bonne estimation des paramètres des optimisations améliore profondément l'accélération qu'elles peuvent apporter. \cite{Yuki} présentent une technique de création des modèles assez précis de prédiction des meilleurs paramètres pour l'optimisation de tuilage de boucles (\textit{TSS})\footnote{\textit{Tile Size Selection} est le problème de sélection du meilleur paramètre pour l'optimisation de tuilage des boucles.} en se basant sur les réseaux de neurones artificiels. La technique a été conçue pour une classe de programmes spécifiques souvent utilisés dans l'algèbre linéaire (des nids de boucles d'une profondeur allant jusqu'à trois dimensions et des matrices de données à deux dimensions). Le principe de cette technique est résumé dans les deux sections suivantes.
\subsubsection{Caractéristiques du code considérées}
\par Le tuilage de boucles vise principalement à maximiser la réutilisation des données et à éviter les accès mémoire coûteux. La technique prend en entrée des caractéristiques des programmes qui décrivent la localité temporelle et spatiale en s'appuyant sur l'aspect du pré-chargement (\textit{prefetching}) qui consiste à récupérer les données avant la demande en parallèle avec le chargement des données voisines, ce qui permet de gagner un temps d'exécution considérable.
\par Les caractéristiques sont basées sur le nombre de références dans les instructions les plus internes du nid de boucle. Les références sont classées en trois types : les références pré-chargées\footnote{(\textit{Prefetched references}) qui sont les références bénéficiant de la localité spatiale créée grâce au prefetcher.}, les références non pré-chargées\footnote{(\textit{Non-prefetched references.}), à savoir les références qui ont besoin de la localité temporelle pour obtenir de bonnes performances.} et les références constantes dans la boucle la plus interne (invariant), c'est-à-dire celles qui sont réutilisées pour toutes les itérations de la boucle la plus interne. Chaque type de référence est ensuite classé selon son mode d'accès : en lecture ou en écriture, ce qui fait un total de six types de caractéristiques. 
\subsubsection{Phases de conception et principe de fonctionnement} 
\par La technique utilisée comporte les quatre étapes suivantes : 
  \begin{itemize}[label=\textendash]
\item  Génération aléatoire des programmes correspondant à la classe de programmes définie. 
\item Collecte de données qui consiste à extraire les caractéristiques des programmes et les exécuter pour concevoir l'ensemble d'apprentissage, et ce, sur différentes architectures. 
\item Entraînement du modèle TSS grâce à l'ensemble d'apprentissage en réduisant l'erreur de prédiction. Les entrées du réseau de neurones sont données à la première couche cachée et les sorties de chaque couche sont données à la couche suivante. Les sorties des couches cachées sont en fonction de la somme pondérée des entrées de cette couche, où la fonction est la tangente hyperbolique. La couche de sortie effectue la somme pondérée des sorties de la dernière couche cachée, mais n'applique pas la fonction de tangente hyperbolique. Compte tenu des pondérations $w$ (coefficients synaptiques) de la couche de sortie, de la dernière couche cachée $h$, de la sortie souhaitée $b$ et du nombre de données d'apprentissage $N$, chaque nœud de la couche de sortie tente à minimiser l'erreur calculée par de l'équation $ \sum_{n=1}^N ((h.w)_n - b_n)^2$. 
\item Une fois le modèle TSS conçu, il peut être utilisé en tant que partie du compilateur pour prédire les tailles de tuilage optimales sur le plan interne, ou en tant que partie d’une méthode d'exploration pour trouver les tailles de tuilage optimales.
\end{itemize}

\subsection{Ithemal, estimateur du coût} \label{ithemal}
\par Ithemal (\textit{Instruction Throughput Estimator using Machine Learning}) est un estimateur de débit d’un ensemble d’instructions de bloc de base à l’aide de l'apprentissage automatique, il peut être intégré dans les compilateurs afin d'estimer le temps d'exécution pour choisir les meilleures optimisations du code.
\par Ithemal utilise une nouvelle approche basée sur un réseau de neurones récurrents sous forme de graphe acyclique dirigé (\textit{DAG}-\textit{RNN}\footnote{\textit{Recurrent Neural Networks.} en anglais, c'est une classe de réseaux de neurones artificiels ayant des connexions récurrentes, qui dote le réseau de mémoire.}). Il modélise l'estimation du débit à l'aide d'un réseau de neurones profonds (\textit{DNN}). Il adopte donc une approche guidée par les données\footnote{\textit{Data Driven Approach} en anglais, regroupe les différentes approches de résolution qui se basent sur l'étude des données.} ce qui permet de basculer facilement d’une microarchitecture à une autre avec un minimum de paramétrage manuel \cite{mendis2018ithemal}.
\subsubsection{Architecture et fonctionnement d'ithemal}
\par
\begin{figure}[ht]
	\begin{center}
		\includegraphics[scale=0.9]{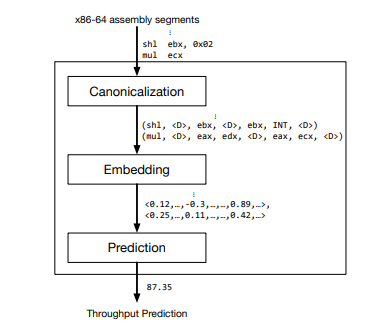}
	\end{center}
	\caption[Architecture globale d'Ithemal.]{ Architecture globale d'Ithemal~\protect\cite{mendis2018ithemal}.}
	\label{fig:IthemalStructure}
\end{figure}

 La technique permet d'estimer le débit d'un bloc de base d'instructions en assembleur. Les blocs passent par trois principales phases : la canonicalisation, le plongement et la prédiction (voir la figure \ref{fig:IthemalStructure}).

\begin{figure}[ht]
	\begin{center}
		\includegraphics[scale=0.8]{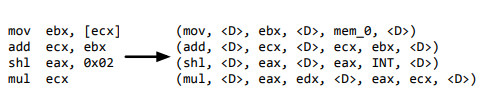}
	\end{center}
	\caption[Phase de la canonicalisation d'Ithemal.]{ La phase de la canonicalisation d'Ithemal~\protect\cite{mendis2018ithemal}.}
	\label{fig:IthemalCode}
\end{figure}

\subsubsubsection{canonicalisation (\textit{canonicalization})} Dans cette phase, Ithemal prend un bloc en assembleur spécifié en tant que texte (syntaxe Intel) et le mappe en tant que texte aussi à une liste d'instructions. Chaque instruction consiste en un code opération, une liste d'opérandes de source et une liste d'opérandes de destination. La canonicalisation rend explicite les opérandes qui sont généralement implicites dans la représentation en code assembleur. La figure \ref{fig:IthemalCode} expose un exemple de cette transformation. 
\subsubsubsection{Plongement (\textit{Embedding})} Cette phase prend un bloc de base canonique et produit une représentation du bloc de base compréhensible par un réseau de neurones. En effet, les réseaux de neurones prennent typiquement en entrée une séquence d’entrées à valeurs réelles. Dans les domaines structurés, tels que le texte ou, comme dans notre domaine, les programmes et les entrées sont de nature discrète (tels que les mots et les blocs de base). Il est donc nécessaire de mapper chaque entrée structurée sur une représentation pouvant être consommée par un réseau de neurones. Ithemal mappe un bloc de base sur un graphe acyclique dirigé avec des vecteurs à valeurs réelles comme contenu de chaque nœud. Chaque nœud du graphe correspond à une instruction du bloc de base. Chaque nœud est associé à un vecteur (à n dimensions) à valeur réelle. Un arc dirigé lie un nœud $n_1$ à un nœud $n_2$ si l'instruction de $n_2$  dépend de $n_1$  (déterminée grâce à l'analyse des opérandes de source de $n_2$ et des opérandes de destination de $n_1$). La création du vecteur à $n$ dimensions pour chaque nœud est basée sur la technique de traitement de langage naturel pour mapper une séquence de jetons textuels (en l’occurrence le code opération, les opérandes de source et les opérandes de destination)  en une représentation détaillée sous la forme d’un vecteur \cite{mendis2018ithemal}.

\subsubsubsection{Prédiction} La phase de prédiction prend en compte le graphe acyclique dirigé d’un bloc de base et prédit son débit. Les réseaux de neurones récurrents (DAG-RNN) sont les plus adéquats pour cette structure en entrée. Le DAG-RNN d’Ithemal parcourt le graphe dans l’ordre topologique, en calculant une représentation vectorielle profonde à valeur réelle de chaque sous-graphe connexe dans le graphe. Étant donné les vecteurs de chaque sous-graphe, Ithemal réduit ces vecteurs en un seul vecteur, puis effectue la prédiction à l'aide d'une régression linéaire. 

\subsubsection{Utilisation du réseau de neurones récurrents (\textit{RNN})} Un RNN prend en entrée une séquence de vecteurs et produit en sortie une séquence de vecteurs. Un RNN est dit récurrent si son exécution est définie de manière récursive tout au long de la séquence. Les nœuds interconnectés interagissent non-linéairement. Les unités sont reliées par des arcs (synapses) qui possèdent un poids. La sortie d'un neurone est une combinaison non linéaire de ses entrées. À chaque étape, le RNN applique une cellule\footnote{\textit{Cell} en anglais  est une fonction avec des paramètres internes apprenables qui, lorsqu’ils sont évalués à la position $i$ dans la séquence, consomment un vecteur d’entrée $v_i$ et un vecteur d’état caché  $h_{i-1}$, à partir de la position précédente, puis produisent un nouveau vecteur caché (vecteur d'état $h_i$).} pour produire le vecteur de sortie de cette étape. Le calcul des cellules dépend du vecteur de sortie de l'étape précédente. Le dernier vecteur qu'un RNN calcule résume donc la séquence entière. Dans son implémentation, Ithemal utilise une cellule LSTM\footnote{\textit{Long Short-Term Memory (LSTM)} est un type de réseaux de neurones récurrents.} \cite{ LSTM} qui mémorise de manière sélective les informations qui lui ont été transmises par la cellule précédente \cite{mendis2018ithemal}.
\par La dernière étape d’Ithemal consiste à calculer l'estimation de débit à l’aide d’une régression linéaire. Etant donné un vecteur d'état caché réduit $h_{DAG}$ obtenu après la réduction de la sortie du RNN,  Ithemal calcule la prédiction du débit par la formule  $w  \times h_{DAG} + b$, où $w$ est le vecteur de paramètres et $b$ est le biais. 

\section*{Conclusion}
\addcontentsline{toc}{chapter}{\textsc{Conclusion}}
\par Dans ce chapitre, nous avons expliqué les différentes approches adoptées pour résoudre les trois principaux problèmes liés à l'optimisation automatique, à savoir la sélection des meilleures combinaisons d'optimisations, l'estimation des meilleurs paramètres pour les optimisations sélectionnées et l'estimation du coût  d'application des optimisations (le temps d'exécution en l'occurrence).

\par Chacune des approches présente des avantages et des défis. Une bonne définition des spécifications du problème donne plus de priorité à certaines contraintes, ce qui oriente le choix de l'approche à utiliser.  
\par Les techniques d'optimisation automatique du code dépendent très souvent du compilateur visé. Dans la partie suivante, nous allons exposer le compilateur Tiramisu qui fera sujet de notre contribution par la suite. 

%=========================================================================
% chapitre 3    

\chapter{\textsc{Le compilateur Tiramisu}}
\label{ch:chapterThree}
\begin{onehalfspace}
\setcounter{footnote}{0}
\section*{Introduction}
\addcontentsline{toc}{chapter}{\textsc{Introduction}}
\par Générer des codes efficaces dédiés aux systèmes de hautes performances devient de plus en plus difficile. Ceci dû aux contraintes d'hétérogénéité des plateformes d'exécution d'une part et de la complexité des algorithmes utilisés d'une autre part.  

\par Afin de remédier à ces difficultés, maints langages spécifiques au domaine (LSD) ont été proposés. Ces langages définissent des spécifications pour répondre en particulier aux contraintes d’un domaine d'application précis. 

Tiramisu est un Langage Spécifiques au Domaine développé par l'équipe de recherche COMMIT de MIT \cite{Baghdadi2018TiramisuV3} capable de générer des codes optimisés dans lesquels les optimisations n’affectent pas le fonctionnement ni la lisibilité du code grâce à la séparation entre les algorithmes et les optimisations appliquées. 
\par Dans ce chapitre nous allons présenter le langage et le compilateur Tiramisu en exposant ses avantages. Nous donnons une vue d'ensemble sur la compilation des codes en Tiramisu afin de mettre en relief l'effet de la représentation du code sur son optimisation et aussi sur sa portabilité. Ensuite, nous expliquons sa logique en exposant comment les algorithmes sont définis et comment les optimisations sur le code sont introduites. 
\section{Tiramisu} 
\par Tiramisu est un LSD (Languages Spécifiques au Domaine) embarqué\footnote{Langage spécifique à un domaine défini en se basant sur un "langage hôte" plus puissant à usage général. il s'appuie sur l'infrastructure du langage hôte (analyse syntaxique, vérification typographique, modularité), le langage hôte peut être utilisé pour la métaprogrammation (l'écriture de programmes manipulant des programmes en DSL).} sur le C++ permettant d'exprimer des algorithmes dits de données parallèles\footnote{Ces algorithmes sont appelés des algorithmes  de données parallèles car leur parallélisme provient d'opérations simultanées sur de grands ensembles de données, plutôt que de plusieurs threads de contrôle.} qui utilisent des tableaux denses et des nids de boucles. Ces algorithems sont souvent utilisés dans des systèmes de hautes performances, à savoir l'algèbre linéaire dense, l'algèbre tensorielle, le traitement d'images et les réseaux de neurones (RNCs). Tiramisu a été conçu afin de couvrir quatre principales caractéristiques d'optimisation de codes \cite{Malek2017ExtendingTiramiu} qui sont :
 \begin{itemize}[label=\textendash]
	\item Cibler différentes architectures matérielles.
	\item Gérer la dépendance de données. En effet l'optimisation d'un programme est limitée par les dépendances entre les représentations des données en mémoire, notamment dans le cas des programmes ciblant différentes architectures.
	\item Générer un code optimisé de hautes performances. En effet, les programmeurs doivent optimiser leurs codes manuellement ou encore automatiquement afin d'obtenir  des résultats comparables aux codes optimisés soigneusement par des experts du domaine d'optimisation. 
   \item Assurer une représentation compréhensible et lisible du code optimisé. Très souvent, l'application des optimisations sur un programme risque d'affecter sa représentation, obombrer sa logique voire même la changer, d’où la nécessité d'une vérification du programme après optimisation.
\end{itemize}

\subsection{Modèle en couche de Tiramisu}
\par Tiramisu est basé sur le modèle polyédrique\footnote{Permet d'avoir une représentation mathématique abstraite pour modéliser
un programme. Chaque instruction du programme est représentée par 3 principales informations : domaine d'itération, relations d'accès (en lecture, écriture) et un \textit{Schedule}. Pour davantage de détails voir \cite{PRADELLE2011polyedrique}.}\cite{polyhedral}. Il utilise une représentation intermédiaire  (RI)\footnote{Représentation intermédiaire que prend le code de haut niveau.} multicouche assurant une séparation complète entre l'algorithme pur, les optimisations de code, le mappage et la structuration des données ainsi que la gestion de la communication et la synchronisation.
L'avantage de cette représentation gît essentiellement dans la séparation entre les couches, ce qui définit un ordre spécifique dans lequel les optimisations sont appliquées. Cette représentation garantit que le compilateur passe d'une couche à une autre sans vérifier les modifications ou annuler des décisions déjà prises dans des couches précédentes.
\par En effet, l'optimisation de code pour différentes architectures est restreinte par certains facteurs liés notamment à la dépendance en mémoire. Toutes les opérations de synchronisation, de communication et de mappage des données vers les différentes hiérarchies mémoire ne doivent pas être effectuées avant l'application des optimisions de code. Pour illustrer cette complexité, prenons l'exemple du mappage des buffers vers la mémoire partagée et la mémoire cache des GPU, les quantités de données à transmettre et la synchronisation des opérations dépendent étroitement des optimisations de code appliquées comme les niveaux de l'optimisation de tuilage de boucles. 

\par Tiramisu sépare la représentation du code en quatre couches que nous détaillons dans les sections suivantes.
 
\subsubsection{Couches d'algorithme abstrait}
\par Dans cette première couche, l'algorithme pur est spécifié indépendamment de la localité temporelle et spatiale : aucun ordre des calculs n'est donné et aucune restriction sur le stockage des données n'est imposée. Les valeurs sont communiquées grâce à une relation producteur-consommateur. 
 
\subsubsection{Couche de gestion des \textit{computations}}
\par Cette couche permet de définir l'ordre d'exécution des calculs, l'architecture d'exécution cible ainsi que les optimisations à appliquer sans pour autant préciser la structuration de données. Ceci, facilite considérablement l'application des différentes optimisations. En effet, les transformations requises pour l'application des optimisations sont plus souples puisque elles ne nécessitent pas des transformations complexes sur la structure de données.
\subsubsection{Couche de gestion de données} 
\par Au niveau de cette couche, les emplacements mémoire pour stocker des valeurs calculées sont définis grâce aux commandes de mappage des données, à savoir l'allocation/libération des tampons et les relations d'accès aux données en lecture ou en écriture pour chaque opération de calcul. Les mappages de données possibles en Tiramisu sont représentés par des structures de tableaux, des tableaux de structures et des tableaux multidimensionnels réduits en tableaux de dimensions minimales ou en scalaires.
\subsubsection{Couche de gestion de communication}
\begin{figure}[ht]
	\begin{center}
		\includegraphics[scale=0.5]{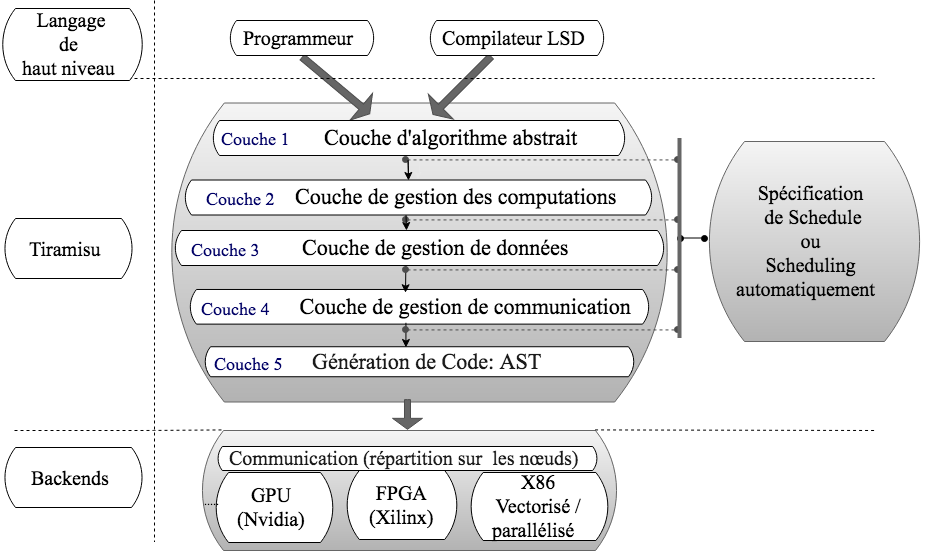}
	\end{center}
		\caption[Vue d'ensemble sur Tiramisu.]{ Vue d'ensemble sur Tiramisu ~\protect\cite{Baghdadi2018TiramisuV3}.}
	\label{fig:exmp_Tiramisu_overView_links}
\end{figure}

Les commandes de synchronisation et de communication sont ajoutées au niveau de cette couche. Il s'agit d'annoter d'une dimension de temps la représentation obtenue après la couche de gestion de données. Ceci est réalisé grâce à des commandes de synchronisation spécifiées manuellement par l'utilisateur. La Figure \ref{fig:exmp_Tiramisu_overView_links} donne une vue globale sur Tiramisu en mettant en relief le flux entre les différentes couches de la représentation intermédiaire en Tiramisu. 

\subsection{Avantages de Tiramisu}
\par Le compilateur Tiramisu permet une application plus souple des différentes optimisations de code et sur des niveaux de code séparés, spécifiques et ordonnés \cite{Malek2017ExtendingTiramiu}. Ceci est grâce à la séparation entre l'algorithme et les optimisations à appliquer sur le code. De ce fait, Tiramisu offre la possibilité de tester les différentes combinaisons d'optimisations sur le même code ou encore d'automatiser cette exploration, ce qui n'est pas du tout facile dans d'autres langages comme le langage C.
\par En effet, la composition de deux optimisations nécessite la réécriture de l'optimisation résultante. Ceci s'avère complexe à fortiori dans le cas de plusieurs optimisations. Tiramisu permet une simple composition des optimisations dans une partie séparée complètement de l'algorithme sous forme de commandes d'optimisations. Cette séparation permet d'empêcher la réécriture du résultat de la composition des optimisations, la gestion se fait automatiquement  par Tiramisu.

\par Pratiquement, tous les autres compilateurs polyédriques imposent des restrictions afin d'assurer que les codes après optimisations sont justes. Tiramisu permet de vérifier la validité des optimisations appliquées grâce à l'analyse de dépendance. De ce fait, il permet d'utiliser sans restrictions  des optimisations souvent considérées difficiles à appliquer notamment sur des espaces d’itération non rectangulaires, ou encore sur des graphes de  flux de données cycliques. 

\par Actuellement, Tiramisu est capable de générer des codes optimisés pour différentes architectures matérielles à savoir, CPU, GPU, FPGA ou encore des systèmes distribués. Ceci en utilisant la même syntaxe tout en s'assurant de tirer profit des avantages qu'offre chaque architecture. 

\section{Programmer en Tiramisu}

\par Tiramisu est un générateur de code. L'objectif d'un programme Tiramisu est de générer des codes censés être appelés à partir d'autres programmes (les programmes utilisateurs). Chaque programme Tiramisu commence par l'initialisation du compilateur Tiramisu. Cela permet aussi de définir le nom de la fonction. Cette fonction sera donc appelée dans un programme appelant (\textit{wrapper}) qui peut être écrit en Tiramisu ou encore dans un autre langage \cite{Ray2018DistributedTiramiu}.
\par
\begin{figure}[ht]
	\begin{center}
		\includegraphics[scale=0.8]{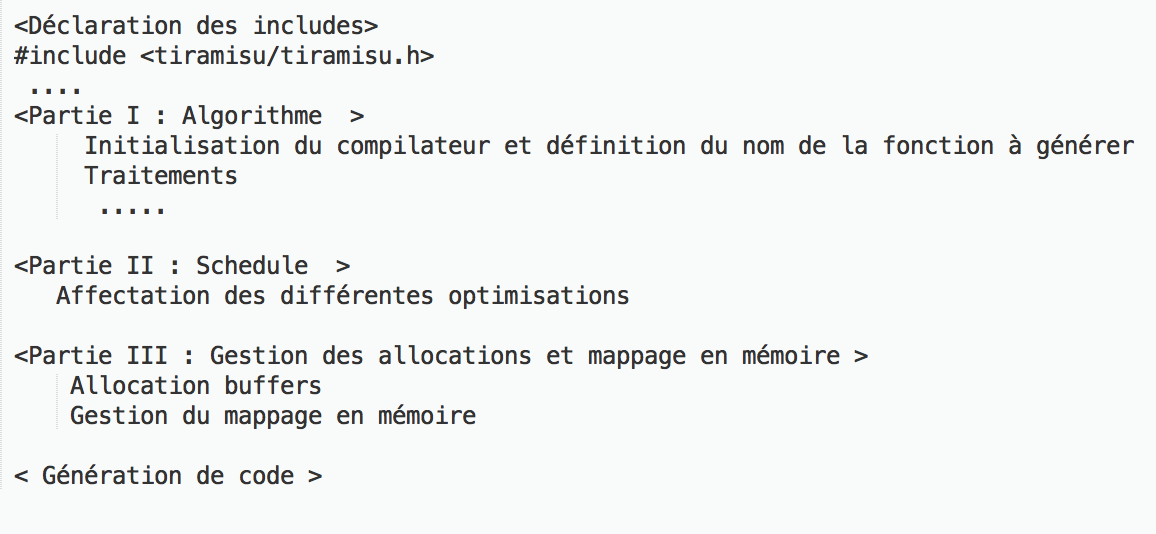} 
	\end{center}
	\caption{Schéma global de code en Tiramisu.}
	\label{fig:exmp_CodeGlobal_links}
\end{figure}

 Afin d'assurer une représentation compréhensible des programmes écrits en Tiramisu, il est recommandé d'organiser le code en deux principales sections correspondant à la forme générale du modèle en couches de la représentation intermédiaire. Dans la première étape, le programmeur définit l'algorithme pur. Dans la deuxième étape, (le \textit{Schedule}\footnote{Équivalent à "Planning" en français, ce terme représente l'ensemble des optimisations à appliquer sur le code, \textit{Schedule} est le terme à utiliser tout au long du rapport.}) il décrit comment le code sera optimisé. Ensuite, il détermine les allocations et le stockage des résultats dans les buffers. Et pour clore chaque programme Tiramisu, il faut lancer la commande de génération du code. Pour décrire une vue globale, un code Tiramisu peut être représenté par le schéma de la figure
\ref{fig:exmp_CodeGlobal_links}.

\subsection{Algorithmes dans Tiramisu}
\par Dans cette première partie, le programmeur décrit la logique de son algorithme grâce à des instructions particulières appelées \textit{computations}. Ces dernières représentent des nids de boucles (voir la section \ref{nidsDeBoucles}) dont la profondeur est égale au nombre des variables (itérateurs) qui lui ont été affectées. Chaque itérateur affecté à un niveau de boucle possède une borne fixée lors de sa déclaration. 

\par
\begin{figure}[ht]
   \centering
   \subfloat[Code C]{{\includegraphics[scale=0.8]{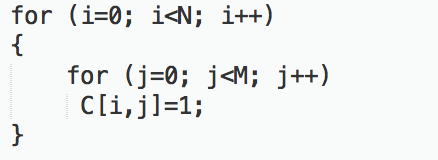} }}
    \subfloat[Code Tiramisu équivalent]{{\includegraphics[scale=0.9]{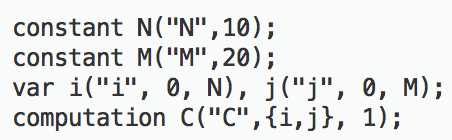} }}

    \caption{Code Tiramisu équivalent à une boucle d'une profondeur de deux.}
 \label{fig:exmp_CodeMulMatrix_links}
\end{figure}

 Une \textit{computation} peut être vue comme étant une expression associée à un domaine d'itération. L'expression représente le calcul à effectuer. Le domaine d'itération est défini grâce aux bornes affectées aux itérateurs du domaine \cite{Baghdadi2018TiramisuV3}. 
Dans la figure \ref{fig:exmp_CodeMulMatrix_links}, la \textit{computation} $C$ permet d'exprimer une boucle d'une profondeur de deux avec $i$ et $j$ comme itérateurs du domaine ayant la borne $N$ et $M$ respectivement.

\par Tiramisu est destiné aux algorithmes de données parallèles qui manipulent les tableaux denses et les nids de boucles. Ainsi, les programmes Tiramisu sont constitués de plusieurs \textit{computations} chacune assure des traitements particuliers dans une boucle "pour" (\textit{for}) imbriquée, souvent de grande profondeur. Notons que seules les boucles "pour" et la structure conditionnelle peuvent être exprimées. Les boucles \textit{while} et les \textit{goto} ne sont pas encore utilisables en Tiramisu, ce qui oriente d'autant la spécialisation au domaine caractérisant les algorithmes exprimés en Tiramisu \cite{Baghdadi2018TiramisuV3}. 

\par L'ordre d'exécution des \textit{computations} est indépendant de l'ordre de leurs déclarations. C'est au niveau de la partie \textit{Schedule} où le véritable ordre est défini et donc les relations entre les \textit{computations} sont prescrites. La définition de l'ordre des \textit{computations} ne doit pas briser les relations producteur – consommateur qui existent entre les \textit{computations}.
\par
\begin{figure}[ht]
   \centering
   \subfloat[Pseudo Algorithme ]{{\includegraphics[scale=0.8]{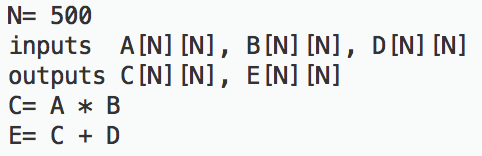} }}
    \subfloat[Code Tiramisu équivalent]{{\includegraphics[scale=0.7]{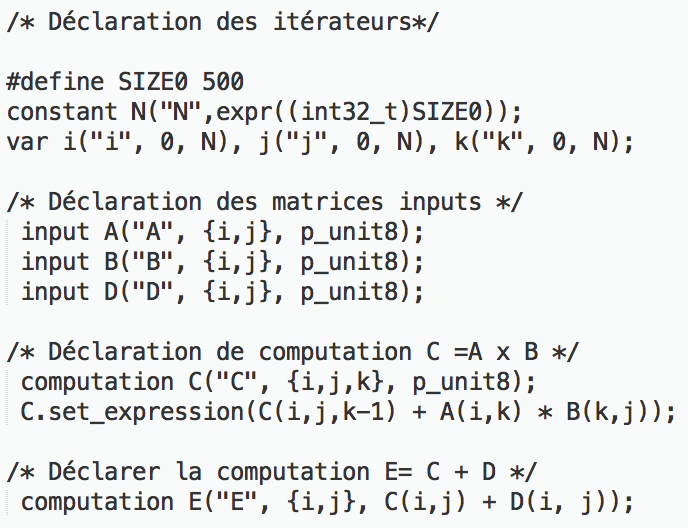} }}
 
    \caption{Code Tiramisu d'un produit suivi d'une somme de deux matrices.}
 \label{fig:exmp_AlgoTiramisu_links}
\end{figure}

 Par exemple, pour calculer le produit de deux matrices $A$ et $B$ soit $(C= A\times B)$, puis calculer la somme du résultat et la matrice D soit $(E= C+D)$. les deux \textit{computations} présentent une relation de producteur – consommateur, avec C comme producteur et E comme consommateur.  La figure \ref{fig:exmp_AlgoTiramisu_links} représente le code Tiramisu (partie algorithme) équivalent. Notons que pour initialiser les inputs, il faut créer des \textit{computations} dédiées de type \textit{Input}. 

\par Dans cette première partie du code, seuls les traitements relatifs à la logique de l'algorithme sont définis, aucune spécification est donnée sur l'ordre d'exécution ni sur la structuration des données ni encore sur les optimisations à appliquer. Pour définir ces critères, l'utilisateur doit les préciser dans la partie \textit{Schedule} \cite{Ray2018DistributedTiramiu}.

\subsection{Schedule dans Tiramisu}

\par
\begin{figure}[ht]
	\begin{center}
		\includegraphics[scale=0.90]{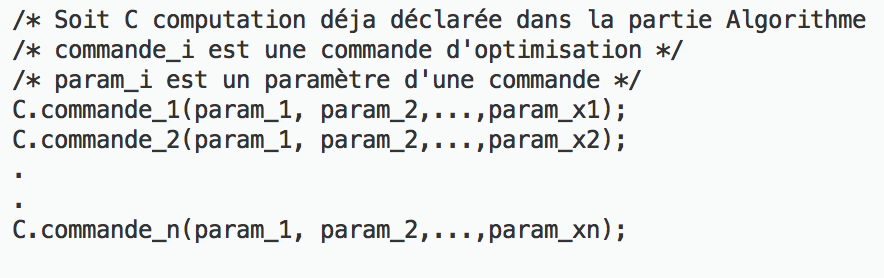}
	\end{center}
	\caption{Forme générale d'application des commandes de Scheduling sur une \textit{computation}.}
	\label{fig:exmp_schedule_general_links}
\end{figure}

 Tiramisu propose un ensemble de commandes de scheduling\footnote{Commandes de planification en Français, nous optons pour le terme "scheduling" pour désigner Planification tout au long du rapport.} de haut niveau pour définir l'ordre des \textit{computations} et optimiser leurs exécutions. Les commandes d'optimisation permettent d'effectuer des transformations sur le domaine d'itération d'une façon transparente pour le programmeur. Cette opération facilite la combinaison de commandes d'optimisation. Le programmeur prescrit pour chaque computation, les commandes d'optimisation à appliquer avec les paramètres nécessaires (voir figure\ref{fig:exmp_schedule_general_links}). 

\par Les commandes de scheduling sont classées en quatre principaux types : les commandes de transformation des nids de boucles, les commandes pour mapper les niveaux des boucles sur l'architecture matérielle, les commandes de manipulation des données et les commandes de synchronisation
\cite{Baghdadi2018TiramisuV3}. Pour davantage de détails, nous proposons dans l'annexe B une description des types de commandes de scheduling. Le tableau \ref{table_Commandes_Tiramisu} dans l'annexe B expose quelques commandes de scheduling, en définissant pour chaque commande sa syntaxe d’application, ses paramètres, son principe et son type également.
\subsubsection {Amélioration de l'optimisation du code en Tiramisu} 
\par L'optimisation de code en Tiramisu est basée sur un ensemble de commandes paramétrables de haut niveau assurant la flexibilité et le contrôle total au programmeur. Or, la spécification d'un \textit{Schedule} optimal manuellement nécessite d'avoir une bonne expertise et d'effectuer plusieurs tests sur les optimisations choisies pour différents paramètres. 
\par La définition automatique du \textit{Schedule} s'avère intéressante pour alléger cette tâche. D'ailleurs, plusieurs compilateurs des langages spécifiques au domaine proposent une gestion automatique des optimisations tel que Pencil \cite{Baghdadi2015Pencil} et Halide \cite{Kelly2013Halide}. Ce dernier, étant proche à Tiramisu puisque il sépare entre l'algorithme et le \textit{Schedule}, propose un auto-scheduler générant automatiquement le \textit{Schedule} susceptible d'être le meilleur. Le compilateur Tiramisu supporte la scalabilité de ses fonctionnalités et permet d'adopter des méthodes pour gérer la partie \textit{Schedule} automatiquement. 

\section*{Conclusion}
\addcontentsline{toc}{chapter}{\textsc{Conclusion}}
\par Tiramisu est un nouveau langage qui offre l'avantage de séparer entre l'algorithme, les optimisations à appliquer et la structuration des données. Ceci permet de gérer des codes rapides visant plusieurs architectures matérielles. \par Dans ce chapitre, nous avons présenté le compilateur Tiramisu, expliqué sa logique et ses fondements. Nous avons aussi exposé les principales notions relatives à la programmation en Tiramisu, à savoir la séparation entre les parties de l'algorithme et de \textit{Schedule}.   
\par La partie état de l'art est close. Nous entamons par la suite la partie contribution, dans laquelle nous allons expliquer en détails les différentes phases de conception, implémentation et tests du système proposé. 
\end{onehalfspace}

% Partie Contribution 
\part*{\textsc{Contribution}} 
\addcontentsline{toc}{part}{Contribution}
\chapter{\textsc{Conception et réalisation}}
\label{ch:conception}
\begin{onehalfspace}
\setcounter{footnote}{0}
\section*{Introduction}
\addcontentsline{toc}{chapter}{\textsc{Introduction}}
\par Notre contribution s’inscrit dans le cadre des projets de recherche lancés par l’équipe COMMIT du laboratoire CSAIL\footnote{CSAIL : laboratoire d’informatique et d’intelligence artificielle (\textit{Computer Science and Artificial Intelligence Laboratory}).} à MIT visant à optimiser le compilateur Tiramisu.
\par L’objectif principal de notre travail est la réalisation d’un modèle de prédiction des meilleurs facteurs de l’optimisation de déroulage (\textit{loop unrolling}) pour des programmes déjà optimisés (manuellement ou automatiquement) ou encore pour des programmes naïfs sans aucune optimisation préalable. Le modèle permet donc d’automatiser le choix du meilleur facteur de déroulage pour faciliter la tâche d’optimisation et améliorer le temps d’exécution du programme. La classe des programmes Tiramisu visée dans notre travail présente un taux assez élevé d’opérations de chargement mémoire considérées comme opérations gourmandes en temps d’exécution. 

\par Dans ce chapitre, une description détaillée du problème traité est donnée suivie des détails sur les différentes phases de conception de notre solution. 
\section{Description du problème }
\begin{figure}[ht]
	\begin{center}
	\includegraphics[scale=0.5]{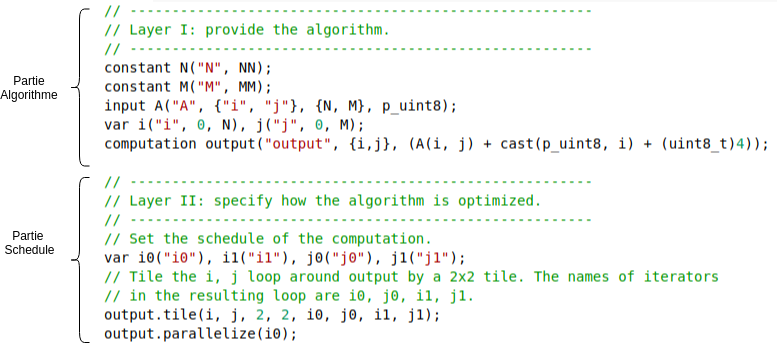}
	\end{center}
	\caption{Structure d'un programme en Tiramisu.}
	\label{fig:descriptionProblem}
\end{figure}

Dans le chapitre \ref{ch:chapterThree}, une explication détaillée de la structuration de programmes Tiramisu est donnée. Pour rappeler, les programmes en Tiramisu sont composés de deux principales parties : la première partie contient le code de l'algorithme et la deuxième partie (\textit{Schedule}) contient l’ensemble des optimisations à appliquer sur le programme (voir la figure \ref{fig:descriptionProblem}). 

\par Différentes combinaisons d'optimisations peuvent être appliquées dans la partie \textit{Schedule}. Les optimisations appliquées peuvent prendre des facteurs comme l’optimisation de déroulage de boucles, l’optimisation de tuilage, etc. Le programmeur doit sélectionner les optimisations ainsi que les meilleurs facteurs à appliquer pour améliorer le temps d’exécution. 
\subsection{Portée de la contribution}
\par Notre contribution se focalise sur l’optimisation de déroulage de boucle (voir section \ref{loopUnrollingSection} et l'annexe A pour plus de détails sur l'optimisation de déroulage de boucles) qui améliore les performances dans pratiquement tous les cas si elle est appliquée d’une manière significative, à savoir appliquée avec un bon facteur \cite{David1994Compilertransformations}.
\par Le tableau \ref{tab:table_unroll_cases} montre la différence entre le temps d'exécution pour trois facteurs de déroulage (8, 16, 32) appliqués sur le même programme présenté dans la figure \ref{fig:descriptionProblem}

\begin{table}[h!]
  \centering
  \caption{Les temps d’exécution d’un programme optimisé avec trois différents facteurs (8, 16, 32) de la transformation de déroulage.  
 }
\noindent\makebox[\textwidth]{%
\begin{tabular}{| c | c | c | c | c |}
  \hline
  \textbf{facteur de déroulage}& sans \textit{unrolling} & 8 & 16 & 32 \\ \hline \hline
    & & & & \\
\textbf{ Temps d'exécution (ms)}& 43.2417 & 40.4118 & 32.4826 & 34.4535  \\ 
 \hline
\end{tabular}
}
\label{tab:table_unroll_cases}
  \end{table}
  
\par
\begin{figure}[ht]
	\begin{center}
		\includegraphics[scale=0.5]{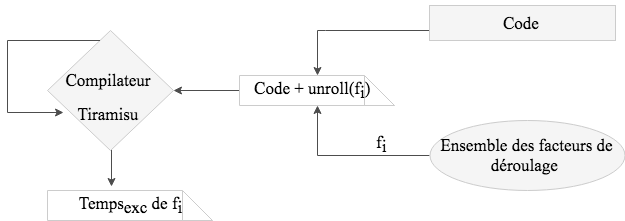}
	\end{center}

	\caption{Processus du choix manuel du meilleur facteur de déroulage.}

	\label{fig:process_manuel}
\end{figure}

 Actuellement, en Tiramisu, la définition du bon facteur de déroulage se fait manuellement. Le programmeur exécute le programme pour différents facteurs de déroulage et choisit le facteur qui donne un temps d’exécution meilleur. L’exécution du programme pour les différents facteurs coûte énormément de temps. Il s’agit d’explorer exhaustivement les différents facteurs et d’exécuter pour chaque facteur plusieurs exécutions afin de mesurer le temps pris pour chaque configuration (voir figure \ref{fig:process_manuel}).
\par L’objectif principal de notre travail est d’automatiser le choix du meilleur facteur de déroulage et d’éviter l’exécution répétitive. La solution proposée doit permettre de prédire le meilleur facteur de déroulage pour un programme donné qui peut être déjà optimisé ou naïf sans aucune optimisation préalable. Pour atteindre cet objectif, il est nécessaire de définir une formalisation (codification) du problème ainsi qu’une fonction objectif. 

\subsection{Formulation du problème}

\par Considérer le problème traité comme étant un problème d’optimisation combinatoire\footnote{Les problèmes d’optimisation combinatoire traitent le choix d’une meilleure alternative dans un ensemble très grand mais fini d’alternatives dites solutions réalisables. La solution permet de satisfaire une fonction objectif. Une évaluation est associée à toute solution réalisable à l’aide d’une fonction dont la minimisation/maximisation définit la fonction objectif.} permet de définir le cadre formel du problème. Soit $P(U, f)$  le problème d’optimisation traité, caractérisé par un ensemble réalisable ou admissible $U$ non-vide et une fonction $f$ qui associe un scalaire dans $\mathbb{R}$  à chaque élément $u \in U$ (solution réalisable) pour un programme donné. Soit $C_{(A, S)}$ le programme à optimiser, $C_{(A, S)}$ est composé de deux partie : la partie $A$ qui constitue l'algorithme  et la partie $S$ qui constitue les optimisations (\textit{Schedule}).
Le programme $C_{(A, S)}$ est représenté par un ensemble de caractéristiques  $X_C(x_1, x_2, x_3, …)$. 

\par $U$ est donc  l’ensemble des facteurs $u$ de l’optimisation de déroulage de boucles (\textit{loop unrolling}) du programme $C_{(A, S)}$, tel que :  $(b_{min} < u < b_{max})$. Les paramètres $b_{max}$ et $b_{min}$ dépendent de $X_C$. Ils donnent les contraintes qui permettent de définir l’ensemble $U$ des solutions réalisables du problème.
 \par La fonction $f$ associe à chaque solution réalisable pour le programme $C_{(A, S)}$ son temps d'exécution. Résoudre le problème $ P(U, f)$ revient à trouver parmi les solutions réalisables, une qui minimise $f$ i.e. trouver une solution $u* \in U$ telle que  $f(C_{(A, S)},u) \geq f(C_{(A, S)},u*)$ pour tout élément $u \in U$. Une telle solution est dite optimale, c’est la solution qui permet de minimiser le temps d’exécution de $C_{(A, S)}$. 

\par La fonction objectif est définie comme suit:
       $$ \min_{u \in U}  f(C_{(A, S)}, u)$$

\subsection{Choix conceptuels étudiés}
 \par Afin de proposer une conception qui vérifie aux objectifs du projet, il est important de répondre à certaines questions pour mieux cerner nos choix conceptuels : 
 \begin{compitemize}
     \item[\textendash] Quel est l’espace de recherche à considérer ?
     \item[\textendash] Quelle approche à adopter pour la prédiction du meilleur facteur de déroulage  ?
     \item[\textendash] Quel est le type de sortie à prédire ?
   \end{compitemize}
   
\subsubsection{Espace de recherche} 
\label{searchSpace}
\par Les facteurs de déroulage sont des valeurs discrètes (des entiers) qui doivent être des multiples de deux. L’optimisation de déroulage fait partie d’un \textit{Schedule} regroupant plusieurs autres optimisations comme le tuilage (\textit{Loop tiling}). Les facteurs de l'optimisation de tuilage appliqué dans Tiramisu sont impérativement des puissances de deux. Si nous appliquons le déroulage avec autres facteurs que les multiples de deux, des problèmes lors de la compilation dans Tiramisu peuvent apparaître. D’autre part, selon les experts, les facteurs de déroulage qui donnent les meilleures performances varient entre 2 et 64. Le facteur 64 est le plus grand facteur exploré manuellement jusqu’à maintenant.
\par Comme nous allons prédire automatiquement, nous pouvons aller au-delà de cette valeur. Cependant, d’après les tests que nous avons effectués, les valeurs qui dépassent 128 détériorent les performances, et risquent de créer des bugs dans certaines architectures d’exécutions.
\par En effet, l’optimisation de déroulage réplique le code plusieurs fois selon le facteur donné, elle crée autant d’instructions que le facteur de déroulage. Donc, de grands facteurs augmente la taille du code. Ainsi, les instructions ne peuvent pas être chargées d’une façon optimale dans le cache, les registres aussi ne seront pas exploités efficacement, ce qui détériore les performances au lieu de les améliorer (pour davantage de détails voire l'annexe A).

\par \textbf{Conclusion} : l’espace de recherche $U$ que nous devons explorer est représenté par les valeurs discrètes paires $u$, avec : $b_{min}$  < $u$<  $b_{max}$  avec  $b_{max}$ = 128  et $b_{min}$  = 0  ( 0 représente le cas où l’optimisation de déroulage n’est pas appliquée). 

\subsubsection{Approche de prédiction du facteur de déroulage}
\par La définition de l’approche d’exploration d’espace de recherche constitue une phase de conception cruciale. Dans la section \ref{sec:Approche_optim_automatic}, plusieurs approches ont été exposées : la recherche exhaustive, les heuristiques et les métaheuristiques ainsi que les approches analytiques. Elles ont été déjà utilisées dans plusieurs travaux traitant ce problème. Chacune présente des avantages et des inconvénients. Le concepteur doit favoriser soit la précision ou le temps d’exécution de la méthode.
\par L’apprentissage automatique a prouvé son efficacité dans la résolution de plusieurs problèmes. Il propose un compromis entre la précision et le temps d’exécution de la méthode. La disponibilité d’un nombre de données suffisant et la puissance de calcul des machines actuelles ont permis d’intégrer ces techniques dans divers problèmes complexes d’optimisation dans les compilateurs. 
\par En effet, \cite{ supervised2005} et \cite{MLAforLoopUnrollingFactorPredictionInHighLevelSynthesis} ont exploré les méthodes d’apprentissage automatique pour le choix du meilleur facteur de déroulage et ils ont pu atteindre une bonne précision (jusqu’à 60\%). Les méthodes précédentes utilisent des techniques basées sur les réseaux de neurones peu profonds (\textit{Shallow Neural Networks}) ou basées sur l'apprentissage automatique classique. 
\par De notre part, nous voulons explorer un nouvel angle du problème de prédiction du meilleur facteur de déroulage par apprentissage automatique afin d’améliorer la précision. Nous avons opté pour les réseaux de neurones profonds qui ont fait preuve d'une bonne précision dans divers domaines. Nous souhaitons donc les explorer pour résoudre notre problème.
\par De plus, contrairement aux techniques classiques d’apprentissage automatique, les réseaux de neurones profonds ont la particularité de créer automatiquement des caractéristiques haut niveau (\textit{high-level features}) à partir des caractéristiques bas niveau (\textit{low-level features}) que nous allons utiliser pour représenter les programmes. 

\par \textbf{Conclusion} : le modèle de prédiction du facteur de déroulage est appuyé sur l'approche basée sur l'apprentissage automatique, plus précisément les réseaux de neurones profonds (pour plus de détails sur les réseaux de neurones voir annexe C).

\subsubsection{Classe de réseau de neurones profond}
\par Le théorème du \textit{no-free-lunch} montre qu’aucun algorithme ou modèle résout parfaitement tous les problèmes. Dans notre cas, le choix du type de réseau de neurones dépend de plusieurs paramètres notamment la nature des contraintes du problème ainsi que les données d’apprentissage. La meilleure approche de sélection du type de réseaux de neurones est d’essayer d’identifier dans le modèle des contraintes présentes dans le problème traité, puis tester les types qui sont plus susceptibles de résoudre le problème. Nous avons effectué une étude comparative entre les trois classes de réseaux de neurones les plus stables et utilisées (voir annexe C), le tableau \ref{tab:table_DNN_types} résume les résultats obtenus. 

\begin{table}[ht]
  \centering
  \caption{Comparaison entre les différentes classes de réseaux de neurones candidates.}
\noindent\makebox[\textwidth]{%
\begin{tabular}{|p{3.7cm}|p{7cm}|p{7cm}|}
  \hline
     \textbf{Classe}   & \textbf{Principales caractéristiques}  & \textbf{Type de données} 
     \\ \hline \hline
  \vspace{0.3in}  \text{Réseaux de neurones} multicouche (MLPs) & 
\begin{compitemize}
 \item [\textendash] Composés d’une ou plusieurs couches
 \item [\textendash] Chaque couche est composée d’un nombre variable de neurones 
 \item [\textendash] Les entrées sont connectées aux sorties à travers des couches intermédiaires
 \item [\textendash] Réseau à propagation directe
   \end{compitemize}
 &
 \begin{compitemize}
 \item [\textendash]Données sous forme tabulaire.
  \item [\textendash]Entrées de taille fixe.
  \end{compitemize} \\
\hline
\vspace{0.3in} \text{Réseaux de neurones} convolutifs (CNNs) & 
\begin{compitemize}
 \item [\textendash]Combinaison de couches, chacune représente une fonctionnalité du réseau : couche de convolution, couche de pooling et couche entièrement connectée 
 \end{compitemize}
 &  
 \begin{compitemize}
 \item [\textendash]Données matricielles.
  \item [\textendash]Données de type image.
  \item [\textendash]Données ayant des relations spatiales\tablefootnote{Les points de données d’une unité de données sont liés de tel sorte qu’ils ne peuvent pas être séparés ni modifiés indépendamment car cela implique que l’unité de donné est corrompue(ex. données audio).}
 \item [\textendash]Traitement d’images et d'audio 
  \end{compitemize}
  \\ 
 \hline
\vspace{0.3in} \text{Réseaux de neurones} récursifs (RNNs) &
\begin{compitemize}
\item [\textendash] \text{Constitués d'unités (neurones)} inter\-connectées interagissant non\-linéairement qui présentent au moins un cycle dans la structure.
 \item [\textendash]Propagation de données  dans les deux sens.
 \end{compitemize}
 &
 \begin{compitemize}
 \item [\textendash]Données séquentielles (ex. texte).
  \item [\textendash]Les entrées peuvent être de tailles variables.
  \item [\textendash]Ils ne sont pas appropriés aux données tabulaires ou de type image.
  \item [\textendash]\text{Traitements de langage naturel (NLP).}
  \end{compitemize}
\\ 
 \hline
\end{tabular}
}
\label{tab:table_DNN_types}
  \end{table}

\par  Les entrées de notre modèle sont de tailles variables, si nous avons par exemple une boucle avec quatre niveaux et une autre avec deux niveaux, nous aurons au moins deux entées supplémentaires pour la boucle à quatre niveaux. 
\par Les RNNs  supportent les entrées de tailles variables, ils  sont utilisés pour des données séquentielles, ce séquencement est défini par le temps. Les RNNs utilisent la sortie prédite par l’entrée précédente et l’entrée actuelle pour produire la sortie actuelle. Cependant nos données ne présentent aucun séquencement, chaque entrée de notre \textit{data\-set} est indépendante de l’autre. En effet, le facteur de déroulage prédit pour le programme$_i$ ne sera pas utilisé avec les caractéristiques du  programme$_{i+1}$ pour prédire son meilleur facteur de déroulage. L’architecture des RNNs ne convient pas à notre problématique.
\par La taille variable des entrées peut être  fixée  à une taille maximale, en utilisant la technique du rembourrage où l’entrée est remplie par des valeurs bidons jusqu’à atteindre la taille maximale. Le rembourrage ou le \textit{padding} permet de résoudre le problème de la taille variable des données en entrée des réseaux de neurones. Le \textit{zéro padding} est la solution la plus reconnue pour la résolution du problème de la taille variable des entrées. Cependant, d’autres valeurs peuvent être utilisées. Généralement les deux valeurs "0" et le "-1". Le « -1 » peut remplacer le "0" si le "0" présente une valeur significative dans les données d'entrée du modèle. 
\par \textbf{Conclusion} : selon le type de problème traité et la représentation des caractéristiques des programmes constituant les données utilisées, les Réseaux de neurones multicouche  MLPs semblent  être la solution la plus convenable. 
\subsubsection{Type de sortie du modèle}
\par Il existe plusieurs types de sorties possibles pour un modèle de prédiction de déroulage. Le modèle peut prédire le temps d’exécution des instances de programmes en prenant en entrée les caractéristiques du programme. Une fonction doit estimer le minimum des temps d’exécution pour les facteurs dans l'espace d'exploration. Le modèle doit aussi apprendre cette fonction. Cependant, dans le cas des réseaux de neurones, la fonction est difficile à entraîner et elle risque très souvent d'être piégée par les optimums locaux.
\par L’autre sortie possible est le facteur de déroulage optimal lui-même ; trouver le meilleur facteur de déroulage directement permet de remédier au problème des optimums locaux. 
\par \textbf{Conclusion} : nous avons décidé de prédire directement le meilleur facteur de déroulage. 

\subsubsection{Classification ou régression ?}
\par Le modèle de prédiction du meilleur facteur de déroulage peut représenter un problème de classification ou un problème de régression.\\
\textbf{a) Classer les programmes selon le facteur optimal de déroulage} : la classification\footnote{Le problème de classification consiste à attribuer à chaque individu (objet) une classe ou une étiquette.} est utilisée lorsque la variable de sortie est une valeur discrète qui représente une catégorie (les différents facteurs de déroulage $u$ : 2, 4, 6, 8, etc. dans notre cas). Cependant, la classification ne permet de classer les programmes que dans un ensemble prédéfini de classes. Il faut avoir suffisamment de programmes qui couvrent les différentes classes afin d’atteindre une bonne prédiction pour des nouveaux programmes.\\
\textbf{b) Utiliser une fonction continue (Régression\footnote{Le problème de régression consiste à prédire une valeur réelle à partir d’un ensemble d’entrées.})} : un problème de régression se pose lorsque la variable de sortie est une valeur réelle ou continue. Dans notre problème, le modèle de régression renvoie des valeurs réelles que nous devons arrondir en valeurs entières multiples de deux (les facteurs doivent être des entiers multiples de deux). Si la sortie est un nombre impair, il faut prendre soit son successeur ou son prédécesseur. En revanche, le nombre choisi risque de ne pas être la valeur la plus convenable pour le programme en entrée ce qui influence négativement la précision du modèle.
\par D’une autre part, la théorie des réseaux de neurones montre qu’ils sont aussi puissants dans la régression que dans la classification. Cependant, dans la pratique, il y a quelques différences en termes de précision. Dans la classification, il faut uniquement décider la classe convenable à l’entrée. Mais, les problèmes de régression sont plus difficiles car il s’agit de prévoir une certaine valeur pour chaque entrée. 
\par Par conséquent, l’utilisation des réseaux de neurones pour les problèmes de régression peut être moins stable comparativement  aux  problèmes de classification. Il est généralement préférable de convertir les problèmes de régression en problèmes de classification lorsque cela est possible. Cependant, les réseaux de neurones restent toujours très puissants pour traiter des problèmes de régression, la difficulté serait dans l’optimisation des traitements du modèle. En effet, quelques techniques d’optimisation s’avèrent difficiles à appliquer dans les cas des modèles de régression (ex.\textit{dropout}\footnote{Le décrochage, ou abandon est une technique de régularisation permet une suppression temporaire de neurones afin d’éviter le surapprentissage.}).
 \par \textbf{Conclusion} :  La classification est la solution la plus convenable à notre problème, il faut juste s’assurer que l’ensemble de données d’apprentissage couvre toutes les valeurs de l’espace de recherche défini.

\section{Conception globale du système }
\par Afin de donner une vue globale de la solution proposée, nous allons exposer l’architecture globale du système proposé, définir la classe des programmes visée et décrire les  différents composants de la solution. 

\subsection{Architechture globale du système}
\par
\begin{figure}[ht]
	\begin{center}
		\includegraphics[scale=0.6]{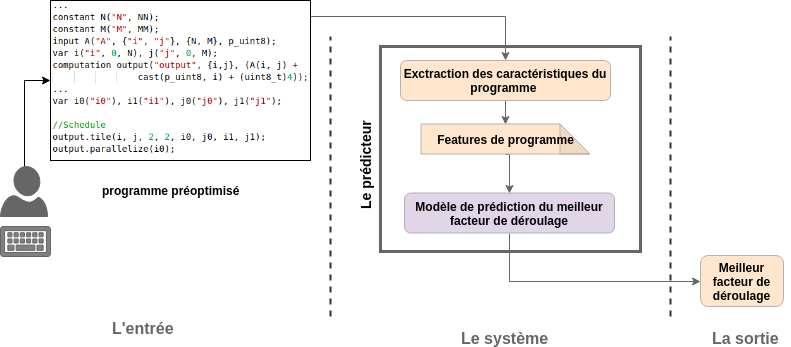}
	\end{center}
	\caption{Architechture globale du système.}
	\label{fig:Architechture_Globale}
\end{figure}

 L’utilisateur définit la partie algorithme $A$ et éventuellement la partie \textit{Schedule} $S$ contenant les différentes optimisations possibles hormis l’optimisation de déroulage de boucle (\textit{loop unrolling}). L’utilisateur appelle le modèle de prédiction proposée, ce dernier  analyse le code $C_{(A, S)}$  et décide le meilleur facteur de déroulage à appliquer. L’architecture globale du système est présentée dans la figure
\ref{fig:Architechture_Globale}.

\begin{itemize}
    \item \textbf{Module d’extraction des caractéristiques de programmes }: ce module permet d’extraire les caractéristiques (\textit{features}) du programme en entrée  $C_{(A, S)}$. L’ensemble des caractéristiques $X_C(x_1, x_2, x_3, …)$ est constitué essentiellement des informations sur les niveaux de boucles, les opérations effectuées ainsi que les caractéristiques des optimisations (\textit{Schedule} $S$) appliquées. Ceci permet de représenter les programmes Tiramisu distinctement. Plus de détails sont données dans la section \ref{ModuleExctractDetail}.
     \item \textbf{Module de prédiction du facteur de déroulage}: le module prend en entrée les caractéristiques $X_C(x_1, x_2, x_3, …)$ extraites par le module précédent  pour prédire le meilleur facteur de déroulage $u*$. Ce facteur est utilisé par le compilateur Tiramisu pour compléter le \textit{Schedule} $S$ du programme.
\end{itemize}
 \subsection{Classe de programmes visée}
 \label{classePgm}
Tiramisu est dédié aux programmes dits de données parallèles qui utilisent des matrices denses et des nids de boucles. Notre équipe travaille sur la génération de programmes optimisés de calculs scientifiques comme l’algèbre linéaire dense. La classe des programmes visée est constituée de boucles à contrôle affine ACLs (\textit{Affine Contol loops}), à savoir les bornes des nids de boucles et les adresses des accès en mémoire sont définies comme étant des fonctions affines des itérateurs de boucles et des paramètres constants. 
\par
\begin{figure}[h!]
	\begin{center}
		\includegraphics[scale=0.6]{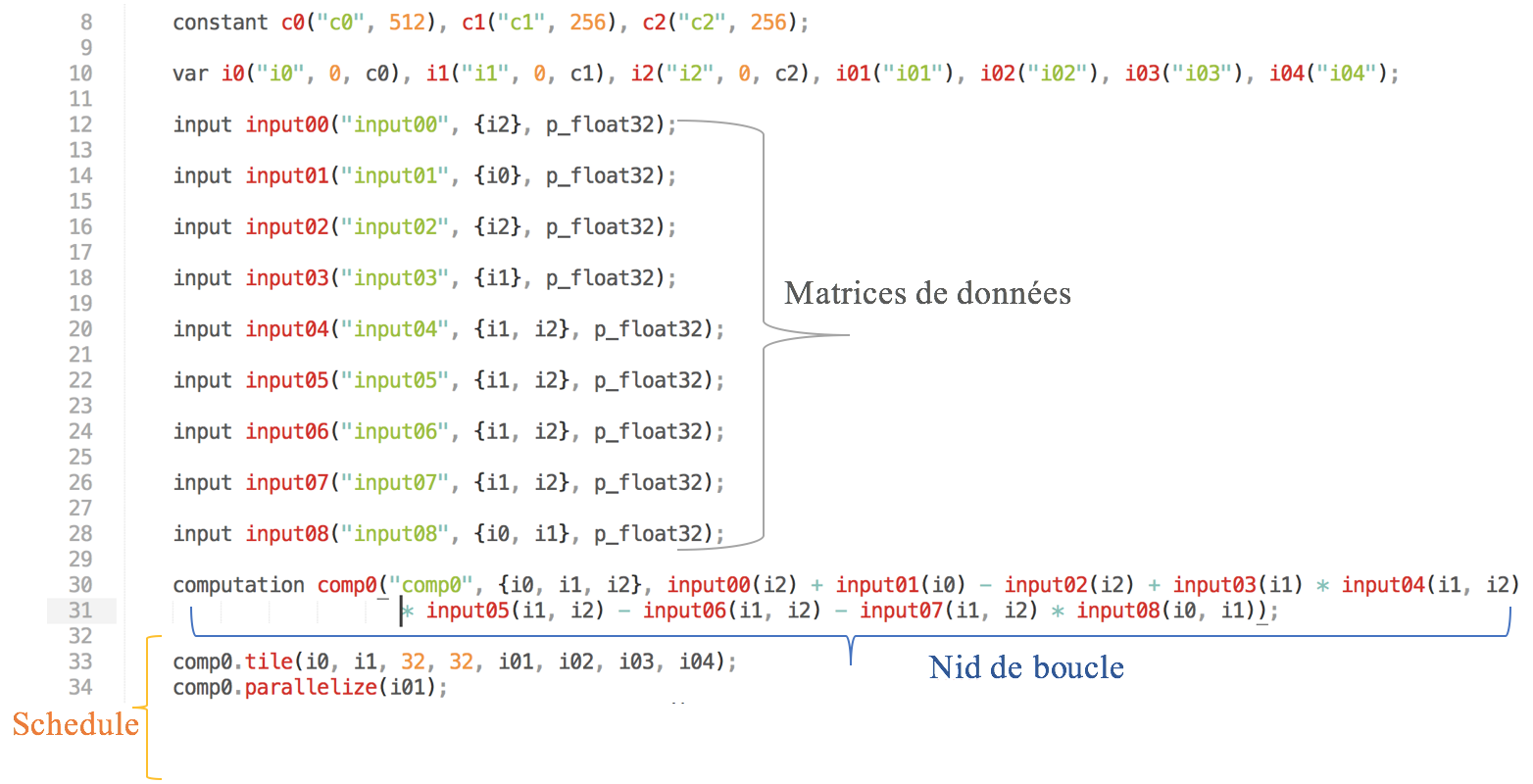}
	\end{center}
	\caption{Exemple de programme Tiramisu appartenant à la classe de code visée.}
	\label{fig:exempleCode}
\end{figure}

 Dans cette classe de programmes, nous avons visé des nids de boucles parfaitement imbriquées (voir section \ref{nidsDeBoucles}) qui présentent des chargements de données en mémoire multiples et intenses. Nous focalisons notre travail sur les programmes composés d'une seule \textit{computation}, à savoir un seul nids de boucles qui n'a pas été fusionné avec d'autres nids de boucles. La figure \ref{fig:exempleCode} donne un exemple de la classe des codes utilisés.
\par Le choix de cette classe de programmes est basé sur la nature des domaines visés par Tiramisu. En effet, cette classe présente des noyaux pour divers programmes dédiés aux calculs scientifiques. De plus, les chargements mémoire sont des opérations coûteuses en temps d’exécution, l’amélioration en temps d’exécution apportée grâce à l’application de déroulage de boucle avec un bon facteur est remarquable. 

\subsection{Caractéristiques des programmes}  
\label{Programefeatures}

\par Un programme en Tiramisu représente un ensemble de nids de boucles parfaitement imbriqués appelés \textit{computations} (voir chapitre \ref{ch:chapterThree}). Le module d’extraction des caractéristiques représente chaque \textit{computation}  par un ensemble de caractéristiques synthétiques $X_C$. Dans certains travaux précédents, un nombre considérable des caractéristiques est utilisé. Ils utilisent des caractéristiques décrivant l’effet de l’architecture d’exécution sur les nids de boucles ce qui augmente le nombre des caractéristiques extraites. Cependant, un grand nombre de caractéristiques risque de nuire la prédiction du modèle. En effet, sélectionner les caractéristiques les plus significatives permet d’optimiser la prédiction du meilleur facteur de déroulage. 
\par Initialement, nous avons opté pour un grand nombre de caractéristiques. Ensuite, nous avons gardé les caractéristiques qui influencent le plus la prédiction. L’abstraction adoptée résume les critères influençant le temps d’exécution d’une \textit{computation} indépendamment de l’architecture d’exécution ce qui offre plus de portabilité au modèle. Elle décrit principalement la structure, les opérations et les optimisations appliquées sur le nid de boucles. 
\par Les caractéristiques des optimisations appliquées sur une \textit{computation} peuvent être classées en deux principales classes : optimisations locales et optimisations globales. Les optimisations locales agissent uniquement sur la structure du nid de boucles sur lequel elles sont appliquées telles que l’optimisation de tuilage de boucles, déroulage de boucles, inversion de boucles, parallélisation, etc. Les optimisations globales agissent sur l’ensemble des \textit{computations}, elles mettent en relation plusieurs \textit{computations} telles que la fusion de nids de boucles, ordonnancement d’ordre de calcul des opérations dans les nids de boucles (\textit{after()}, \textit{before()}, \textit{compute\_at()}, etc.).
\par Le tableau \ref{table_Features} donne un sous-ensemble des caractéristiques des programmes considérées.

\begin{table}[H]
\caption{Sous-ensemble des caractéristiques données par le module d’extraction des caractéristiques d'une \textit{computation}.}
\noindent\makebox[\textwidth]{%
\begin{tabular}{|l|}
 \hline 
 \multicolumn{1}{|l|}{\textbf{Caractéristiques de la structure du nid de boucles} } \\
   \hline \hline
 \multicolumn{1}{|l|}{Nombre de niveaux du nid de boucles.} \\
  \multicolumn{1}{|l|}{L’étendu de chaque niveau de boucle. } \\
   \multicolumn{1}{|l|}{Nombre de dépendances entre les niveaux du nid de boucles. } \\
   %%% Liste
    \multicolumn{1}{|l|}{Liste des dépendances entre les niveaux du nid de boucles.\tablefootnote{Pour chaque niveau de boucle, exprimer les dépendances sous forme d'un vecteur d'itérateurs, si la définition du domaine d'itération (représentée par les contraintes sur les bornes de l'itérateur) présente des dépendances avec  d'autres itérateurs.} } \\
  \multicolumn{1}{|l|}{ L’estimation de l’étendu de la boucle s'il n’est pas une valeur constante.} \\
  \multicolumn{1}{|l|}{La précédence du niveau de boucle par un prédicat de test (\textit{if statement}). } \\
  \hline \hline
 \multicolumn{1}{|l|}{\textbf{Caractéristiques des opérations du nid de boucles}  } \\
   \hline
\multicolumn{1}{|l|}{Le niveau de boucle dans lequel  l'opération est effectuée.} \\
\multicolumn{1}{|l|}{Le rang d'exécution de l'opération dans le niveau de boucle qui lui a été affecté. } \\
\multicolumn{1}{|l|}{Nombre de variables/invariants utilisés dans les opérations. } \\
\multicolumn{1}{|l|}{Histogramme des opérations par type de données.\tablefootnote{L'histogramme des opérations représente le nombre d'opérations par type de données. C'est une matrice dont les lignes sont les types d'opérations (opérations arithmétiques, opérations d'accès mémoire, max, etc.) et les colonnes sont les types de données utilisés (\textit{Integer}, \textit{Float} et \textit{Boolean}, avec les différentes modalités pour chaque type).} }\\
\multicolumn{1}{|l|}{Histogramme des chargement/sauvegarde en mémoire par type de données. } \\
\multicolumn{1}{|l|}{Liste des niveaus de boucles définissant chaque accès mémoire.\tablefootnote{Pour chaque opération d'accès mémoire, nous définissons la succession des itérateurs des niveaux de boucles utilisés. Par exemple, l'accès $Input(i_2,i_0,i_1)$ est représenté par la liste des itérateurs $[i_2,i_0,i_1]$ .}} \\
\multicolumn{1}{|l|}{Nombre d’unités de données (Quantité de données) chargées par niveau. } \\
\multicolumn{1}{|l|}{Nombre des appels externes dans chaque niveau de boucle.  } \\
 \hline \hline
 \multicolumn{1}{|l|}{\textbf{Caractéristiques du \textit{Schedule} (Optimisations)} } \\
   \hline
  \hline
\multicolumn{1}{|l|}{Les niveaux de boucle sur lesquels l’optimisation est appliquée. } \\
   \multicolumn{1}{|l|}{Facteurs utilisés pour chaque optimisation.} \\
    \multicolumn{1}{|l|}{Liste des  \textit{computations} dans le cas des optimisations globales.} \\
    \hline
\end{tabular}}
\label{table_Features}
\end{table} 
%\footnotetext{}

\section{Conception détaillée }
\par Dans cette section, nous allons exposer les détails de la conception des modules composant notre solution. En effet, la solution proposée est basée sur deux principaux modules : le module d’extraction des caractéristiques des programmes et le module de prédiction du facteur de déroulage.
\subsection{Module d'extraction des caractéristiques des programmes}
\label{ModuleExctractDetail}
\par Ce module permet d’extraire les caractéristiques des unités composant les programmes Tiramisu, à savoir les nids de boucles (\textit{computations}). L’extraction des caractéristiques d’une \textit{computation} passe par deux phases, la première phase consiste à extraire les caractéristiques décrivant la structure du nid de boucles, l’ensemble des opérations effectuées ainsi que les types des données. La deuxième phase consiste à enregistrer les caractéristiques des optimisations appliquées (\textit{schedule}) sur le nid de boucles pour ensuite, les utiliser dans la mise à jour des caractéristiques de la structure du nid de boucles. \\
\subsubsection{Extraction initiale des caractéristiques de la \textit{computation}} 
\begin{figure}[H]
	\begin{center}
		\includegraphics[scale=0.5]{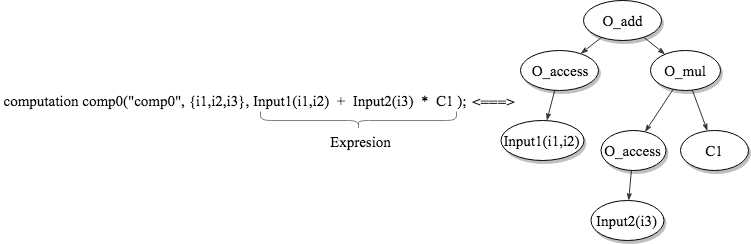}
	\end{center}
	\caption{représentation d'une expression en Tiramisu.}
	\label{fig:express_tree}
\end{figure}

Afin d’extraire initialement les caractéristiques définissant la structure d’une \textit{computation}, le module commence par parcourir la liste des itérateurs puis les expressions (opérations) Tiramisu associées au nid de boucles. L’expression associée à une \textit{computation} en Tiramisu est un arbre n-aire dont les nœuds sont des expressions aussi (voir figure \ref{fig:express_tree}). Dans notre cas, les expressions qui nous intéressent sont de type opération (les opérations arithmétiques et les opérations d’accès mémoire). L’exploration de l’arbre qui représente l'expression permet d’extraire les caractéristiques des opérations effectuées dans le nid de boucles.
\par L’optimisation de déroulage est fortement sensible à l’étendu de chaque niveau du nid de boucle, il influence également l’effet des opérations effectuées notamment les chargements mémoire\footnote{Les chargements mémoire représentent la classe d’opérations la plus utilisée dans les programmes visés.}. Pour accentuer cette relation nous avons défini des caractéristiques décrivant la quantité de donnée ("mots mémoire selon le type de données) chargée pour chaque variation d’un niveau de boucle. D'autre part, ces caractéristiques, décrivant les accès mémoire, dépendent de l'ordre des itérateurs lors des accès comparativement à l'ordre des niveaux de la \textit{computation}(voir l’algorithme \ref{algo2}). 
\begin{algorithm}[h]
\SetAlgoLined
\KwResult{ définir la caractéristique de quantité de données pour chaque niveau de boucle de la $computation\_c$ en entrée }
\For {(iterator $it_i$ dans $computation\_iterators$) }{ 
\For {(access\_operation $access_i$ dans $computation\_operations$)  }{ 
  $var: \hspace{5 mm} acces\_load =1$ \;
  soit $access_{i}\_iterators$  la liste des itérateurs de l'opération $access_i$ \;
  soit $sup\_level\_iterators$  la liste des itérateurs de la $computation\_c$ dont le niveau est supérieur au niveau de $it_i$ \;
\For {(iterator $it_k$ dans $sup\_level\_iterators$)  }{ 
\If{ ($it_{k}  \in sup\_level\_iterators$ ) }{
$acces\_load = acces\_load \times ( it_{k}.upper\_bound - it_{k}.lower\_bound)$  \;}
}
}
  /* sauvegarder le nombre total d'accès de $it_i$ */

 $it_{i}.data\_loaded = it_{i}.data\_loaded + acces\_load$ \;

}
\caption{Pseudo algorithme de l'extraction de la caractéristique de quantité de données chargées par niveau de boucle}
\label{algo2}
\end{algorithm}
\subsubsection{Mise à jour et exportation des caractéristiques de la \textit{computation}}      
\par Suite à l’application de chaque optimisation dans la partie \textit{Schedule} du programme, le module d’extraction doit enregistrer cette optimisation en gardant trace de son type, ses facteurs et les niveaux de boucles sur lesquels elle a été appliquée. L’optimisation modifie la structure du nid de boucles d’où la nécessité de mettre à jour les caractéristiques de la \textit{computation}. Par exemple l’application de l’optimisation de tuilage de boucle change d’une part, les étendus des niveaux des boucles ainsi que leurs ordres. D’une autre part, elle modifie les caractéristiques relatives à l’opération d’accès.
\par Après la mise à jour des caractéristiques, le module les prépare sous un format utilisable par le réseau de neurones. Il s'agit d'une représentation vectorielle où chaque caractéristique unitaire (nombre de niveaux, nombre des opérandes, type de donnée, etc.) représente une entrée (\textit{input}) au réseau de neurones.     
\par
\begin{figure}[H]
	\begin{center}
		\includegraphics[scale=0.4,frame]{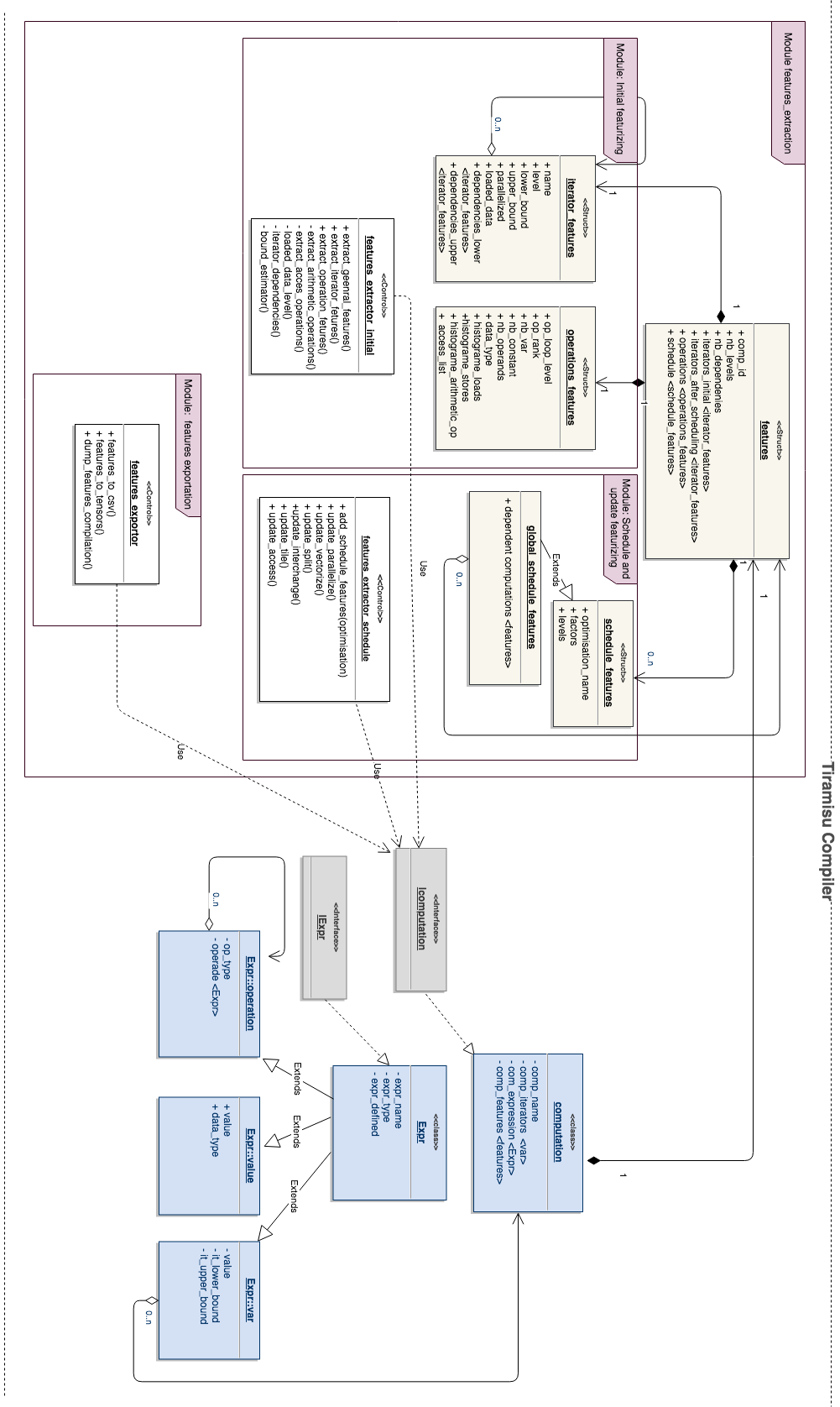}
	\end{center}
	\caption{Diagramme de classe du module d'extraction des caractéristiques des programmes.}
	\label{fig:diagrammeClasseFeatures}
\end{figure}

 Le module d’extraction est composé de trois sous modules : un module d’extraction initiale des caractéristiques de chaque \textit{computation}, un second module pour l’extraction des caractéristiques du \textit{schedule} et la mise à jour des caractéristiques de la \textit{computation} après l’application du \textit{schedule} et un troisième module pour l’exportation des caractéristiques sous le format utilisable par le réseau de neurones (voir figure \ref{fig:diagrammeClasseFeatures}) 
\newpage
\newpage
\subsection{Modèle de prédiction du meilleur facteur de déroulage}
\par Nous avons conçu notre modèle de prédiction d’une manière itérative. Nous avons commencé par un modèle de réseau de neurones de base, puis nous avons ajouté des améliorations à chaque itération afin d’atteindre une bonne précision. Les améliorations apportées ont été à base des tests et des résultats de la précision du modèle à chaque fois. Le processus d’amélioration est arrêté une fois la précision désirée est atteinte. \\ Dans cette section, nous allons présenter les principales étapes suivies. 

\subsubsection{Architecture du réseau de neurones }
\par
\begin{figure}[h!]
	\begin{center}
		\includegraphics[scale=0.75]{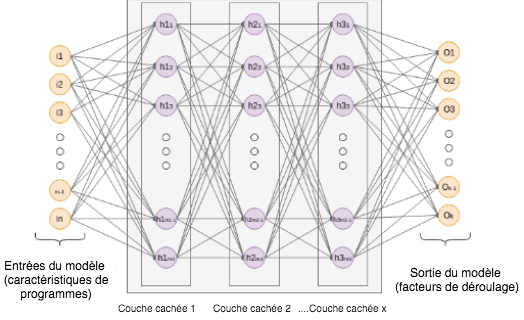}
	\end{center}
	
	\caption{Architecture de base du modèle.} 
	\label{fig:architechture_NN_phase1}
\end{figure}

 Nous avons utilisé le réseau de neurones pour construire un modèle de classification supervisée. Il permet la prédiction du meilleur facteur de déroulage. Dans un modèle de classification, les sorties (les classes) sont prédéfinies. Il reçoit un ensemble de données d'apprentissage (\textit{training}) étiquetées pour apprendre à classer les nouveaux programmes en entrée. L'architecture du réseau de neurones est basée sur l'architecture typique des réseaux de neurones multicouche (voir figure \ref{fig:architechture_NN_phase1}). Nous définissons dans les sections suivantes le nombre des couches cachés et le nombre de neurones dans chaque couche (voir \ref{fig:architechture_NN_phase1}). \\ Le modèle doit prédire la sortie parmi les classes définies qui représentent la plage de valeurs possibles du facteur de déroulage (voir section \ref{searchSpace}).

\par Dans les sections qui suivent, nous présentons les notions décrivant le fonctionnent du réseau de neurones proposé (voir l'annexe C pour plus de détails sur les réseaux de neurones). Nous détaillons les différentes étapes pour la génération du modèle de prédiction du facteur de déroulage.

\subsubsection{Propagation de l’information}
\par
\begin{figure}[h!]
	\begin{center}
		\includegraphics[scale=0.25,frame]{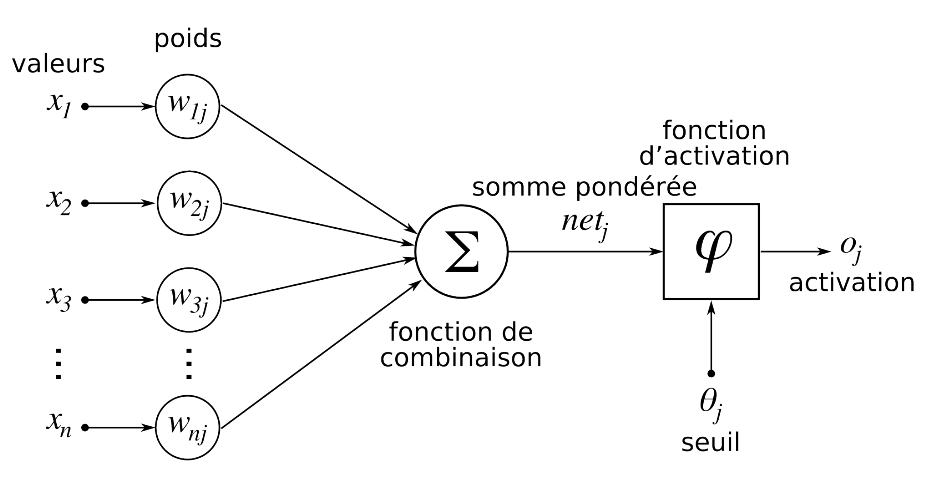}
	\end{center}
	\caption[La structure d'un neurone artificiel.]{La structure d'un neurone artificiel~\protect\cite{Goodfellow}.}
	\label{fig:Strucurneurone}
\end{figure}

 La couche d’entrée du réseau est alimentée par les caractéristiques du programme (voir section \ref{Programefeatures}), ses sorties sont attribuées à la première couche cachée qui, à son tour, passe ses sorties à la couche suivante et ainsi de suite. Ce processus est appelé la propagation de l’information. La sortie de chaque couche cachée est le résultat de l'application d'une fonction à la somme pondérée des sorties des couches précédentes à laquelle un biais est ajouté. Cette fonction est appelée la fonction d’activation (voir la figure \ref{fig:Strucurneurone}). 

 Essentiellement, les fonctions d’activation (ou fonctions de transfert) convertissent le signal d’entrée en un signal de sortie. Elles produisent une sortie non linéaire à partir d’une entrée linéaire. Ceci permet de représenter des fonctions plus complexes. Il existe différents types de fonctions d’activation, pour notre modèle nous considérons la fonction \textit{Relu} comme fonction d’activation pour les couches cachées, elle retourne le max entre le $0$ et la valeur d’entrée $x$ ($ReLu = max (0, x)$ ). 
 \par La sortie de chaque couche est donnée par  $\displaystyle ReLU(\omega_0 + \sum_{n=1}^{N} \omega_nx_n)$. 
 La sortie du réseau est fournie directement par  $y= \displaystyle ( \sum_{n=1}^{N} \omega_nx_n)$, tel que $w_n$ sont les poids, $x_n$ sont les entrées et $w_0$ est le vecteur de biais associé. 
\par Comme il s'agit d'un problème de classification en classes multiples (nombre de classes > 2), nous utilisons la fonction $Softmax$ pour la couche de sortie. La fonction $Softmax$ permet de calculer la distribution de probabilités des différentes classes. Initialement, le modèle effectue le traitement avec des valeurs de pondération aléatoires qui sont mises à jour à chaque itération afin de minimiser l’erreur. 

 \par La fonction de perte (\textit{loss function}) permet de calculer l'erreur de prédiction. Le choix de cette fonction  dépend de type du problème traité. Dans le problème de classification, nous utilisons \textit{Cross-Entropy}\footnote{Connue aussi sous le nom \textit{log loss}.}, elle  permet de mesurer l’erreur de probabilité dans le cas où les classes sont mutuellement exclusives. Cela signifie que chaque entrée appartient à une et une seule classe. Dans notre cas, l’optimisation de déroulage peut avoir un seul facteur optimal. L’erreur commise est généralement mesurée pour un ensemble de données (\textit{dataset}), donc nous calculons la moyenne de l'erreur commise pour l’ensemble d’apprentissage fourni.

\subsubsection{Rétropropagation de l'erreur}
Une fois le processus de propagation de l’information est terminé, le réseau de neurones procède à la correction des poids afin de minimiser l’erreur autant que possible. Le gradient de l’erreur calculée est rétro-propagé pour mettre à jour les poids en fonction du degré de leur contribution à l’erreur. Nous appelons ce processus la rétropropagation du gradient \cite{Goodfellow}.
\par La valeur des poids mis à jour est contrôlée par un paramètre appelé le taux d’apprentissage (\textit{learning rate}). La modification apportée aux poids du réseau pour une erreur donnée est souvent de l’ordre $10^{-1}$ ou $10^{-2}$ ou encore moins. Il existe plusieurs algorithmes utilisés pour mettre à jour les poids et donc minimiser l’erreur (les algorithmes d’optimisation) tels que l’algorithme de la descente de gradient stochastique (SGD), l’algorithme ADAM et l’algorithme RMSprop \cite{NNOptimizers}. En fonction de l’algorithme choisi, certains paramètres doivent être ajustés tels que le taux d’apprentissage. Nous avons opté pour \textit{ADAM Optimiser} \cite{AdamOptimizer} comme algorithme d’optimisation, car il est connu pour sa robustesse et son efficacité. Le taux d’apprentissage pris pour cet algorithme est $10^{-3}$ suite à plusieurs tests.
\par Les poids dans les réseaux de neurones peuvent être mis à jour à partir des erreurs calculées pour chaque ligne de données de traitements. Il s’agit de l’apprentissage en ligne. Cala permet de faire des mises à jour rapides mais parfois cause des effets chaotiques au réseau. Une autre alternative consiste à enregistrer l’erreur au niveau de tous les exemples du \textit{dataset} d’entraînement. Ensuite, les mises à jour sont effectuées vers la fin. Il s’agit de l’apprentissage par lots (\textit{batch learning}) qui est souvent plus stable. Étant donné que le nombre d’exemples dans le \textit{dataset} est très grand, la taille du lot est réduite afin d’augmenter l’efficacité de calcul. Nous considérons des lots \textit{batch} de taille de 100. 

\par Le processus est répété pour toutes les données du  \textit{dataset} d’entraînement. Chaque passe sur l'intégralité de l’ensemble de données pour entraîner le réseau de neurones  est appelée une itération (\textit{epoch}). Le réseau de neurones peut être entraîné des dizaines, centaines, voire même des milliers d'itérations afin d’améliorer la précision. Le nombre d’itérations est un paramètre à choisir soigneusement. Nous considérons un nombre d’itérations pour lequel aucune amélioration n’est apportée au modèle. 

\subsubsection{Génération et préparation de données}
 \label{GenData}
\par Le réseau de neurones est alimenté par un ensemble de données dont chaque élément $(X,u)$ représente les caractéristiques $X(x_1, x_2,..., x_n)$ d'un programme donné $C(A,S)$ et son facteur de déroulage optimal $u$. Ce dernier est obtenu en exécutant le programme pour toutes les valeurs possibles dans l'espace d'exploration $U$. 
\par Cependant, le temps d'exécution d'un programme est influencé par plusieurs évènements externes relatifs au système et à l'architechture d'exécution (ordre d'instructions choisi par la machine, taille de cache alloué, les attentes entrées/sorties, etc.). Nous considérons alors une estimation de la moyenne du temps d'exécution pour chaque programme. Selon la loi des grands nombres\footnote{La loi affirme que la moyenne empirique, calculée sur les valeurs d’un échantillon, converge vers l’espérance lorsque la taille de l’échantillon tend vers l’infini. Dans notre cas, ceci signifie que pour un grand nombre N d'exécutions, la moyenne des temps d'exécution enregistrés tend vers la moyenne réelle du temps d'exécution du programme}, le nombre d'exécutions (soit $N$) doit être grand pour avoir une bonne estimation. Nous avons considéré la valeur minimale de $N$ qui permet d'avoir une stabilité du facteur de déroulage optimal $u$. Après avoir effectué plusieurs tests, nous avons pris $N=30$.
\par La génération des programmes se fait grâce à l'outil 
\textit{Tiramisu\_Code\_Generator} (voir la section \ref{envirement}). Il s'agit de générer aléatoirement des programmes appartenant à la classe des programmes visée (voir section \ref{classePgm}). Les \textit{schedules} des programmes Tiramisu générés contiennent des combinaisons d'optimisations suivantes : tuilage de boucles (à deux et à trois niveaux), inversion de boucles et la parallélisation. Le module d'extraction des caractéristiques est ensuite utilisé pour extraire le vecteur de caractéristiques $X(x_1, x_2,..., x_n)$ pour chaque programme généré.

\par Les données fournies aux réseau de neurones doivent être numériques. Elles peuvent être redimensionnées (normalisées) dans la plage comprise entre 0 et 1. Les données peuvent être aussi standardisées de tel sorte que chaque colonne soit centrée et réduite. Les données de notre \textit{dataset} sont toutes numériques mais nécessitent une mise à l’échelle, nous allons donc les normaliser. Le \textit{Dataset} est divisé en trois parties, chacune est utilisée pour une phase de création du modèle.
\begin{itemize}[label=\textendash]
\item \textbf{Dataset d’entraînement} ou \textit{Training set}  ($60\%$ de l'ensemble de données d'origine). Nous utilisons ce \textit{dataset} pour entraîner notre modèle (trouver les bons poids et biais pour le modèle). 
\item \textbf{\textit{Dataset} de validation}  ($20\%$ de l’ensemble de données d’origine) : ce \textit{dataset} est utilisé pour minimiser le surapprentissage. Il ne contribue pas directement dans la modification des poids, il permet uniquement de vérifier que toute augmentation de la précision par rapport à l'ensemble de données d'apprentissage entraîne une augmentation de la précision par rapport à un ensemble de données qui n'a pas encore été montré au réseau, ou au moins le réseau ne l’a pas encore utilisé pour s'entraîner. Si la précision sur l'ensemble de données d'entraînement augmente, mais la précision sur l'ensemble de données de validation reste la même ou diminue, cela implique que le modèle sur-apprend et donc nous devons arrêter l’entraînement.
\item \textbf{\textit{Dataset} de test} ou \textit{Test set}  ($20\%$ de l’ensemble de données d’origine) : nous l’utilisons une fois le modèle finit l’entraînement. Ce \textit{dataset} est utilisé uniquement pour tester la solution finale afin de confirmer la précision réelle du réseau. 
\end{itemize} 
\par
\begin{figure}[H]
	\begin{center}
		\includegraphics[scale=0.65]{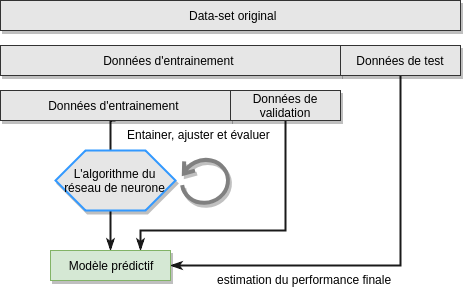}
	\end{center}
	\caption{Utilisation des différents \textit{datasets} pour entraîner le modèle final.}
	\label{fig:data_set_parts}
\end{figure}

 Il est fortement déconseillé que la phase de test soit ignorée. En effet, l’algorithme qui prédit bien pendant la phase de validation ne signifie pas forcément que c’est le modèle le plus convenable au problème traité. Au cours de la phase de test, le modèle final est utilisé pour prédire de données non déjà vue. Donc, si la précision du modèle est très mauvaise pendant le test, tout le processus de conception du modèle doit être remis en cause (voir la figure \ref{fig:data_set_parts}). 
%%% 

\subsection{Itérations de construction du modèle de prédiction}
\par Nous avons suivi un processus itératif afin d'arriver à la conception finale de notre modèle. La conception passe par trois principales itérations.
\subsubsection{1ère itération : modèle de réseau de neurones de base}
\par Le but de cette étape est de bien cerner les entrées et les sorties du modèle. Nous avons expliqué dans la section \ref{Programefeatures} qu'initialement, le nombre de caractéristiques décrivant chaque nid de boucle est considérable. Il faut définir les caractéristiques les plus significatives pour la prédiction. D’autre part, le nombre de classes de sortie du modèle est grand. Nous avons fait une étude à priori sur un échantillon de programmes pour définir les classes qui ne seront pas utiles dans notre problème (voir section \ref{searchSpace}).
\par
\begin{figure}[H]
	\begin{center}
		\includegraphics[scale=0.7]{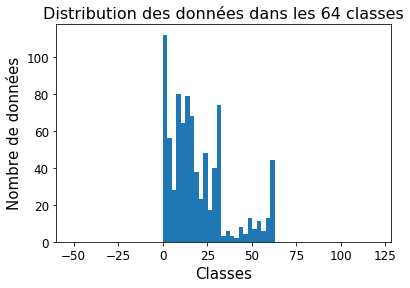}
	\end{center}
	\caption{Distribution des données du \textit{dataset} sur les 64 classes.} 
	\label{fig:64ClassDistribution}
\end{figure}

 Le premier entraînement du modèle a été effectué sur un \textit{dataset} de 1500 éléments générés toute au long de deux semaines suite à une exploration exhaustive de tous les \textit{Schedules} possibles pour 20 programmes(partie algorithme). Évidement, La précision enregistrée a été médiocre car le \textit{dataset} considéré est très petit et ne présente que 20 fonctions avec leurs \textit{schedules}. En revanche, nous avons pu constater des points très importants.
\begin{itemize}[label=\textendash]
\item Quelques colonnes (\textit{features}) du \textit{dataset}  ont la même valeur pour toutes les lignes comme la colonne qui représente le niveau d’application de l’optimisation de parallélisation, elle ne prend que la valeur '1'\footnote{La valeur '1' veut dire que la parallélisation est appliquée sur ce niveau de boucle} sur toute la colonne ou que le '0'\footnote{La valeur '0' veut dire que la parallélisation n’est pas appliquée sur ce niveau} car le niveau d’application de l’optimisation de parallélisation est le même (le niveau le plus profond). En effet, ce genre de colonnes ne porte aucune information utile au modèle, il faut donc les enlever.

\item Les valeurs de certaines colonnes présentent une distance considérable  par rapport à d’autres (d’ordre de $10^5$). Par  exemple, la colonne du nombre d’opération d'accès mémoire (unitaire) par rapport à celle de l’étendue de boucles. Même en effectuant la normalisation, la distance entre les colonnes reste la même. Pour remédier à ce problème, il faut redimensionner ces colonnes en les divisant par  $10^3$ par exemple pour avoir des valeurs plus petites. 
\item Pour chaque programme (partie algorithme), les lignes qui le représentent dans le  \textit{dataset} ont plusieurs colonnes communes. La différence entre ces lignes touche seulement les colonnes décrivant la partie \textit{schedule}. Ce qui diminue la diversification dans le \textit{dataset} et influence ainsi la précision du modèle.

\item Le nombre de classes est très grand (64 classes) par rapport aux données que nous pouvons générer. Plus le nombre de classe est grand, plus la taille du \textit{dataset} nécessaire et qui couvre toutes les classes est grand. Générer un \textit{dataset} aussi immense nécessite des super calculateurs. Ceci impose des contraintes sur le nombre des classe à considérer, a fortiori, si le modèle enregistre une mauvaise précision, il serait difficile de changer la stratégie et lancer encore la génération pour avoir un nombre suffisant de données.\\
De ce fait, nous avons effectué une étude statistique afin de déterminer la marge de classes à éliminer. La figure \ref{fig:64ClassDistribution} montre la distribution des données du \textit{dataset} sur les $64$ classes. 

\end{itemize}

\subsubsection{2ème itération : sélection de bons hyperparamètres du modèle }
\par Les résultats de la première itération nous ont mené à remettre en question l’architecture du modèle ainsi que la nature de données à générer. Concernant le problème de diversification des données nous avons décidé de réduire le nombre de \textit{Schedules} générés pour le même programme (partie algorithme) à 10 \textit{Schedules} aléatoires.  Le nombre de classes pose également un grand problème par rapport au nombre de données possible à générer compte tenu du temps disponible. Nous avons décidé de restreindre le nombre de classes à sept classes. Il s'agit des puissances de deux {0, 2, 4, 8, 16, 32, 64}. Ce sont les facteurs de déroulage les plus utilisés par les experts. Les tests que nous avons effectués dans la phase précédente le montre également (voir figure \ref{fig:64ClassDistribution}).
\par D'autre part, nous avons effectué une analyse des caractéristiques (\textit{features}) les plus significatives en utilisant l'outil \textit{features\_selctor} (voir la section \ref{envirement}). L'outil propose la suppression de 5 colonnes telles que le nombre de constantes et le nombre de niveaux de boucle\footnote{Il s'agit d'une information déductible à partir de certaines autres colonnes (le nombre de colonnes qui présentent les étendus des niveaux du nid de boucle)}. Nous avons effectué des tests pour s'assurer que la précision augmente après la suppression des 5 colonnes. 
\par Dans cette itération, nous allons relancer les tests sur un \textit{Dataset} plus large pour sélectionner les bons hyperparamètres du modèle, à savoir le nombre de couches, le nombre de neurones dans chaque couche, l’algorithme d’optimisation à utiliser, etc. (pour davantage de détails, voir l'annexe D). Pour chaque hyperparamètre, nous effectuons un ensemble de tests, la valeur qui donne la meilleure précision du modèle est maintenue. 

\par Nous avons commencé par choisir le nombre de couches et le nombre de neurones dans chaque couche. Nous avons constaté que quatre couches cachées avec 500, 400, 250 et 100 neurones dans chaque couche respectivement, donne la meilleure précision. En effet, nous avons testé pour 12 couches au maximal\footnote{Les tests effectués pour un nombre de couches supérieur à 12 a donné une précision très basse, et ce, pour les différents cas de nombre de neurones.}, puis nous diminuons le nombre de couches si la précision dégrade. Pour chaque couche nous avons testé dichotomiquement les cas du nombre de neurones. D'abord, nous avons défini une borne minimale et maximale au nombre de neurones. Ensuite, nous définissons la valeur divisant l'ensemble de cas de tests en deux sous ensembles. Nous continuons l'exploration du sous ensemble qui donne une précision meilleure.\\
Concernant les autres hyperparamètres du modèle, les tests effectués (voir section \ref{ModelRealisation}) donnent les résultats résumés dans le tableau \ref{tab:HyperParamsTable}.

\begin{table}[h!]
  \centering
  \caption{hyperparamètres choisis pour le modèle.}
\noindent\makebox[\textwidth]{%
\begin{tabular}{|p{5.5cm}|p{8cm}| }
  \hline
 \textbf{Type} & \textbf{hyperparamètre choisi}       \\ \hline \hline
 \mbox{Algorithme d'échelonnement} & Standardisation \\  \hline 
 Fonction d'activation & ReLu                  \\  \hline 
 Algorithme d’optimisation & ADAM    \\  \hline
 Taux d’apprentissage      &  $10^{-3}$         \\  \hline
 \mbox{Algorithme d’initialisation} des poids & Algorithme de Random\_uniform  \\  \hline
 Nombre d’itérations & technique d'arrêt précoce (\textit{Early stopping}) avec une patience de 10.   \\  \hline
\end{tabular}
}
\label{tab:HyperParamsTable}
  \end{table}
Le choix des hyperparamètres est suivi par une phase d’optimisation (voir l'annexe D). Pour remédier au problème du sousapprentissage, nous avons utilisé la technique d’optimisation \textit{dropout} avec les facteurs (0.12, 0.1, 0.04 et 0.07) respectivement. Quant à la régularisation, nous l'avons pas appliquée car elle a influencé négativement sur la précision du modèle.

Par contre, nous avons appliqué la technique de \textit{batch-normalization} pour améliorer la vitesse d'entraînement (d'ordre de 10 fois), améliorer la précision\footnote{L'application de la technique de \textit{batch-normalization} a amélioré la précision de 5\%} et la stabilité du modèle. 

\begin{comment}
\par \textbf{Sélection des bons \textit{features}} : les caractéristiques des programmes ne sont pas toutes en faveur de contribuer dans la précision du modèle. Nous avons utilisé un outil \textit{features\_selctor} (voir la section \ref{envirement}) pour sélectionner les bons \textit{features}. L'outil propose de supprimer 5 colonnes : la colonne du nombre de constantes dans le programme, la colonne de type de données (entier ou nombre flottant), la colonne de nombre de niveau de boucle, la colonne représentant l'étendu de la boucle du 7ème niveau et la colonne du chargement relatif à ce dernier niveau. La colonne de nombre des constantes jugée de zéro importance car elle ne contiens qu'une seule valeur. La colonne du type de données (entier ou nombre flottant) et celle du nombre de niveaux de boucle jugé moins importante par l'outil. Enfin, les deux colonnes d'étendue de la boucle du 7ième niveau et les chargements relatifs à ce niveau à cause du manque des valeurs. Ce qui est clair, car nous avons fixé le nombre de niveaux à sept pour remédier au problème de la taille variable, et donc pour une boucle à 3 niveaux toutes les colonnes relatives au niveau supérieur à 3 vont avoir la valeur nul. Nous avons gardé ces deux dernières colonnes comme elles sont d'importance égale à celle d'autres colonnes.
\end{comment}

\subsubsection{3ème itération : l’entraînement du modèle sur les données finales}

\par Tout au long de deux mois, nous avons pu générer un \textit{dataset} de taille réduite\footnote{La taille du \textit{dataset} est de $36 012$ éléments} relativement aux  tailles nécessaires des \textit{datasets} utilisés pour l'apprentissage profond. En effet, pour chaque programme (élément du \textit{dataset}), nous effectuons 30 exécutions pour avoir une estimation de la moyenne réelle du temps d'exécution du programme (voir section \ref{GenData}). Ceci consomme énormément du temps. Il faut presque une année de génération de donnée pour collecter autant de données que nécessite notre problème.
\par
\begin{figure}[ht]
	\begin{center}
		\includegraphics[scale=0.65]{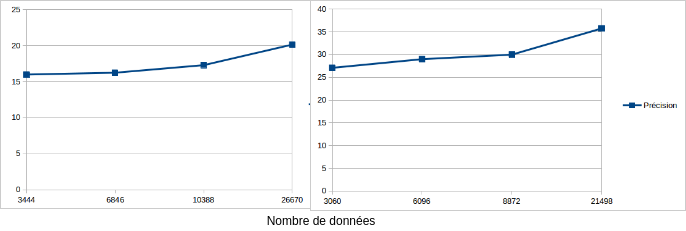} 
	\end{center}
	\caption{L'amélioration de la précision du modèle au cours de génération de données dans le cas de 7 classes à gauche et 4 classes à droite.}
	\label{fig:prec_am}
\end{figure}

 Toutefois, la précision du modèle augmente tout au long du processus de génération de données. L'augmentation du nombre de données générées à chaque fois n'est pas aussi grande pour donner des changements considérables, mais la précision du modèle s'améliore graduellement et elle enregistre une augmentation relativement remarquable. (voir la figure \ref{fig:prec_am}).

\par
\begin{figure}[H]
	\begin{center}
		\includegraphics[scale=0.6]{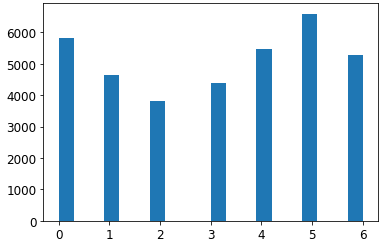}
	\end{center}
	\caption{Distribution de données dans le \textit{dataset} final sur les différentes classes du modèle}
	\label{fig:dis_data}
\end{figure}

 Nous avons constaté également que la distribution des données du \textit{dataset} sur les classes du modèle n'est pas équilibrée (voir la figure \ref{fig:dis_data}). La distribution doit être uniforme pour permettre à toutes les classes d'avoir le même degré de contribution dans l'apprentissage. De ce fait, nous avons appliqué la technique d'équilibrage de classes en définissant un seuil minimal d'éléments dans chaque classe. Certes, la taille du \textit{dataset} a diminué (devient 26670 éléments), or, la précision du modèle s'est améliorée.

\par Nous lançons l'entraînement du modèle sur le \textit{dataset} final. Le but est d'extraire et de sauvegarder les paramètres finaux du modèle (derniers poids et biais avec l'architecture du modèle) et par la suite, l'intégrer dans Tiramisu.
%%% 
\begin{comment}
\par Nous avons pu atteindre une précision de 20\% avec ce \textit{dataset}. Ce résultat est très attendu par rapport au nombre de données générées. Cependant, cette précision dépasse le choix aléatoire, c'est-à-dire la probabilité qu'un exemple appartiens à une classe ($1/7$ ou 14\% dans notre cas). Ce qui montre que le modèle apprend réellement. D'autre part, la précision du modèle au cours de génération de données a toujours augmenté. Le nombre de données générer à chaque fois n'est pas assez grand pour donner un changement considérable, mais il est quand même très remarquable (voir la figure \ref{fig:prec_am}). Ceci est un signe très positive que si nous continuons à générer plus les données, nous pouvons améliorer considérablement la précision. Citant comme exemple projet de Facebook qui consiste automatisation du choix du \textit{schedule} pour un programme écrit sur Halide\footnote{Est un langage destiné au traitement d'image. Il a une logique très similaire à celle du Tiramisu.}. Dans ce projet qui se base sur les réseaux de neurones le modèle de prédiction a été besoin de 3 millions de donnée pour qui'il a pu atteindre une précision de 93\%, selon ce qu'il nous a apporté l'un des chercheurs qui ont travaillé sur ce projet. L'ingénierie du \textit{features} (\textit{feature engineering}) et la collecte données leur ont pris plus d'une année, et ceci avec l'utilisation du super calculateur de Facebook. 
\end{comment}

\section{Synthèse de conception}

Dans cette  première partie de ce  chapitre , nous avons détaillé les différentes étapes de conception de notre solution. D'abord, nous avons discuté les choix conceptuels concernant les entrées, les sorties et l’approche d’optimisation automatique à considérer pour concevoir la solution. L’architecture conceptuelle de notre solution est basée sur deux principaux modules : le module d’extraction des caractéristiques et le modèle de prédiction du meilleur facteur de déroulage. Ce dernier est basé sur les réseaux de neurones, nous avons expliqué et justifié tous les choix considérés pour le concevoir.

\end{onehalfspace}
\end{onehalfspace}
%%%%%%%%%%%%%%%%%%%%%%%%%Begin Rélaisation%%%%%%%%%%%%%%%%%%%%%%%%%%
\label{ch:realisation}
\begin{onehalfspace}

\par  Dans cette deuxième partie, nous détaillons l'implémentation des choix conceptuels considérés dans les sections précédentes. 
\par D'abord, nous exposons les phases de réalisation du système ainsi que son architecture technique. Ensuite, nous définissons l'environnement du développement en citant les technologies, les outils et les plateformes utilisés pour développer notre solution. Tous les choix techniques considérés seront également justifiés. 

\section{Architecture technique du système}
\par Notre système de prédiction du meilleur facteur de déroulage représente un composant de tout un système d'optimisation automatique dans le compilateur Tiramisu\footnote{Le système d'optimisation automatique dans Tiramisu est un projet en cours de développement, notre contribution touche principalement l'optimisation de déroulage de boucle.}. Ce dernier utilise une interface fournie permettant de gérer les appels à notre système prédicteur.

 \begin{figure}[H]
	\begin{center}
		\includegraphics[scale=0.65]{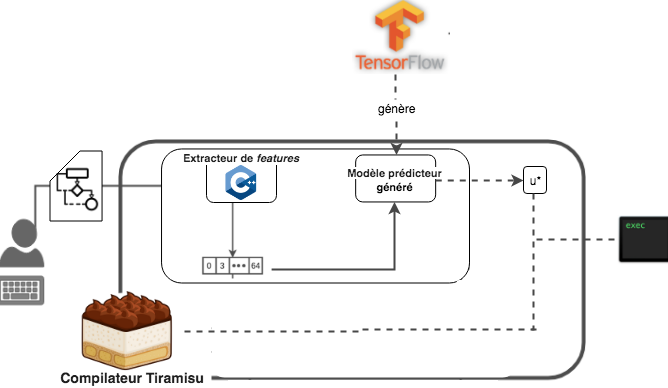} 
	\end{center}
	\caption{Architecture technique globale du système.}
	\label{fig:Rea_arch_tech}
\end{figure}
\par Lorsque l’utilisateur de Tiramisu  lance la commande \textit{automatic\_unrolling()} pour automatiser le choix du facteur de déroulage, le compilateur fait appel à la première couche du système, à savoir l'extraction des caractéristiques du programme. Ces derniers sont ensuite passées au modèle de prédiction pour prédire le bon facteur de déroulage. Enfin, Tiramisu exécute le programme en considérant le facteur prédit (voir la figure \ref{fig:Rea_arch_tech}).

\section{Environnement de développement}
\label{envirement}
\par Nous avons choisi les technologies suivantes pour implémenter notre solution.
\begin{itemize}[label=\textendash]
  \item \textbf{Tiramisu\footnote{\url{https://github.com/Tiramisu-Compiler/tiramisu}}} : est le système hôte de notre solution : les modules implémentés sont intégrés comme étant des sous systèmes en Tiramisu. Il est utilisé pour compiler et exécuter les programmes. 
  
  \item \textbf{Tiramisu\_Code\_Generator}\footnote{\url{https://github.com/Tiramisu-Compiler/tiramisu/tree/master/utils/code_generator}} : afin de construire notre \textit{Dataset}, nous avons généré aléatoirement  des codes appartenant à la classe de programmes visée (voir section \ref{classePgm}). Le générateur permet d'introduire des caractéristiques définissant certaines classes de programmes, il permet également de personnaliser la génération aléatoire de codes Tiramisu (parties algorithme) et les \textit{schedules} associés.  
  
  \item \textbf{TensorFlow}\footnote{\url{https://www.tensorflow.org/}} : est un Framework utilisé pour développer le modèle de prédiction du bon facteur de déroulage. Il est open source, bien documenté et soutenu par une très grande communauté active et par le géant du développement Google. C'est  l’outil de développement de solutions d’apprentissage automatique le plus utilisé. D'autre part, TensorFlow facilite l'exportation et le déploiement des modèles générés vers d'autres plateformes cibles, le C++ en l'occurrence (Tiramisu est embarqué sur le C++). 

  \item \textbf{Keras}\footnote{\url{https://keras.io/}} : ce Framework haut niveau écrit en Python, facilite l'implémentation des algorithmes du \textit{machine learning}. Nous l'avons utilisé pour comparer notre modèle implémenté sous TensorFlow notamment pour la sélection des bons hyperparamètres.
  \item \textbf{Feature\_Selctor}\footnote{\url{https://github.com/WillKoehrsen/feature-selector}.} : nous avons utilisé cet outil open source pour identifier les \textit{features} qui ne contribuent pas significativement à l'apprentissage du modèle. Il se base sur plusieurs principes pour identifier les \textit{features} à supprimer tels que la colinéarité des caractéristiques. cet outil offre la possibilité de visualiser les résultats sous forme de tableaux ou de graphes. Il donne également la liste directe des \textit{features} à supprimer.
  
\item \textbf{Pandas} : une bibliothèque open source de Python qui permet de manipuler et d’analyser les données. En particulier, elle offre plusieurs structures de données (\textit{DataFrame}) qui facilitent la manipulation des données numériques. Elle dispose également de nombreuses méthodes utilisées pour l'analyse de données, ce qui s'avère très utile lorsque nous travaillons sur des problèmes d’apprentissage automatique sur Python. Nous avons utilisé la structure \textit{DataFrame} pour lire et manipuler notre \textit{dataset} sauvgardé dans les fichiers CSV\footnote{CSV signifie "valeurs séparées par des virgules". Nous l'avons choisi comme format de fichiers pour sauvegarder les données d’entraînement (caractéristiques des programmes).}.  
  
\item \textbf{TensorBoard} : est un outil de visualisation de TensorFlow. Il facilite la compréhension et l’optimisation des modèles implémentés sous TensorFlow. Il permet de visualiser le graphe d’exécution et les différents paramètres du modèle (les poids, les biais, fonction de perte, etc.).

\item \textbf{Matplotlib} : est une bibliothèque de visualisation rapide et facile à manipuler. Nous l'avons utilisée pour visualiser les différents résultats du modèle. Ceci nous a permis de comparer l'évolution de la fonction de perte pour différents types d'hyperparamètres (différents algorithmes d'optimisation, taux d'apprentissage, etc.). 

 \item \textbf{scikit-learn} : est une bibliothèque Python open source qui offre une solide implémentation de plusieurs algorithmes d’apprentissage automatique (SVMs, K-means, etc.). Nous l’avons utilisée principalement pour diviser notre \textit{Dataset} en trois parties (entraînement, validation et test).
 
 \item \textbf{Lanka}: le \textit{cluster} est géré exclusivement par MIT. Il est composé de 24 machines. Il comporte au total 48 nœuds chacun dispose de 12 cœurs de 128GB de RAM. Nous avons utilisé ce \textit{cluster} pour lancer les scripts de génération de codes et d'extraction des caractéristiques des programmes. 

  \item \textbf{GoogleColab} : ou Colaboratory est un service du Cloud totalement gratuit. C'est un environnement portable (qui ne nécessite aucune configuration) qui permet de développer des applications d'apprentissage automatique en utilisant les bibliothèques populaires telles que Keras, TensorFlow, PyTorch et OpenCV. Il facilite l'entraînement des modèles sans se soucier des problèmes d’installation et de puissance de calcul. Nous l’avons utilisé comme un IDE pour développer notre modèle afin de remédier au problème en puissance de calcul de nos machines.

\begin{comment}
  \item \textbf{CSV} :  nous avons choisi de sauvegarder les données d’entraînement (caractéristiques des programmes) dans un fichier CSV, comme il est un format de fichier qui est très facile à manipuler. Il est utilisé pour sauvegarder les données tabulaires ce qui convient au type de notre données et il est le format le plus utilisé pour le stockage du \textit{dataset} dans l’apprentissage automatique.  CSV signifie "valeurs séparées par des virgules". Ses champs de données sont le plus souvent séparés ou délimités par une virgule.
  \end{comment}

\end{itemize}

\section{Phases de réalisation du système}
\label{ModelRealisation}
\par Après avoir cerné les objectifs du système et défini une conception de base, nous avons commencé la réalisation pour valider la conception. Initialement nous avons préparé l’environnement de développement, à savoir l’installation des technologies et des outils ainsi que la préparation de la plateforme Lanka. Les différentes phases de réalisation ont été lancées parallèlement suivant un processus itératif pour tester à chaque itération les décisions prises.
\subsection{Implémentation du module d’extraction des caractéristiques} 
Le module d’extraction est implémenté directement sur le compilateur Tiramisu, Il a un accès direct aux différents composants et classes de Tiramisu. Le langage utilisé est le C$++$ qui est le plus approprié car Tiramisu est embarqué sur C$++$. Le module permet d’exporter les caractéristiques sous forme d’un fichier CSV pour pouvoir construire l’ensemble de données. 
\subsection{Préparation de données}  Nous avons configuré le générateur Tiramisu\_code\_Générator  pour donner des programmes appartenant à la classe des programmes visée : le générateur prend un fichier de configuration en entrée dans lequel nous avons introduit les caractéristiques de la classe des programmes visée. Nous avons introduit d'autres paramètres définissant des bornes supérieures pour certains \textit{features} telles que le nombre maximal des niveaux dans les nids de boucles, le nombre maximal des \textit{Inputs}\footnote{Inputs en Tiramisu est une matrice multidimensionnelle pour charger les données.}, etc.\\ Nous avons défini également les différentes optimisations de boucles ainsi que l’ensemble de facteurs à explorer lors de la génération des programmes.
\par Nous avons lancé les scripts d’extraction des caractéristiques en parallèle sur 10 nœuds du cluster Lanka. Chaque nœud traite un ensemble de programmes, pour chaque programme, le générateur donne tous les \textit{schedules} possibles (en se basant sur les paramètres définis dans le fichier de configuration de générateur) et pour chaque \textit{Schedule} possible, le générateur donne le code pour chaque facteur de déroulage possible. 

\par Dans la première itération de réalisation, les facteurs de déroulage sont les multiples de 2 (0, 2, 4, … min(étendu de la boucle/2, 128)). Ensuite dans la seconde itération, les facteurs explorés sont les puissances de deux (0, 2, 4, 8 … 64). \\ 
Afin d'avoir une bonne précision des temps d’exécution, le programme généré pour chaque facteur de déroulage doit être exécuté 30 fois et prendre la moyenne des temps d’exécution enregistrés. Cette phase a coûté énormément de temps. En effet, si par hypothèse le temps d’exécution d'un programme en moyenne prend un temps $t_{ex}$, avec $nb_s$ le nombre des \textit{Schedules} en moyenne pour chaque programme et $nb_u$ est le nombre de facteurs de déroulage de boucles. La génération au niveau de chaque nœud prend pour chaque programme (partie algorithme commune) \hspace{4mm}
 $T_n = 30 \times t_{ex} \times nbr_s \times nbr_u$  
\subsection{Implémentation du modèle de réseau de neurones}  Le modèle est implémenté en utilisant le FrameWork TensorFlow sur l'environnement de développement \textit{GoogleColab}. Nous avons implémenté toutes les méthodes qui permettent de construire le modèle et d'effectuer les traitements nécessaires. Nous nous sommes appuyés sur certaines bibliothèques de Python(Pandas, Sklearn) pour implémenter d'autres opérations de pré-traitement essentiellement(la manipulation du \textit{dataset} sous le format "CSV" et sa division en trois parties).
\par Pour définir les paramètres du modèle, nous effectuons des tests pour chaque paramètre, la valeur qui donne la meilleure précision du modèle est maintenue.
\par
\begin{figure}[H]
  \centering
   \subfloat[Tests des taux d'apprentissage.]{{	\includegraphics[width=8.5cm]{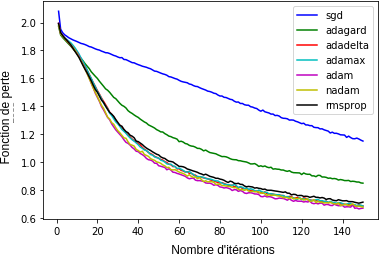}  }}%
  
    \subfloat[Tests des algorithmes d'optimisation]{{\includegraphics[width=8.5cm]{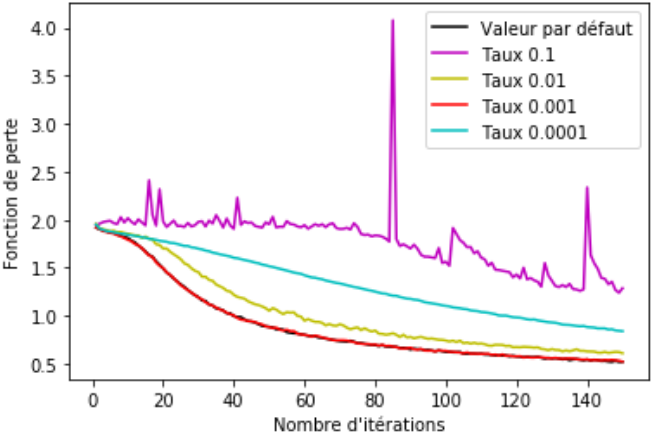}}}%

    \caption{Variation de la fonction perte durant les tests des algorithmes d'optimisation et des taux d'apprentissage.}%
 \label{fig:test_optimizer}%
\end{figure}

 Pour l’algorithme d’optimisation, nous avons testé cinq algorithmes les plus utilisés : le SGD, ADAM, NADAM et RMSProp  \cite{NNOptimizers}.\\ 
Quant au taux d’apprentissage, il existe plusieurs méthodes de test permettant de choisir sa meilleure valeur. Nous avons adopté une approche efficace et simple qui consiste à essayer une valeur relativement élevée ( nous avons choisi 0.1\footnote{Haut-delà de cette valeur la fonction de perte ne converge pas.}), puis la réduire avec un facteur de 10 à chaque étape du test empirique. Pendant ces tests, nous varions le taux d’apprentissage tout en gardant les autres paramètres fixes afin de voir le vrai impact de ce paramètre sur le modèle. Nous choisissons vers la fin, l’algorithme d’optimisation avec le temps d’apprentissage permettant d’avoir plus de précision. La figure \ref{fig:test_optimizer}.(a) montre la variation de la fonction perte (\textit{loss}) pour les données de validation en fonction de nombre d'itérations pour les différents algorithmes d'optimisation. Le test est effectué en gardant l'initialisation par défaut de chaque algorithme\footnote{Le taux d'apprentissage par défaut de chaque algorithme}. 

Nous avons choisi ADAM comme algorithme d'optimisation. En effet, l'algorithme ADAM et ses deux versions optimisées (Nadam et Adamx) surmonte significativement l'algorithme SGD et Adagrad. Quant aux autres algorithmes, ils ont pu enregistrer des résultats relativement compétitifs, mais l'algorithme ADAM converge plus rapidement et minimise mieux le taux d'erreur de prédiction. 
\par Après avoir choisi l'algorithme d'optimisation, nous  procédons aux tests de différents taux d'apprentissages. Les résultats obtenus sont présentés dans le graphe de la figure \ref{fig:test_optimizer}.(b). Le graphe permet de constater que pour des taux élevés (à partir de $10^{-1}$) le modèle diverge au fil des itérations. Nous constatons également que les taux bas (à partir de $10^{-4}$), la fonction perte commence à diminuer plus lentement ce qui montre que le taux d’apprentissage du modèle devient trop faible. En se basant sur ces résultats, nous choisissons la valeur $10^{-3}$ comme taux d'apprentissage pour le modèle.

\par
\begin{figure}[ht]
	\begin{center}
		\includegraphics[scale=0.45]{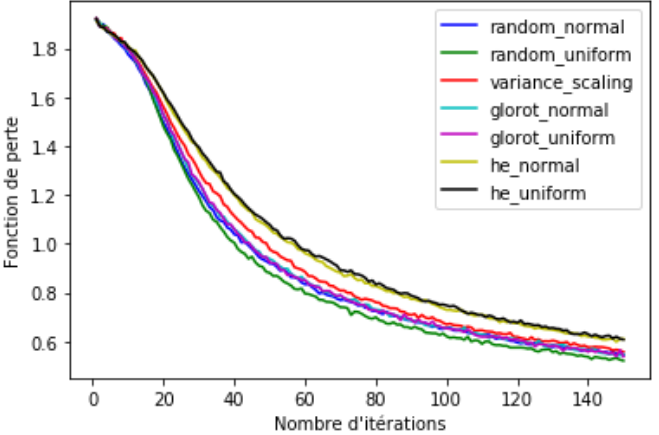} 
	\end{center}
	\caption{La variation de la fonction perte pour les différents algorithmes d'initialisation de poids.}
	\label{fig:test_weight}
\end{figure}

 Pour trouver les bonnes valeurs initiales à attribuer au poids $w_{i}$, nous avons testé plusieurs algorithmes d'initialisation de poids (voir la figure \ref{fig:test_weight}). Les résultats des tests affirment que l'algorithme d'initialisation uniforme ((\textit{Random unifrom initializer}) minimise le plus la fonction perte du modèle. Il permet d'initialiser les poids avec une distribution uniforme\footnote{La distribution uniforme est le type de distribution de probabilité dans lequel tous les valeurs ont la même probabilité d'apparaître.}.
\par Quant au nombre d’itérations, nous avons effectué la méthode d'arrêt précoce (\textit{Early stopping}) qui permet d'arrêter l'entraînement du modèle une fois l'erreur est stable et aucune amélioration n'est apportée au réseau. La patience\footnote{Le nombre d'itérations effectuées après avoir atteint le stade où aucune amélioration n'est apportée au modèle} est un paramètre important à définir pour cet algorithme. Selon les tests effectués nous posons ce paramètre à 10.
\subsection{Entraînement du modèle de prédiction} L'entraînement du modèle est fait sur l'environnement de développement \textit{Colaboratory} en deux étapes : la première consiste à entraîner le réseau de neurones pour ajuster le modèle sur les bons hyperparamètres. La deuxième étape consiste à relancer le processus d'entraînement mais  en considérant les meilleurs hyperparamètres choisis dans la première étape, et ce, en utilisant l'ensemble de données final. Les paramètres finaux obtenus à la fin du traitement(les dernières valeurs du poids, biais) sont sauvegardés afin de les utiliser directement pour exporter le modèle. 
\subsection{Génération du modèle et intégration à Tiramisu } Après avoir sauvegarder la dernière architecture et paramètres du réseau, le modèle est prêt pour l'exporter sous forme de script pour l'intégrer à Tiramisu. Le modèle de prédiction  peut être appelé directement par la méthode \textit{"automatic\_unrolling()"} afin de prédire le meilleur facteur de déroulage pour de nouveaux programmes écrits en Tiramisu.

\begin{comment}
\par Au cours de notre projet nous avons développé deux modules essentiels, modèle d’extraction des caractéristiques du programme Tiramisu et le modèle de prédiction du bon facteur de déroulage :  
\begin{itemize}
 \item \textbf{Module d’extraction de features} : ce module développé en Tiramisu permet d’extraire les caractéristiques de n’importe quel programme écrit par Tiramisu, il est extensible et peut\-être utiliser comme extracteur de features pour résoudre n’importe quel problème d’optimisation ; problème de sélection de paramètre d’optimisation et le problème de sélection de bonne combinaison d’optimisations.
 \item \textbf{Modèle de prédiction du bon facteur de déroulage} : ce modèle développé en TensorFlow permet de prédire le facteur de déroulage pour un nid de boucle pour les programmes écrits en Tiramisu. Il peut être adapté et modifier pour être utiliser pour prédire d’autre paramètre pour d’autres optimisation et même pour d’autre langage.
\end{itemize}
\par Ces deux modules forment ensemble la méthode “Automatic\_unrolling” qui permet de prédire et appliquer le bon paramètre de déroulage sur les programmes écrits par l’utilisateur en Tiramisu. Si le paramètre renvoyé est le “0” cela veut dire qu’il ne faut pas utiliser cette optimisation donc la méthode elle ne va pas appliquer cette optimisation sur le code. L’utilisateur n’a qu’à appeler la méthode pour les nids de boucle qu’il veut optimiser par cette transformation puis la méthode fait le travail demandé toute seule. 
 \end{comment}

\section*{Conclusion}
\addcontentsline{toc}{chapter}{\textsc{Conclusion}}
\par Dans ce  chapitre , nous avons détaillé les différentes étapes de conception de notre solution. Nous avons discuté les différents choix considérés. Dans une deuxième partie, nous avons présenté l'environnement de développement ainsi que les phases de réalisations suivies. Dans le chapitre suivant, nous allons tester le modèle avec  différents scénarios  pour confirmer que le modèle est en mesure de bien prédire sur des cas réels. 
\end{onehalfspace}
%%%%%%%%%%%%%%%%%%%%%%%%%%% Test %%%%%%%%%%%%%%%%%%%

\chapter{\textsc{Tests et évaluation}}
\label{ch:tests}
\begin{onehalfspace}
\section*{Introduction}
\setcounter{footnote}{0}
\addcontentsline{toc}{chapter}{\textsc{Introduction}}
\par Après avoir détaillé notre conception et les étapes de la réalisation, nous exposons le processus d'évaluation du modèle prédicteur du meilleur facteur de déroulage. D’abord, nous avons comparé la précision de notre modèle avec d'autres modèles du \textit{machine learning}. Ensuite, il est évalué sur un ensemble de benchmarks. Nous comparons entre les résultats obtenus par notre modèle et ceux obtenus par la recherche exhaustive. 

\par Dans ce chapitre, nous commençons par décrire la plateforme matérielle de tests, les critères d’évaluation considérés et les benchmarks utilisés. Ensuite nous exposons les résultats obtenus pour enfin donner une synthèse sur les tests effectués.  

\section{Plateforme de tests}
\par Afin d’évaluer le prédicteur du meilleur facteur de déroulage, nous avons utilisé deux plateformes de tests suivant les deux phases de tests considérées.
 
\par
\begin{figure}[ht]
	\begin{center}
		\includegraphics[scale=0.6]{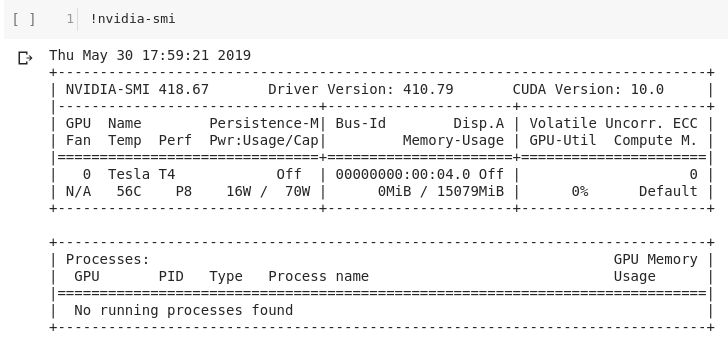} 
	\end{center}
	\caption{Les caractéristiques du GPU offert par la plateforme GoogleColab.}
	\label{fig:colab_gpu}
\end{figure}

 D’abord, pour l’évaluation initiale, nous avons évalué les perfomramnces du modèle en comparant sa précision avec d'autres algorithmes du \textit{machine learning}. Cette évaluation est effectué sur la même plateforme GoogleColabe (voir section \ref{envirement}). Cette plateforme offre un environnement de développement qui ne nécessite pas des configurations complexes ni des installations de bibliothèques. De plus, GoogleColabe offre actuellement l’accès à un processeur graphique GPU de type T4 avec une RAM de 16 GiB\footnote{Voir les caractéristiques du GPU sur \url{https://www.nvidia.com/fr-fr/data-center/tesla-t4/}.} (voir la figure \ref{fig:colab_gpu}). Nous avons lancé l'exécution du modèle sur le processeur GPU pour bénéficier de la puissance en calcul offerte et diminuer le temps d'entraînement du modèle.

\par Durant la deuxième phase de tests basée sur les benchmarks, nous avons utilisé les machines du cluster Lanka (voir section \ref{envirement}). Ce cluster dispose de plusieurs nœuds ayant les mêmes caractéristiques (voir tableau \ref{tab:Lanka_caracteristiques}). 
 Nous avons déployé notre système intégré au compilateur Tiramisu. Pour chaque benchmark, nous avons exécuté exhaustivement sur Lanka les instances de codes avec les différents facteurs de déroulage possibles (utilisés lors d'entraînement du modèle) pour définir le meilleur facteur. Utilisant la même plateforme, nous avons également exécuté les benchmarks pour la prédiction automatique par le modèle implémenté.

\begin{table}[h!]
  \centering
  \caption{Caractéristiques de l'architecture de test (évaluation par benchmarks).}
\noindent\makebox[\textwidth]{%
\begin{tabular}{| c | c | }
  \hline
 \textbf{Caractéristique} & \textbf{Performance}       \\ \hline \hline
 Processeurs &   Intel Xeon E5-2695 v2   \\  \hline
 Nombre des coeurs (par noeud)     &  12  \\  \hline
Fréquence  &   2.40GHz     \\  \hline
 Taille de la RAM  &  120GB    \\  \hline
  Taille du cache L1 (données/instructions)   &  32 Ko    \\  \hline
  Taille du cache L2  &  256 Ko    \\  \hline
  Taille du cache L3  &  18432 Ko   \\  \hline
Nombre de noeuds utilisés  &  10 à 15   \\  \hline
\end{tabular}
}
\label{tab:Lanka_caracteristiques}
  \end{table}

\section{Évaluation initiale du modèle}
\par L'évaluation initiale consiste à comparer les résultats que donne notre modèle avec d'autres modèles d'apprentissage automatique supervisé présentes dans des travaux précédents. Il s'agit de l'algorithme des plus proches voisins et les arbres de décision. 

\begin{itemize}[label=\textendash]
    \item \textbf{ L'algorithme de K plus proches voisins \textit{KNN}} : l'application de cet algorithme pour classer les programmes (individus) selon leurs facteurs de déroulage consiste à chercher des individus qui leurs ressemblent dans le \textit{dataset}. Deux programmes sont proches (ont le même facteur de déroulage) si leurs caractéristiques se ressemblent.
    \item \textbf{Les arbres de décision} : Ils sont plus rapides et plus efficaces comparés à l'algorithme de K plus proches voisins \textit{KNN}. La section (\ref{MLExploration}.\ref{arbre_decision}) donne plus de détail sur l'utilisation de cet algorithme dans le problème de sélection des optimisations.
\end{itemize}
\par La métrique d'évaluation considérée est la précision du modèle sur les données de tests. La précision du modèle représente le rapport entre le nombre des prédictions correctes et le nombre total des éléments dans le \textit{dataset} du test. Il s'agit de la même métrique utilisée pour évaluer les différents modèles générés lors de la phase d'entraînement afin de sélectionner le meilleur.\\  Le but est d'évaluer l'amélioration apportée par l'utilisation des réseaux de neurones pour résoudre le problème de sélection du facteur de déroulage. La précision de notre modèle est comparée aux deux algorithmes du \textit{machine learning} choisis (voir le tableau \ref{tab:compar_model}). Les résultats exposées présentent la moyenne des précisions suite à plusieurs exécutions effectuées. Après un certain nombre d'exécutions, la précision converge vers la même valeur.
\begin{table}[h!]
  \centering
  \caption{Comparaison de la présion de notre modèle de réseau de neurones avec les deux autres modèles de machine learning (KNN et arbre de décsiion).}
\noindent\makebox[\textwidth]{%
\begin{tabular}{| c | c | c| c |}
  \hline
 \textbf{Modèle} & Réseau de neurones &  \textit{KNN} & Arbre de décision \\ \hline \hline
 \textbf{Précision} & 20,39\% & 19.70\% & 19.23\% \\  \hline
\end{tabular}
}
\label{tab:compar_model}
 \end{table} 
 
\subsection{Analyse des résultats}
 \par Notre modèle est compétitif aux modèles de \textit{machine learning} utilisés dans des travaux précédents. Nous remarquons que les réseaux de neurones apportent une légère amélioration. Ceci prouve que les réseaux de neurones peuvent être utilisés pour résoudre le problème de prédiction du meilleur facteur de déroulage. \\ Nous rappelons que le but principal de notre projet est d'explorer la technique du \textit{deep learning} pour le problème du choix du meilleur facteur de déroulage. En effet, cette technique n'a pas été utilisée auparavant pour résoudre ce problème. Nous rappelons aussi que la précision atteinte  est jugée très acceptable malgré le manque significatif de données dû aux contraintes techniques.
\par Dans cette partie nous avons évalué l'effet apporté par l'utilisation de réseaux de neurones pour résoudre le problème de sélection de meilleur facteur de déroulage. Nous avons pu conclure que les réseaux de neurones peuvent être utilisés pour prédire le meilleur facteur de déroulage. Avec un nombre suffisant de données, le modèle peut atteindre un seuil de précision important. Dans la prochaine section, nous évaluons notre modèle sur un ensemble de benchmarks réels.
\section{Évaluation par Benchmarks}
\par Dans une deuxième phase, le système est évalué grâce à un ensemble de benchmarks (appartiennent à la classe de programmes choisie) que nous avons implémentés nous même.  \\
 Nous avons implémenté d'abord la méthode d'exploration exhaustive des facteurs de déroulage. Elle représente une référence pour comparer les résultats de prédiction que donne le modèle.
\par Pour chaque benchmark, nous avons lancé l'exploration exhaustive des facteurs de déroulage afin de définir le meilleur facteur et le comparer avec le facteur  prédit par le modèle. Nous avons répété ce processus pour différentes tailles de données d'entrées (voir tableau \ref{tab:modalitySize}) car la taille de données est un facteur déterminant qui influence fort la perdition. \\ Nous avons également répété les tests de chaque benchmarks pour différents \textit{schedules} possibles.  
\begin{table}[h!]
  \centering
  \caption{Modalités de la taille des données en entrée.}
\noindent\makebox[\textwidth]{%
\begin{tabular}{| c | c | }
  \hline
 \textbf{Modalité de la taille} & \textbf{Taille}  \\ \hline \hline
 Petite($T_p$)  &  256    \\  \hline
 Moyenne($T_m$)  & 1024    \\  \hline
 Grande ($T_g$)  & 2048  \\  \hline
\end{tabular}
}
\label{tab:modalitySize}
 \end{table}   
\par Pour évaluer les résultats, nous avons considéré les deux métriques d'évaluation suivantes :
 \begin{compitemize}

  \item [\textendash] \textbf{Le coût de prédiction $PC$} (\textit{prediction cost}) qui représente le rapport entre le temps d'exécution dont le facteur est obtenu par l'exploration exhaustive ($optimal\_exec$), avec le temps d'exécution dont le facteur de déroulage est prédit automatiquement($predit\_exec$). 
   \begin{center}
    PC = $ optimal\_exec / predit\_exec$ 
   \end{center}
 \item [\textendash] \textbf{Accélération $SP$} (\textit{speedup}) qui est le rapport entre le temps d'exécution sans application d'optimisation de déroulage ($sans\_exec$) et le temps d'exécution obtenu en appliquant le facteur prédit ($predit\_exec$).  
    \begin{center}
           SP = $sans\_exec / predit\_exec$ 
    \end{center}
 \end{compitemize}
\subsection{Benchmarks de tests}
\par Dans cette section, nous exposons les résultats obtenus de test sur les benchmarks. Nous commençons d’abord par présenter chaque benchmark et donner ses caractéristiques. Ensuite, nous exposons l'évaluation des performances suivie par une analyse des résultats obtenus. Le tableau \ref{tab:abv_benchmarks} représente la liste des benchmarks utilisés.
%%% Tableau banchmarks: caractéristiques 
\begin{table}[h!]
  \centering
  \caption{Liste des benchmarks d'évaluation.}
\noindent\makebox[\textwidth]{%
\begin{tabular}{| c | c | c | }
  \hline
 \textbf{Benchmark} & \textbf{Abréviation} & \textbf{Domaine} \\ \hline \hline
  Multiplication de Matrices $\times$ Matrice &  MM$\times$M & Algèbre linaire  \\  \hline
  Somme Matricielle  &  SMM & Algèbre linaire  \\  \hline
  Conversion Image RGB\_Gris  &  RGB\_gray & Traitement d'images\\  \hline
  Flouter Image  &  Blur & Traitement d'images  \\  \hline
  Convolution & Conv\_layer & Réseaux de neurones convolutifs  \\  \hline
 
\end{tabular}
}
\label{tab:abv_benchmarks}
 \end{table}   
\subsection{Benchmark MM$\times$M}
\par La multiplication de deux matrices constitue un noyau pour divers programmes scientifiques (voir l'algorithme \ref{algoMul}). MM$\times$M est considéré comme un benchmark gourmand en temps d'exécution. Ceci est dû notamment à la contrainte de localité de données qu'il impose. En effet, les accès en mémoires ne sont pas contigus. Ce fait influence l'efficacité du cache considérablement, et ce, pour les deux modes d'accès aux éléments de matrices que peut adopter l'architecture d'exécution\footnote{Ligne par ligne (\textit{row-major access}) ou colonne par colonne \textit{column-major access}.}. \\
L'effet qu'apporte l'optimisation du benchmark est remarquable. Le tableau \ref{tab:featuresMMM} présente un sous ensemble des caractéristiques (\textit{features})
du benchmark MM$\times$M données par le module d'extraction des \textit{features}. Le facteur \textit{Msize} peut prendre trois valeurs (voir le tableau \ref{tab:modalitySize}) selon le cas de test. \begin{algorithm}[h!]
\SetAlgoLined
\KwResult{Matrice produit de M1$\times$M2 }
 input M1("input00", {i0, i1});

 input M2("input01", {i0, i1});
computation mul\_init("mul\_init", {i0,i1},expr(0));

computation mul("mul", {i0, i1, i2}, mul\_init(i0, i1)+ M1(i0, i2) $\times$ M2(i2, i1));

\caption{Partie Algorithme (Tiramisu) du benchmark MM$\times$M }
\label{algoMul}
\end{algorithm}
%%% Tableau features
\begin{table}[h!]
  \centering
  \caption{Sous ensemble de caractéristiques du benchmark \textit{MM$\times$M}.}
\noindent\makebox[\textwidth]{%
\begin{tabular}{| c | c | }
  \hline
  Nombre de niveaux de boucle &  3  \\  \hline
  Étendu (\textit{span}) du niveau de boucle (i0, i1, i2) & \textit{Msize} \\
  \hline
  Nombre de chargements (unitaire)  &  3  \\  \hline
   Nombre d'opérandes & 6  \\  \hline
  Quantité de données chargées pour le niveau i0  & $3 (Msize)^2 $\\
   Quantité de données chargées pour le niveau i1 & $(Msize)^2 + 2 Msize $ \\
   Quantité de données chargées pour le niveau i2 & $2 Msize $ \\
  \hline
\end{tabular}
}
\label{tab:featuresMMM}
 \end{table}  
 
\textit{Msize} représente la taille de matrices (soupons que les deux matrices sont carrées de taille égale).
 
\subsubsection{Résultats de tests}
\par Les tests sont lancés pour les trois modalités de tailles de données. Pour chaque cas, nous avons également proposé un \textit{schedule} (voir tableau \ref{tab:testsMMM}). 
\begin{table}[H]
  \centering
  \caption{Résultats de tests sur le benchmark \textit{MM$\times$M}.}
\noindent\makebox[\textwidth]{%
\begin{tabular}{| c | c | c | c | c | c |}
  \hline
 \textbf{cas de test} & \textbf{predit\_exec\tablefootnote{Temps d'exécution du benchmark avec le facteur de déroulage prédit.} \begin{tiny}(ms) \end{tiny}} & \textbf{optimal\_exec\tablefootnote{Temps d'exécution du benchmark avec le facteur de déroulage optimal (par exploration exhaustive)}\begin{tiny}(ms) \end{tiny}} & \textbf{sans\_exec\tablefootnote{Temps d'exécution du benchmark sans application de déroulage.}\begin{tiny}(ms) \end{tiny}} &\textbf{PC\tablefootnote{Résultat du rapport de $optimal\_exec / predit\_exec$.}} &\textbf{SP\tablefootnote{Résultat du rapport de $sans\_exec / predit\_exec$.}} \\  \hline \hline
 \textbf{\textit{schedule0}}  & 0.152787 & 0.026849 &0.181856 & 0.176 & 0.966
 \\  \hline
\textbf{\textit{schedule1}}   &\ul{1.56327} &1.56327 &2.13072 &\ul{1} & \ul{1.362}\\  \hline
 \textbf{\textit{schedule2}}  &\ul{3.79004} &3.79004 &6.47257& \ul{1} &\ul{1.708}
\\  \hline
\end{tabular}
}
\label{tab:testsMMM}
 \end{table} 
Nous avons opté pour l'optimisation de tuilage de boucles (\textit{tile}) appliquée sur les deux niveaux ( i0 et i1). Ensuite, la parallélisation est toujours appliquée sur le niveau le plus externe. Les cas de tests sont les suivants: 
 \begin{compitemize}
  \item [\textendash] \textbf{\textit{schedule0} (pour le cas de taille $T_p$ des entrées): } le facteur de tuilage est 16.
 \item [\textendash] \textbf{\textit{schedule1} (pour le cas de taille $T_m$ des entrées): } seule l'optimisation de parallélisation est appliquée.
  \item [\textendash] \textbf{\textit{schedule2} (pour le cas de taille $T_g$ des entrées):} le facteur de tuilage est 32.
\end{compitemize}

%%%%%%% 2nd benchmark
\subsection{Benchmark \textit{SMM}}
\par Il représente la formule de somme matricielle générale $\alpha M + \beta N$ (voir l'algorithme \ref{algoAdd}). Nous avons sélectionné, dans le tableau \ref{tab:featuresSMM}, un sous ensemble de caractéristiques du benchmark SMM que donne le module d'extraction de \textit{features}.
\begin{algorithm}[h!]
\SetAlgoLined
\KwResult{Matrice de la somme de $\alpha M1 + \beta M2$ }
 input M1("input00", {i0, i1});

 input M2("input01", {i0, i1});
 
computation add("add", {i0, i1}, $\alpha M1(i0, i1) + \beta M2(i0, i1)$);

\caption{Partie Algorithme (Tiramisu) du benchmark SMM }
\label{algoAdd}
\end{algorithm}

\begin{table}[h!]
  \centering
  \caption{Sous ensemble de caractéristiques du benchmark \textit{SMM}.}
\noindent\makebox[\textwidth]{%
\begin{tabular}{| c | c | }
  \hline
  Nombre de niveaux de boucle &  2  \\  \hline
  Étendu (\textit{span}) du niveau de boucle (i0, i1) & \textit{Msize} \\
  \hline
  Nombre de chargements (unitaire)  &  2  \\  \hline
  Nombre d'opérandes & 6  \\  \hline
  Quantité de données chargées pour le niveau i0  & $2 (Msize)^2 $\\
   Quantité de données chargées pour le niveau i1  & $2 Msize $ \\
  \hline
\end{tabular}
}
\label{tab:featuresSMM}
 \end{table}  
 \textit{Msize} représente la taille de matrices (nous soupons également que les deux matrices sont carrées de taille égale).
\subsubsection{Résultats de tests}
\par Nous appliquons dans chaque cas de test (voir tableau \ref{tab:testsSMM}) l'optimisation de tuilage de boucles (\textit{tile}) sur les deux niveaux ( i0 et i1). Ensuite, l'optimisation de l'inversion de boucles (\textit{interchange}) entre le deuxième et le troisième niveau. L'optimisation de parallélisation est toujours appliquée sur le niveau le plus externe. Les cas de tests sont les suivants: 
 \begin{compitemize}
  \item [\textendash] \textbf{\textit{schedule0} (pour le cas de taille $T_p$ des entrées):} seule l'optimisation de déroulage est appliquée.
 \item [\textendash] \textbf{\textit{schedule1} (pour le cas de taille $T_m$ des entrées):} le facteur de tuilage est 16 avec l'application de l'inversion de boucles. 
  \item [\textendash] \textbf{\textit{schedule2} (pour le cas de taille $T_g$ des entrées):} le facteur de tuilage est 32 avec application de l'inversion de boucles.
\end{compitemize}

\begin{table}[h!]
  \centering
  \caption{Résultats de tests sur le benchmark SMM.}
\noindent\makebox[\textwidth]{%
\begin{tabular}{| c | c | c | c | c | c |}
  \hline
 \textbf{cas de test} & \textbf{predit\_exec\begin{tiny}(ms) \end{tiny}} & \textbf{optimal\_exec\begin{tiny}(ms) \end{tiny}} & \textbf{sans\_exec\begin{tiny}(ms) \end{tiny}} &\textbf{PC} &\textbf{SP} \\  \hline \hline
 \textbf{\textit{schedule0}}  &0.274662 & 0.025656 & 0.026884 &0.093 & 0.098
 \\  \hline
\textbf{\textit{schedule1}}   &0.0739955& 6.6859 & 0.056336 & 0.385 & 0.761\\  \hline
 \textbf{\textit{schedule2}}  &0.080874 & 0.080841 & 0.081542 &0.999 &\ul{1.008} \\  \hline
\end{tabular}
}
\label{tab:testsSMM}
 \end{table}
 %// Analyse des résultats 
 %%%%%%% 3rd benchmark
\subsection{Benchmark \textit{RGB\_gray}}
\par Le benchmark effectue la transformation d'un image constituée de trois couches : une couche rouge (R), une couche verte (G), une couche bleue (B) vers une image en niveaux de gris. (voir l'algorithme \ref{algogris}). Dans le tableau \ref{tab:featuresGray}, nous avons sélectionné un sous ensemble de caractéristiques définissant le benchmark \textit{RGB\_gray}.
\begin{algorithm}[h!]
\SetAlgoLined
\KwResult{Image en entrée convertie en niveaux de gris}
 input r\_input("r\_input",{x,y});
 
 input g\_input("g\_input",{x,y} );
 
 input b\_input("b\_input",{x,y});
 
computation griser("griser",{x,y},  f\_red * r\_input(x, y) +  f\_green *  g\_input(x, y) +  f\_blue * b\_input(x, y) );

\caption{Partie Algorithme (Tiramisu) du benchmark \textit{RGB\_gray}}
\label{algogris}
\end{algorithm}
%%% Tableau features
\begin{table}[h!]
  \centering
  \caption{Sous ensemble de caractéristiques du benchmark \textit{RGB\_gray}.}
\noindent\makebox[\textwidth]{%
\begin{tabular}{| c | c | }
  \hline
  Nombre de niveaux de boucle &  2  \\  \hline
  Étendu (\textit{span}) du niveau de boucle $(x, y)$ & \textit{Isize} \\
  \hline
  Nombre de chargements (unitaire)  &  3  \\  \hline
  Nombre d'opérandes & 13  \\  \hline
  Quantité de données chargées pour le niveau $x$  & $3 (Isize)^2 $\\
   Quantité de données chargées pour le niveau $y$  & $3 Isize $ \\
  \hline
\end{tabular}
}
\label{tab:featuresGray}
 \end{table}  
\par L'image est représentée sous forme de trois matrices carrées chacune ayant la taille  \textit{Isize}.
\subsubsection{Résultats de tests}
\par Nous appliquons dans chaque cas de test un \textit{schedule} donnée (voir tableau \ref{tab:testsgris}). L'optimisation de tuilage de boucles (\textit{tile}) est appliquée sur les deux niveaux (x et y). L'optimisation de parallélisation est toujours appliquée sur le niveau le plus externe. \\

\begin{table}[h!]
  \centering
  \caption{Résultats de tests sur le benchmark \textit{RGB\_gray}.}
\noindent\makebox[\textwidth]{%
\begin{tabular}{| c | c | c | c | c | c |}
  \hline
 \textbf{cas de test} & \textbf{predit\_exec\begin{tiny}(ms) \end{tiny}} & \textbf{optimal\_exec\begin{tiny}(ms) \end{tiny}} & \textbf{sans\_exec\begin{tiny}(ms) \end{tiny}} &\textbf{PC} &\textbf{SP} \\  \hline \hline
 \textbf{\textit{schedule0}}  &0.297978& 0.0197& 0.0241585 & 0.066& 0.081 \\  \hline
\textbf{\textit{schedule1}}  &0.042133 &0.0187 &0.0192395& 0.444 &0.457
\\  \hline
 \textbf{\textit{schedule2}}  &0.0926365 &0.0617655 &0.0621665 &0.667 &0.671
\\  \hline

\end{tabular}
}
\label{tab:testsgris}
 \end{table}
 \par Les cas de tests sont les suivants: 
 \begin{compitemize}
  \item [\textendash] \textbf{\textit{schedule0} (pour le cas de taille $T_p$ des entrées):} seule la parallélisation est appliquée.
 \item [\textendash] \textbf{\textit{schedule1} (pour le cas de taille $T_m$ des entrées):} le facteur de tuilage est 32.
  \item [\textendash] \textbf{\textit{schedule2} (pour le cas de taille $T_g$ des entrées):} le facteur de tuilage est 64.

\end{compitemize}

 %// Analyse des résultats 
 %%%%%%% 4th benchmark
\subsection{Benchmark \textit{Blur}}
\par Ce benchmark permet de flouter une image (en niveaux de gris). Il s'agit de calculer la moyenne des valeurs du voisinage pour chaque pixel (voir l'algorithme \ref{algoblur}). Dans le tableau \ref{tab:featuresblur}, nous exposons un sous ensemble de caractéristiques définissant le benchmark \textit{Blur}.
\begin{algorithm}[h!]
\SetAlgoLined
\KwResult{L'image en entrée grisée}
 input b\_input("c\_input",{x,y,c});

 computation blur\_x("blur\_x",{x,y,c}, (b\_input(x, y, c) + b\_input(x+1, y, c) + b\_input(x+2, y, c))/3);
 computation comp0("comp0",{x,y,c},( blur\_x(x, y, c) + blur\_x(x, y+1, c) + blur\_x(x, y+2, c)) /3 );

\caption{Partie Algorithme (Tiramisu) du benchmark \textit{Blur}}
\label{algoblur}
\end{algorithm}
%%% Tableau features
\begin{table}[h!]
  \centering
  \caption{Sous ensemble de caractéristiques du benchmark \textit{Blur}.}
\noindent\makebox[\textwidth]{%
\begin{tabular}{| c | c | }
  \hline
  Nombre de niveaux de boucle &  3  \\  \hline
  Étendu (\textit{span}) du niveau de boucle $(x, y, c)$ & \textit{Isize} \\
  \hline
  Nombre de chargements (unitaire)  &  3  \\  \hline
  Nombre d'opérandes & 10  \\  \hline
  Quantité de données chargées pour le niveau $x$  & $3 (Isize)^3 $\\
   Quantité de données chargées pour le niveau $y$  & $3 (Isize)^2 $ \\
   Quantité de données chargées pour le niveau $c$  & $3 Isize $ \\
  \hline
\end{tabular}
}
\label{tab:featuresblur}
 \end{table}  
L'image est représentée sous forme d'une matrice carrée dont la taille est  \textit{Isize}.
\subsubsection{Résultats de tests}
\par Le tableau \ref{tab:testsblur} représente les cas de tests effectués sur le benchmark \textit{Blur}. Nous avons appliqué seulement l'optimisation de parallélisation (sur le niveau le plus externe). Les cas de tests dépendent donc de la taille de données en entrée. 
\begin{table}[h!]
  \centering
  \caption{Résultats de tests sur le benchmark \textit{Blur}.}
\noindent\makebox[\textwidth]{%
\begin{tabular}{| c | c | c | c | c | c |}
  \hline
 \textbf{cas de test} & \textbf{predit\_exec\begin{tiny}(ms) \end{tiny}} & \textbf{optimal\_exec\begin{tiny}(ms) \end{tiny}} & \textbf{sans\_exec\begin{tiny}(ms) \end{tiny}} &\textbf{PC} &\textbf{SP} \\  \hline \hline
 \textbf{\textit{Isize}= $T_p$ }  &\ul{1.19716} & 1.19716 &1.32359 &\ul{1} & \ul{1.106}
 \\  \hline
\textbf{\textit{Isize}= $T_m$ }   &2.36202 &1.42873 &2.28374 &0.605 & 0.967
 \\  \hline
 \textbf{\textit{Isize}= $T_g$ }  &\ul{4.72417} &4.7241 &4.97623 &\ul{1} & \ul{1.053}
 \\  \hline
\end{tabular}
}
\label{tab:testsblur}
 \end{table}
 
%\par  // Analyse des résultats  
 %%%%%%% 5th benchmark
 
\subsection{Benchmark \textit{Conv\_layer}}
\par Ce benchmark\footnote{L'implementation  est inspiré du benchmark implémenté par l'équipe Tiramisu \url{https://github.com/Tiramisu-Compiler/tiramisu/tree/master/benchmarks/DNN/layers/convolution/}.} représente l'opération de convolution effectuée aux niveaux des couches de réseaux de neurones convolutifs \textit{CNN}.  L'algorithme reçoit en entrée une matrice de quatre dimensions et un filtre de trois dimensions. Le benchmark \textit{Conv\_layer} agit sur un lot de données (\textit{batch\_size}) dont la taille représente la dimension la plus externe de la matrice d'entrée.
\par La convolution est similaire à une opération de filtrage. Pour chaque position du filtre, les valeurs des deux matrices en superposition (filtre et image à traiter) sont multipliées. Chaque valeur ainsi inférée est projetée dans une nouvelle matrice (voir l'algorithme \ref{algoconv}).
\begin{algorithm}[h!]
\SetAlgoLined
\KwResult{L'image en entrée grisée}
input c\_input("c\_input",{n, z, y, x});

input filter("filter", {z1, k\_z , k\_y, k\_x});
   
 computation conv\_init("conv\_init",{n, z, y1, x1}, expr(0));
 
computation conv("conv",{n, z, y1, x1, k\_z, k\_y, k\_x }, conv\_init(n, z, y1, x1) + filter(z, k\_z, k\_y, k\_x) * c\_input(n, k\_z, y1 + k\_y, x1 + k\_x));

\caption{Partie Algorithme (Tiramisu) du benchmark \textit{Conv\_layer}}
\label{algoconv}
\end{algorithm}
 
\subsubsection{Résultats de tests}
\par Les cas de tests (voir le tableau \ref{tab:testsconv}) sont définis selon les tailles de la matrice d'entrée et le filtre. En effet, la taille du lot de données (\textit{batch\_size}) influence sur les performances. Nous considérons trois valeurs possibles de la taille du lot (64, 32, 8), et ce, selon la taille du  \textit{Data\_set} (petit, moyen, grand respectivement). \\
Nous avons appliqué seulement l'optimisation de parallélisation (sur le niveau le plus externe).
\begin{table}[h!]
  \centering
  \caption{Résultats de tests sur le benchmark \textit{Conv\_layer}.}
\noindent\makebox[\textwidth]{%
\begin{tabular}{|p{3.2cm}| c | c | c | c | c |}
  \hline
 \textbf{cas de test} & \textbf{predit\_exec\begin{tiny}(ms) \end{tiny}} & \textbf{optimal\_exec\begin{tiny}(ms) \end{tiny}} & \textbf{sans\_exec\begin{tiny}(ms) \end{tiny}} &\textbf{PC} &\textbf{SP} \\  \hline \hline
 \textbf{\textit{DataSet} petit}  &25.6503 &17.602 &17.602 & 0.686 & 0.687 \\  \hline
\textbf{\textit{DataSet} moyen}   &13.1555 &6.6859 & 6.6859 & 0.508 &0.508 \\  \hline
 \textbf{\textit{DataSet} grand}  &1.31926 &0.43521 & 0.43521&0.330 & 0.330 \\  \hline
\end{tabular}
}
\label{tab:testsconv}
 \end{table}
\section{Analyse et synthèse des tests}

\par Notre solution est testée sur un ensemble de 5 benchmarks que nous avons implémentés manuellement. Pour chaque benchmark nous avons proposé trois cas de tests qui dépendent de \textit{schedules} affectés ou de la taille de données en entrée.

\par
\begin{figure}[H]
	\begin{center}
		\includegraphics[scale=0.9, frame]{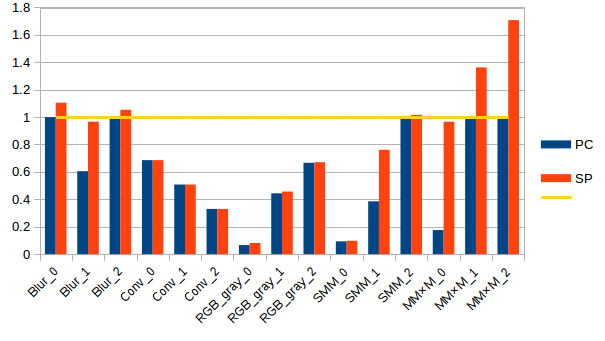}
	\end{center}
	\caption{Résultats du test sur les différentes cas du schedules du benchmarks.}
	\label{fig:test_synth}
\end{figure}

 Le modèle proposé arrive à prédire correctement le meilleur facteur de déroulage dans 4/15 des cas de tests effectués. Nous avons enregistré une amélioration (accélération) du temps d'exécution dans 5/15 des cas. La figure \ref{fig:test_synth} synthétise les résultats enregistrés.

\par Nous rappelons que $PC$ (coût de prédiction) représente le rapport entre le temps d'exécution dont le facteur est obtenu par l'exploration exhaustive (facteur optimal), avec le temps d'exécution dont le facteur de déroulage est prédit automatiquement. Quant à $SP$ (\textit{speedup}), c'est le rapport entre le temps d’exécution sans application d’optimisation de déroulage et le temps d’exécution obtenu en appliquant le facteur prédit. 

\par Nous remarquons dans la figure \ref{fig:test_synth} que les taux $PC$ et $SP$ varient d'un benchmark à un autre. En effet, pour le benchmark MM$\times$M et \textit{Blur}, nous avons enregistré des taux assez positifs. Ceci signifie que le modèle arrive à apprendre de bonnes prédictions pour gérer le problème de localité de données qui s'impose dans les deux benchmarks. 

\par Pour certains cas de tests, le modèle permet d'avoir une accélération ($SP$) suite à l'application de l'optimisation de déroulage. Dans d'autres cas, nous avons enregistré un $SP <1$ comme dans les trois cas de tests du benchmark  \textit{Conv\_layer}. Ceci prouve que l'optimisation de déroulage peut ne pas apporter d'amélioration.

\par Pour synthétiser, le modèle arrive à apprendre des caractéristiques de haut niveau (\textit{high level features\footnote{Caractéristiques plus complexes liées au comportement dynamique du programme tel que la gestion des chargements/écritures de données en mémoire et en cache, la gestion des registres, etc.}}) à partir de caractéristiques bas niveau (\textit{low level features\footnote{Des caractéristiques statiques basiques  qui décrivent le programme.}}). Il donne des prédictions du meilleur facteur de déroulage pour les nouveaux programmes\footnote{les programmes appartiennent à la  classe visée} avec une précision qui atteint 20\%. Ceci montre que le modèle apprend et il dépasse la prédiction aléatoire (dont la précision est de 14\%).
\par La précision de notre modèle est compétitive par rapport à d'autres modèles basés sur des algorithmes déjà explorés dans des travaux précédents. Ceci nous permet de déduire que les réseaux de neurones peuvent être utilisés pour le problème de prédiction du facteur de déroulage. \\
D'autre part, nous avons démontré que la précision de la prédiction augmente avec l'augmentation de la taille du \textit{dataset} utilisé pour entraîner le modèle. 

\section*{Conclusion}
\addcontentsline{toc}{chapter}{\textsc{Conclusion}}
\par Nous avons montré à travers ce chapitre que le modèle proposé est compétitif aux algorithmes du \textit{machine learning} utilisés pour résoudre le problème de sélection du facteur de déroulage. De ce fait, la solution basée sur les réseaux de neurones peut être explorée davantage pour améliorer la précision. 
\par Nous avons également évalué notre modèle sur un ensemble de banchmarks. Les résultats obtenus affirment que le modèle arrive bien à apprendre la prédiction du facteur de déroulage. La précision des prédictions s'améliore avec l'augmentation de la taille du \textit{dataset} utilisé pour l'entraînement. Nous avons pu générer un \textit{dataset} relativement petit, ceci empêche la précision de dépasser le seuil obtenu.

\end{onehalfspace}
\cleardoublepage
\thispagestyle{plain}
\addcontentsline{toc}{chapter}{\textbf{Conclusion générale et perspectives}}
\vspace{0.5cm}
  \mbox{\Huge{\textbf{Conclusion générale et perspectives}}}
\vspace{0.7cm}

\begin{onehalfspace}

\par Notre projet de fin d'étude s’inscrit dans le cadre du projet lancé par l'équipe COMMIT de MIT qui vise à améliorer le compilateur Tiramisu. Notre contribution consiste à concevoir et réaliser un modèle basé sur les réseaux de neurones pour automatiser le choix du meilleur facteur associé à l'optimisation de déroulage de boucles. Le modèle est conçu pour prédire sur des programmes écrits en Tiramisu (programmes déjà optimisés ou non optimisés) appartenant à une classe de programmes souvent utilisés dans Tiramisu (voir section \ref{classePgm}). Les architectures d'exécution considérées sont à base de CPU. 
\par Le projet vise également à explorer l'utilisation des réseaux de neurones profonds pour la résolution du problème de sélection du facteur de déroulage.

\par L’optimisation de code est une étape cruciale pour assurer une meilleure exploitation des ressources matérielles. Dans certains domaines, les systèmes conçus sont critiques ou nécessitent un calcul immense, l'optimisation devient indispensable.  L'optimisation de code est l'ensemble des techniques utilisées afin d'améliorer les performances d'un programme. En effet, la réduction du temps d’exécution constitue l’objectif principal de l’optimisation de codes. Maintes optimisations appliquées sur les boucles sont proposées car les boucles consomment plus de $90\%$ du temps d’exécution d’un programme. Chaque optimisation de boucles vise à améliorer certaines métriques influençant le temps d’exécution, à savoir l’utilisation du cache, le pré-chargement de données, l’utilisation des registres, etc.

\par Nous avons exposé les différentes optimisations de boucles. Nous nous sommes restreints aux optimisations de boucles utilisées dans le compilateur Tiramisu. L'optimisation de codes nécessite une expertise approfondie, du temps et beaucoup d'effort. C'est une tâche fastidieuse qui nécessite plusieurs tests pour trouver les meilleures combinaisons d'optimisations. Elle est aussi une tâche critique, car elle risque de dégrader les performances du programme au lieu de les améliorer. En effet, l'optimisation dépend de plusieurs paramètres comme les interactions entre les transformations et l'architecture matérielle d'exécution. L'automatisation de cette tâche permet de décharger le programmeur du travail fastidieux.
\par Plusieurs compilateurs intègrent un module d'optimisation automatique. Cependant, l'optimisation automatique du code présente plusieurs défis et reste contraignante. Les principaux problèmes d'optimisation automatique se résument dans le problème de sélection de bonnes combinaisons d'optimisations, le problème de sélection des meilleurs facteurs affectés aux optimisations sélectionnées et le problème d'estimation du temps d'exécution. En effet, il s'agit d'un problème NP-complet : la sélection de bonnes combinaisons d'optimisations de boucles à appliquer ainsi que leurs facteurs se fait sur un ensemble dont le cardinal dépend exponentiellement du temps de résolution. 
\newpage
\pagestyle{fancy}
\thispagestyle{plain}
\par Pour concevoir des techniques d'optimisation automatique, plusieurs approches sont adoptées. D'abord, l'approche exploratrice qui consiste à parcourir un ensemble d'optimisations puis tester et mesurer chacune des combinaisons afin de renvoyer celle qui minimise le temps d'exécution du programme. Cette approche est la plus précise, mais elle impose des restrictions sur la taille du problème, car elle nécessite un temps d'exécution considérable. Ensuite, l'approche analytique pour la prédiction du temps d'exécution. Elle se base sur un modèle défini en fonction de plusieurs paramètres souvent relatifs à l'architecture d'exécution. Or, ces modèles doivent être assez complexes pour donner de bons résultats. L'approche basée sur l'apprentissage automatique, permet de générer des modèles capables de prédire les bonnes optimisations à appliquer, leurs facteurs ou encore le temps d'exécution des programmes. La précision de cette approche dépend de l'algorithme d'apprentissage automatique choisi, la définition des hyperparamètres du modèle et de la taille du \textit{dataset} d'apprentissage dans le cas d'apprentissage supervisé.
\par Tiramisu représente le système hôte de notre modèle. C'est un nouveau langage et compilateur qui permet de générer des codes très rapides et de cibler différentes architectures matérielles (multicore, GPU, FPGA et systèmes distribués).

\par La conception de notre solution traite deux principaux modules : l'extraction automatique des caractéristiques (\textit{features}) de programmes et le modèle de prédiction du meilleur facteur de déroulage. Nous avons suivi un processus de développement itératif. Dans chaque phase, nous apportons des améliorations sur la conception du système en se basant sur les tests des métriques de performances.
\par Le module d'extraction automatique des caractéristiques de programmes que nous avons proposé agit en mode online, ce qui veut dire que l'extraction de \textit{features} se fait au moment de l'exécution. Ce mode permet d'avoir des informations nécessaires accessibles en cours d'execution. Le module est complètement intégré dans Tiramisu, ceci accélère les traitements d'extraction des caractéristiques. 

\par L'entraînement du modèle de réseau de neurones proposé nécessite un \textit{dataset} de taille immense (d'ordre de $10^6$) afin d'atteindre une précision élevée. Après deux mois de génération de données, nous avons pu construire un \textit{dataset} relativement petit ($3.6\times 10^4$). Ceci est dû aux contraintes de temps et de puissance de calcul des machines. En effet, la génération de \textit{datasets} immenses pour les utiliser dans les réseaux de neurones profonds consomme énormément de temps et nécessite des machines puissantes. Par exemple, le projet lancé par Facebook qui traite également le problème de sélection du facteur d'une optimisation de boucles (le tuilage) pour le langage Halide\footnote{Halide partage une logique très similaire à Tiramisu}. La collecte de données (\textit{dataset} de taille  $3 \times 10^6$ éléments) et l'ingénierie des caractéristiques (\textit{feature engineering}) ont pris plus d'une année, et ce, en utilisant les supercalculateurs de Facebook\footnote{Ces informations sont données par notre promoteur qui a travaillé sur ce projet}.   
 \newpage
\pagestyle{fancy}
\thispagestyle{plain}
 \par Pour évaluer notre modèle, nous avons comparé la précision de notre modèle avec deux autres algorithmes de \textit{machine learning} (K plus proches voisins et les arbres de décision) utilisés dans des travaux précédents pour le problème de sélection du bon facteur de déroulage. Les résultats permettent de vérifier que l'utilisation d'un modèle basé sur les réseaux de neurones profonds dans notre problème donne des résultats compétitifs. Ceci encourage à continuer la recherche afin d'améliorer davantage le modèle.\\  
 Nous avons également testé les performances du modèle sur un ensemble de benchmarks implémentés manuellement en Tiramisu. Dans $4/15$ des cas de tests, le modèle a prédit correctement le meilleur facteur de déroulage. Dans $5/15$ des cas, il a amélioré le temps d'exécution des programmes sur lesquels l'optimisation de déroulage n'a pas été appliquée.

\par Les résultats exposés dans la partie évaluation confirment que le modèle apprend à prédire, car la précision atteinte dépasse la prédiction aléatoire (14\%). Cependant, la précision de notre modèle est influencée par le manque de données. En effet, nous avons démontré que la précision de notre modèle augmente avec l'augmentation de la taille du \textit{dataset}.

\par Les résultats obtenus nous encouragent à améliorer davantage notre modèle. Nous avons défini les perspectives suivantes:
\begin{itemize}[label=\textendash]
    \item Générer plus de données pour améliorer la précision du modèle.
    
    \item Diversifier la génération de programmes en introduisant des nouvelles opérations (comparaison, max, etc.). D'une autre part, le générateur de codes se restreint sur les optimisations locales suivantes tuilage, interversion, déroulage de boucles et la parallélisation. Il faut améliorer le générateur de codes pour qu'il puisse générer toutes les optimisations locales de Tiramisu telles que la torsion de boucles (\textit{loop skewing}). Cette diversification des éléments du \textit{dataset} permet d'améliorer la précision du modèle.
    
    \item Étendre la solution pour traiter une classe de programmes plus large. En effet, notre solution résout le problème de sélection du facteur de déroulage pour une seule \textit{computation} (voir \ref{classePgm}). Des optimisations globales telles que la fusion de boucles permettent de regrouper un ensemble de nids de boucles (qui ne sont pas parfaitement imbriqués).

    \item Tester la solution sur le problème de sélection des paramètres pour d'autres optimisations de boucles tels que l'optimisation de tuilage et la vectorisation.

    \item Nous visons à long terme, l'extension de notre solution pour cibler d'autres plateformes notamment le GPU. Ceci permet d'explorer d'autres optimisations de boucles qu'offre Tiramisu.
 \end{itemize}   
\par Pour conclure, ce projet de fin d'études nous a permis d'explorer une nouvelle approche de résolution du problème de sélection du meilleur facteur de déroulage. Le modèle de prédiction basé sur les réseaux de neurones profonds donne des résultats encourageants pour continuer ce chemin et réaliser des versions plus améliorées.

\end{onehalfspace}
\part*{\textsc{Annexes}}
\fancyhead{}
\addcontentsline{toc}{part}{Annexes}
\pagestyle{fancy}
\renewcommand{\thechapter}{\Alph{section}}
\appendix

\fancyhf{}
\renewcommand{\footrulewidth}{0.4pt}
\cfoot{\thepage}

\begin{onehalfspace}

\chapter{ Détails sur l’optimisation de déroulage et de tuilage}

\label{appendix:A}
\section{Optimisation de déroulage}
 \par L'idée derrière l'optimisation de déroulage est de répliquer les instructions à l'intérieur de la boucle plusieurs fois. Le nombre de réplication est appelé le facteur de déroulage. Le nombre d'itération est divisé par ce facteur. Par exemple, si on réplique une fois le corps d'une boucle dont l'étendue est 100, le nombre d'itérations peut être réduit de 100 à 50.
 \par Cette transformation est clairement utile pour diminuer le nombre de branchements exécutés ainsi que les instructions de maintenances (avancement des adresses et comptage d'itérations). Pour montrer ça, considérant le code "C" ci-dessous qui permet de calculer la somme de 100 entrées du vecteur A.
 \begin{figure}[ht]
	\begin{center}
		\includegraphics[scale=0.8, frame]{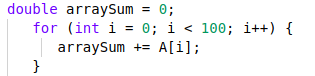}
	\end{center}
	\caption{Somme d’un vecteur de 100 entrées en "C".}
	\label{fig:annexA_unr_1}
\end{figure}

 \par
\begin{figure}[ht]
	\begin{center}
		\includegraphics[scale=0.67, frame]{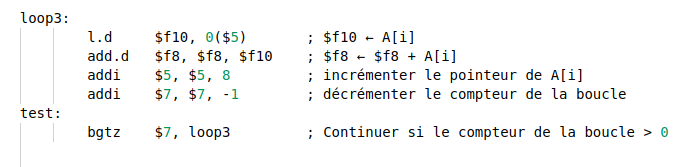} 
	\end{center}
	\caption[Le code assembleur équivalent au code C de la somme d’un vecteur de 100 entrées.]{Le code assembleur équivalent au code C de la somme d’un vecteur de 100 entrées~\protect\cite{umn_unrolling}.}
	\label{fig:annexA_unr_3}
\end{figure}

 Le code assembleur équivalent est donné dans la figure \ref{fig:annexA_unr_3} (Initialiser le compteur de la boucle (\$7) à 100, initialiser le vecteur arraySum (\$f10) à 0, initialiser le pointeur de A[i] (\$5) sur l'adresse de base de A). Les instructions “addi” dans ce code sont les instructions de maintenances de la boucle.

 \par
\begin{figure}[ht]
	\begin{center}
		\includegraphics[scale=0.7, frame]{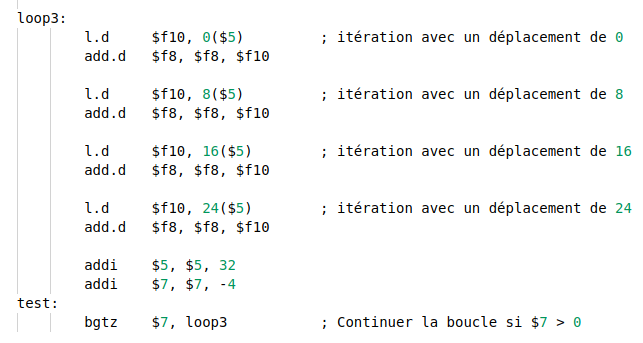}
	\end{center}
	\caption[Le code assembleur équivalent au code C de la somme d’un vecteur de 100 entrées après l’application de déroulage avec un facteur de 4.]{Le code assembleur équivalent au code C de la somme d’un vecteur de 100 entrées après l’application de déroulage avec un facteur de 4~\protect\cite{umn_unrolling}.}
	\label{fig:annexA_unr_2}
\end{figure}

 L’application de déroulage  sur le code précédent avec un facteur de 4 donne le code résultat de la figure \ref{fig:annexA_unr_2}. Notez que les déplacements dans les chargements (\textit{loads}) sont incrémentés par le compilateur dans les instructions répliquées (la deuxième, la troisième et la quatrième copies des instructions du corps de la boucle d'origine). La boucle avance avec un pas de 4, les valeurs immédiates dans les instructions "addi" ont été multipliées par 4, de telle sorte que l'effet d'une itération dans la boucle dont nous appliquons la transformation de déroulage est identique à celui de 4 itérations de la boucle d'origine. Les branchements et les instructions de maintenance de boucles ont été réduites d'un facteur de 4 \cite{umn_unrolling}.

 \par La transformation de déroulage s'applique facilement aux boucles de traitement de tableaux séquentiels où le nombre d'itérations est connu avant l'exécution de la boucle. Évidemment, il faut bien choisir le facteur de déroulage pour ne pas dégrader les performances au lieu de les améliorer.
 
\subsection{Avantages de l'optimisation de déroulage}
 \par L'optimisation de déroulage a des effets directs et indirects sur le code que le compilateur peut produire sur une boucle donnée. Les performances finales dépendent de ces deux effets. Les avantages directs peuvent être la réduction des instructions de test et de branchement et la réduction du trafic mémoire en créant des réutilisations dans le corps de la boucle. Parmi les effets indirects \cite{science_direct_unrolling} :
\begin{itemize}
  \item L'augmentation du nombre d'opérations indépendantes dans le corps de la boucle peut améliorer le \textit{schedule} des instructions. Avec plus d'opérations, le planificateur (\textit{scheduler}) a plus de chances de garder plusieurs unités fonctionnelles occupées et de masquer la latence des opérations de longues durées telles que les instructions de branchement et les accès à la mémoire.
  \item Le déroulage peut déplacer des accès mémoire consécutifs dans la  même itération de boucle, où le compilateur peut les \textit{scheduler} ensemble, ce qui permet d’améliorer la localité. 
  \item Cette optimisation peut améliorer la réutilisation des registres ce qui affecte radicalement la vitesse du code.
\end{itemize}

\subsection{Inconvénients du déroulage}
 \par L'optimisation du déroulage est très utile comme elle améliore indirectement nombreux aspects de performance de système (l'architecture du registre, système de mémoire, etc). Puisque cette transformation peut avoir des effets secondaires, il est difficile de décider quand l'application de cette transformation est appropriée. Superficiellement, cette optimisation apparaît comme une optimisation qui est toujours bénéfique. Cependant, elle peut nuire aux performances dans certains cas. Les inconvénients possibles de cette optimisation sont \cite{supervised2005} :
\begin{itemize}
  \item L'inconvénient le plus avéré de cette optimisation est qu’elle peut provoquer l'expansion du code ce qui peut dégrader les performances du cache d'instructions.
  \item Il est possible que la boucle déroulée demande plus de registres que l'originale. Dans le cas ou cette demande entraîne des \textit{spills}\footnote{S'il n'y a pas assez de registres pour contenir toutes les variables, certaines variables peuvent être déplacées vers et depuis la RAM. Ce processus s'appelle "renverser" (\textit{spills}) les registres.} (stockage et rechargement) supplémentaire, le trafic mémoire résultant peut annuler les avantages potentiels de l'optimisation. 
  \item Si le compilateur ne peut pas déterminer qu’une  boucle peut prendre une sortie précoce, il faudra qu’il ajoute un flux de contrôle\footnote{ Flux d'instructions ou \textit{Control flow} : sont en particulier des instructions de contrôles : if, for, while, break, appels de fonctions, etc.} (\textit{control flow}) pour la boucle déroulée, ce qui risque de nier ou neutraliser les avantages de l’optimisation.
\end{itemize}

\section{Optimisation de tuilage}
\par Le tuilage est une optimisation qui transforme une boucle en une autre boucle, qui effectue le même calcul, mais dans un ordre différent. Elle divise les données de la boucle en sous-blocs de petites tailles déterminées par le facteur de tuilage. Ce partitionnement en blocs permet aux éléments de données requises par un bloc de tenir dans la mémoire locale. Ceci permet d'améliorer la localité de données et d'exploiter le parallélisme. Chaque boucle est transformée en deux boucles : une itère à l'intérieur de chaque bloc (intra-tuile) et l'autre itère à travers les blocs (inter-tuile). Cependant, le choix du facteur de tuilage est un problème critique comme il affecte considérablement les performances du programme. Si le facteur de tuilage est trop petit, il conduit à une mauvaise exploitation de ressources (ex : cache de données). Tandis qu'un facteur de tuilage trop grand augmente les défauts de cache et affaiblit ou dégrade les performances.
\par Le déroulage consiste en deux étapes : le découpage en bandes  de la boucle (\textit{strip mining}) et la permutation de boucles. Durant le découpage en bandes, la boucle est divisée en intra-tuile et inter-tuile (\textit{intra-tile} et \textit{inter-tile}) boucles\footnote{La première (intra-tile) itère à l’intérieur des blocs et la dernière (inter-tile) itère à travers les boucles.}. Pendant la phase de permutation, les boucles inter-tuiles sont déplacées vers l’extérieur. 

\par
\begin{figure}[ht]
	\begin{center}
		\includegraphics[scale=0.60]{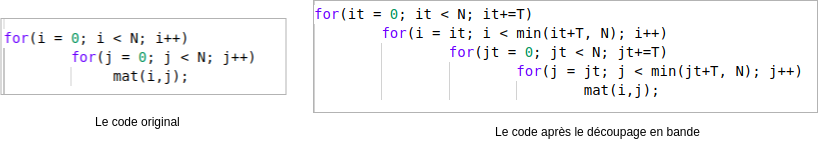} 
	\end{center}
	\caption[Découpage en bande du code original avec un facteur de 4]{Découpage en bande du code original avec un facteur de 4~\protect\cite{hTSS2017}.}
	\label{fig:annexA_tile3}
\end{figure}

 Dans la figure \ref{fig:annexA_tile3}, nous avons découpé la boucle originale "i" en intra-tuile boucle "i" et inter-tuile boucle "it". La boucle originale "j" est aussi découpée en intra et inter-tuile boucles "j" et "jt" respectivement. 
\par
\begin{figure}[ht]
	\begin{center}
		\includegraphics[scale=0.65, frame]{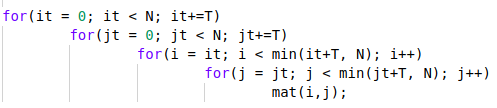} 
	\end{center}
	\caption[Le code après avoir effectué la permutation.]{Le code après avoir effectué la permutation~\protect\cite{hTSS2017}.}
	\label{fig:annexA_tile5}
\end{figure}

 L'étape suivante est de permuter les boucles résultantes de telle sorte que les boucles inter-tuile soient déplacées à l'extérieur et les boucles intra-tuile soient déplacées à l'intérieur (voir la figure \ref{fig:annexA_tile5}).
 \par
\begin{figure}[ht]
	\begin{center}
		\includegraphics[scale=0.65]{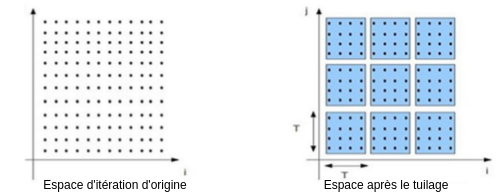} 
	\end{center}
	\caption[Illustration de la division de l’espace de données après l’application de tuilage.]{Illustration de la division de l’espace de données après l’application de tuilage~\protect\cite{hTSS2017}.}
	\label{fig:annexA_tile6}
\end{figure}

 Dans le code ci-dessus, "T" est le facteur de tuilage. Le choix du facteur "T" détermine si le cache sera bien ou mal utilisé. Le tuilage de la boucle originale peut être présentée comme dans la figure \ref{fig:annexA_tile6}.

\subsection{Aspects à considérer dans la conception du modèle de tuilage}

\par Plusieurs modèles sont conçus pour prédire le facteur de tuilage le plus optimal. Toutefois, la conception de tels modèles est très complexe. La difficulté de sélection de la taille des blocs (\textit{Tile Size Selection}) est dû à la hiérarchie de mémoire, complexité du cache, préfecheur matériel (\textit{Hardware prefetcher}), optimisations complexes du compilateur et l'évolution de l'architecture matérielle. Pour construire un modèle plus précis différents aspects doivent être pris en considération \cite{Rahmani2010TSS} :
\begin{itemize}
    \item la réutilisation de données dans l'exécution d'un bloc.
    \item La réutilisation de données entre les blocs.
    \item La disposition de données utilisées par un bloc dans la mémoire.
    \item La pénalité relative au défaut de cache dans chaque niveau de la hiérarchie, qui dépend de la machine.
    \item la politique de remplacement du cache.
    \item l'interaction avec d'autres unités, tel que le préfecheur.
    \item L’interaction avec d’autres optimisations (la vectorisation) pour en profiter de cette dernière.
\end{itemize}

\subsection{Approches adoptées de choix du facteur de tuilage optimal}
\par Différentes approches proposées pour trouver le bon facteur de tuilage. Elle sont classées en plusieurs catégories\cite{hTSS2017} :
\begin{itemize} 
    \item \textbf{Approches basées sur les modèles statiques} : ce sont des modèles analytiques développés en se basant sur l'observation de l'architecture matérielle, certains détails du logiciel et du comportement d'un ensemble de programmes. Ils donnent directement en sortie le facteur de tuilage optimal pour le programme donné sur une architecture spécifique. L'inconvénient majeur de ce type de modèles, c'est qu'ils ne sont pas portables entre les différents types d'architectures. L'évolution rapide dans les architectures matérielles rend l'utilisation de ces modèles plus difficile.
    
    \item \textbf{Approches basées sur des modèles conduits par la recherche empirique} : ces approches exécuteent le code pour chaque facteur de tuilage dans l'espace de recherche, le facteur de tuilage qui donne le meilleur temps d'exécution est sélectionné. L'avantage de ces modèles est qu'ils sont portables entre les différentes architectures. Cependant, ces approche ne peuvent pas être adoptées lorsque l'espace de recherche est très grand, comme elles doivent explorer toutes les combinaisons possibles. Par exemple, il est impossible d'utiliser ces approches pour trouver les tailles de tuilages rectangulaires optimales (des blocs de tailles rectangulaires), car l'espace de recherche de ces tailles est exponentiellement plus grand que celui des tuilages carrés.
    
    \item \textbf{Approche basés sur l'apprentissage automatique} : les modèles basées sur ces approches deviennent plus populaires grâce à leurs portabilités, ainsi qu'ils peuvent être utilisés pour les deux types de tuilage (rectangulaire et carré). Ces modèles sont entraînés avec un petit sous-ensemble de l'espace de recherche global, et donc ils peuvent être utilisés pour le problème de sélection de facteur de tuilage rectangulaire.
\end{itemize}
\newpage

\chapter{Commandes de scheduling en Tiramisu }
Les commandes de scheduling sont classées en quatre principaux types : les commandes de transformation des nids de boucles, les commandes pour mapper les niveaux des boucles sur l'architecture matérielle, les commandes de manipulation des données et les commandes de synchronisation
\cite{Baghdadi2018TiramisuV3}.
\section{Commandes de transformation des nids de boucles}
\par
\begin{figure}[ht]
	\begin{center}
		\includegraphics{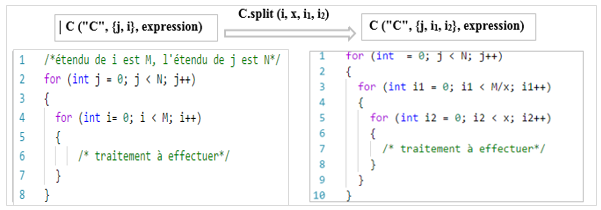}
	\end{center}
	\caption{transformation suite à l'application de la commande split.}
	\label{fig:exmp_split_Tiramisu_links}
\end{figure}

 Il s'agit des commandes d'optimisation qui engendrent des transformations sur la structure du nid de boucles, à savoir éliminer des niveaux de la boucle, les diviser ou les dérouler. Par exemple, soit la \textit{computation} $C$ ("C", {$j$, $i$}, $expression$), le fait de lui appliquer la commande de découpage en bandes $C$.split ($i$, $x$, $i{_1}$, $i{_2}$) change sa structure en éliminant le niveau de boucle $i$ et en produisant deux nouveaux niveaux de boucle $i{_1}$, $i{_2}$, avec l’étendue de $i_2$ est $x$ et celui de $i{_1}$ est égal à l'étendu de $i$ divisé par $x$.  La figure \ref{fig:exmp_split_Tiramisu_links} montre la transformation de la computation $C$.

\section{Commandes du mappage des niveaux des boucles sur l'architecture matérielle}
\par Ces commandes permettent de définir comment les niveaux de boucles seront lancées (parallélisés ou vectorisés) selon l'architecture cible, à savoir les niveaux à exécuter sur des blocs GPU ou sur des nœuds donnés dans le cas distribué.\\ Par exemple, la commande C.gpu ($i{_0}$, $i{_1}$, $i{_2}$) permet de spécifier que les niveaux $i{_0}$, $i{_1}$, $i{_2}$ seront exécutés sur le GPU.
\section{Commandes de manipulation des données}
\par Utilisées pour la gestion et la structuration de données. Il permet principalement les manipulations suivantes:
\begin{compitemize}
	\item [\textendash] allocation des tableaux (tampons). 
	\item [\textendash] définition des propriétés des tampons à savoir, s'ils sont stockés dans la machine hôte, dans un périphérique, dans la mémoire partagée ou encore dans la mémoire locale (GPU).
	\item [\textendash] copier les données (entre les différents niveaux hiérarchiques la de mémoire ou entre les nœuds dans le cas de distribué).
   \item [\textendash] configuration des accès aux tampons.
\end{compitemize}
\par Le programmeur utilise principalement un ensemble de commandes de haut niveau. Cependant, si ces dernières ne sont pas assez précises pour exprimer le niveau de détail souhaité, le programmeur fait recours aux commandes de bas niveau.

\section{Commandes de synchronisation}
\par Il s'agit des directives assurant la communication et la synchronisation. Par exemple, la commande barrier\_at (p, i)  permet de définir une barrière de synchronisation\footnote{Selon Michel Riveill (dans le cours Communication, Coopération et Concurrence entre processus, 2007) les barrières permettent de synchroniser les processus d’un groupe à un endroit donné du programme. Lorsque c’est le cas, chacun des processus peut reprendre son travail.} de la boucle $p$ au niveau $i$. En revanche, pour gérer automatiquement les domaines d'itération pour les opérations de synchronisation, le programmeur utilise barrier\_at () sans préciser les niveaux de synchronisation ce qui le dispense de calculer les domaines d'itération, notamment quand il explore plusieurs \textit{Schedules} pour en tirer le meilleur\cite{Baghdadi2018TiramisuV3}. 
  
\par Il est important de signaler que l'appel de certaines commandes telles que cache\_shared\_at(), cache\_local\_at(), allocate\_at(), copy\_at(), barrier\_at(), etc. retourne une opération Tiramisu\footnote{\textit{Operation} en Tiramisu est un type héritant de \textit{computation} (il peut être donc planifié dans la patrie Schedule) mais il retourne aucune valeur.} sur laquelle il est possible d'appliquer les commandes de scheduling.
\par Nous exposons quelques commandes de scheduling en Tiramisu. Supposons que $C$ et $D$ sont deux \textit{computations}, $b$ est un tampon et $i$ et $j$ sont des itérateurs de boucle.
\newpage
\begin{table}[H]
\caption{Exemple de Commandes de la parite \textit{Schedule} de Tiramisu}
\noindent\makebox[\textwidth]{%
\begin{tabular}{|p{4.5cm}|p{9cm}|p{5cm}|}
 \hline
\textbf{Commande} &  \textbf{Description}  & \textbf{Nom équivalent}  \\ 
 \hline
 \rowcolor{gainsboro}\multicolumn{3}{|c|}{\textbf{Commandes de transformations des nids de boucle}} \\
   \hline
C.split (i, s, $i{_0}$, $i{_1}$)
 & 
Découper le niveau de boucle $i$ en produisant deux nouveaux niveaux de boucle $i{_0}$, $i{_1}$. Où l’étendue de $i{_1}$ est $s$.
 &
 Découpage en bandes \\ 
  \hline
C.interchange ($i$, $j$) &
Invertir les niveaux de la boucle entre $i$ et $j$   &
Inversion de boucle  \\
  \hline
C.tile (i, j, $t{_1}$, $t{_2}$, $i{_0}$, $j{_0}$, $i{_1}$, $j{_1}$) &
Découper les niveaux ($i$, $j$) de $C$ en produisant des nouveaux niveaux ($i{_0}$, $j{_0}$, $i{_1}$, $j{_1}$) avec ($i{_0}$, $j{_0}$) comme niveaux externes et ($i{_1}$, $j{_1}$) comme niveaux internes ayant l'étendu $t{_1}$ et $t{_2}$respectivement.  & 
Tuilage de boucle  \\
  \hline
C.unroll ($i$, $n$) & 
Dérouler le niveau $i$ de la boucle $C$ avec un facteur de $n$.  & 
Déroulage de boucle   \\
  \hline
C.skew ($i$, $j$, $i{_1}$, $j{_1}$) & 
Appliquer la torsion de boucles sur les niveaux $i$ et $j$ de $C$. $i{_1}$ et$j{_1}$ sont les noms des nouveaux niveaux. &
Torsion de boucles \\
 \hline
C.after ($D$, $i$) & 
$C$ doit être exécutée après $D$ au niveau $i$ ($C$ et $D$ ont le même ordre d'exécution avant le niveau $i$). &
\text{Ordonnancement de boucles}\\
 \hline
C.compute\_at (D, $j$) & 
Calculer $D$ dans le nid de boucle $D$ au niveau $j$, cela pour assurer que les valeurs à consommer par $D$ sont prêtes. Ceci peut engendrer des calculs redondants.   &
Calcul redondant   \\
 \hline
 \rowcolor{gainsboro}\multicolumn{3}{|c|}{\textbf{Commandes pour mapper les niveaux des boucles sur l'architecture matérielle}}\\
  \hline
C.parallelize ($i$) & 
Paralléliser la boucle C au niveau $i$. &
Parallélisation de boucle.   \\
 \hline
C.vectorize ($i$, $n$) & 
Vectoriser le niveau $i$ par un vecteur de taille $n$.   &
Vectorisation de boucle.   \\
 \hline
C.tile\_gpu ($i{_0}$, $i{_1}$, $t{_1}$, $t{_2}$) & 
Tuilage de C aux niveaux $i{_0}$ et $i{_1}$ avec un facteur de $(t{_1}\times$ $t{_2})$ et les mapper au GPU.  &
Tuilage de boucles (GPU).   \\
 \hline
C.distribute ($i$) & 
Distribuer C au niveau $i$ (marquer $i$ comme nœud).  &
\text{Distribution de boucles}
\text{(systèmes distribués).}   \\
 \hline
 \rowcolor{gainsboro}\multicolumn{3}{|c|}{ \textbf{Commandes de manipulation des données}}\\
 \hline
C.store\_in (b {$i$, $j$})& 
Sauvegarder le résultat de $C$ ($i$, $j$) dans b [i, j].  &
Copie des résultats dans les tampons (commande de haut niveau).   \\
 \hline
 \rowcolor{gainsboro}\multicolumn{3}{|c|}{ \textbf{Commandes de synchronisation} } \\
 \hline
barrier\_at(C, $i$))& 
Crée une barrière au niveau $i$ de $C$.&
Barrière de synchronisation\\
 \hline
\end{tabular}}

\label{table_Commandes_Tiramisu}
\end{table}

%%%%%%% Annexe C
\newpage
\chapter{ Principaux concepts du Réseaux de neurones}

\par
\begin{figure}[H]
	\begin{center}
		\includegraphics[scale=0.5]{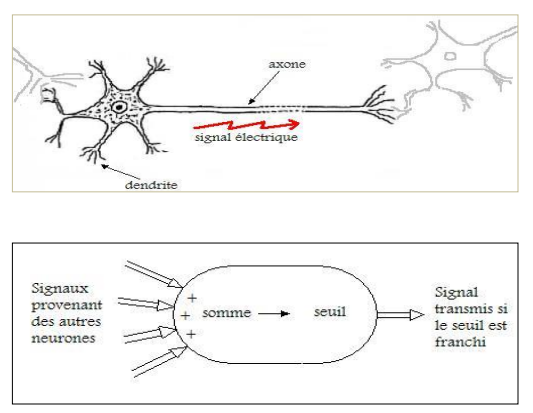}
	\end{center}
	\caption{Comparaison entre la structure d'un réseau de neurones artificiel et un réseau de neurones biologique.} 
	\label{fig:Ann_bio}
\end{figure}

 Les réseaux de neurones artificiels sont des systèmes inspirés du fonctionnement des neurones biologiques (voir la figure \ref{fig:Ann_bio}). L'un des types de réseaux de neurones les plus fameux est le perceptron multicouche dit également réseaux de neurones multicouches (\textit{MLPs} en anglais).
\par Dans cette partie, nous allons voir les notions de base du perceptron multicouche. Les MLPs sont un cas particulier de réseaux de neurones, mais très souvent lorsque nous disant réseaux de neurones nous pensons aux MLPs comme il est le cas le plus simple et le plus répondu. Bien qu'il présente un autre type de réseau de neurones comme les réseaux de neurones convolutifs, récurrents, etc.

\section{Principe général}
\par L'idée générale derrière les réseaux de neurones est la séparation de l'espace de données pour fournir une sortie. Grâce à cette séparation, qu'ils peuvent pour une nouvelle donnée de décider à quel ensemble doit-elle appartenir. Si nous avons deux classes par exemple, le modèle à partir des entrées essaye d'apprendre la différence entre eux. Ensuite, il tracera une droite de séparation entre ces deux types de données. Chaque nouvelle entrée sera positionner soit à droite ou à gauche de la droite.
\par Le réseau de neurones multicouches est un ensemble de neurones organisés en couche. Chaque couche se propage le signal d'entrée à la couche suivante jusqu'à la sortie, en activant ou non au fur et à mesure les neurones. 
\par Une fois la sortie est fournie par le modèle, elle est comparée avec la sortie attendue, puis les liaisons entre les neurones sont mises à jour pour améliorer les résultats à chaque fois.
\section{Neurone}
\par
\begin{figure}[H]
	\begin{center}
		\includegraphics[scale=0.75]{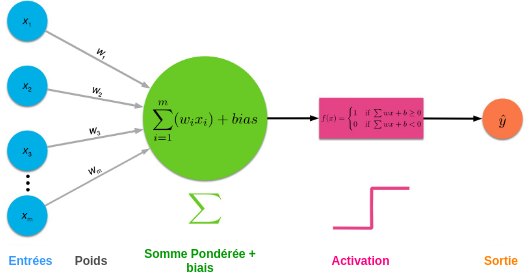}
	\end{center}
	\caption{La structure d'un réseau de neurones artificiel.} 
	\label{fig:AnnexC_structureNeurones}
\end{figure}

 Les neurones sont l’unité principale qui construit les réseaux de neurones. Ce sont des simples unités de calcul qui ont des signaux d'entrée pondérés. Il produisent un signal de sortie à l'aide d'une fonction d'activation (voir la figure \ref{fig:AnnexC_structureNeurones}).
\subsection{Poids (\textit{weights})}  
\par Des signaux ($x_0$, $x_1$, …, $x_n$) arrivent au neurone par chaque neurones des couches précédentes. Chaque lien qui amène le signal est pondéré, respectivement $w_0$, $w_1$, … ,$w_n$ . C'est ces poids qui vont être adaptés tout au long de l'apprentissage pour permettre au réseau de prédire efficacement (en général il prend des valeurs entre $0$ et $1$ ou $-1$ et $1$). La somme de tous ces signaux pondérés est calculé après ($ sum_{i=1}^{N}  w_i * x_i $) et un certain biais ($b$) est ajouté.
\subsection{Biais (\textit{bias})}
\par Le biais peut être vue comme un neurone externe supplémentaire. La valeur du biais permet de décaler la fonction d'activation vers la gauche (désactiver le neurone) ou vers la droite (activer le neurone). Ceci peut être essentiel pour l'apprentissage.
\subsection{Activation (\textit{activation})}
\par Les entrées pondérées sont additionnées et transmises via une fonction d'activation (\textit{activation function}), parfois appelée fonction de transfert, pour obtenir le signal de sortie. Elle est appelée une fonction d'activation, car elle régit le seuil d'activation du neurone.
\begin{itemize}
    \item La formule de sortie des neurones cachés est de la forme $y= factivation(b + sum_{i=1}^{N}  w_i * x_i)$ 
    \item Aucun calcule n’est effectuée aux neurones d’entrée.
    \item La formule des neurones de sortie est : $y= sum_{i=1}^{N}  w_i * x_i)$. 
\end{itemize}
\par Dans la pratique, les poids et les biais sont initialisés au hasard lors de la création du réseau de neurones. Au niveau de la fonction d'activation, aucun calcule n'est effectuée pour les neurones d'entrée. Pour les couches cachées plusieurs fonctions d'activation peuvent être considérées (sigmoïde, relu, …). La sortie est fournie directement, mais quelques fonctions peuvent être appliquées comme “\textit{softmax}” au cas de classification.
\section{Couches du MLP}
\begin{figure}[H]
	\begin{center}
		\includegraphics[scale=0.5]{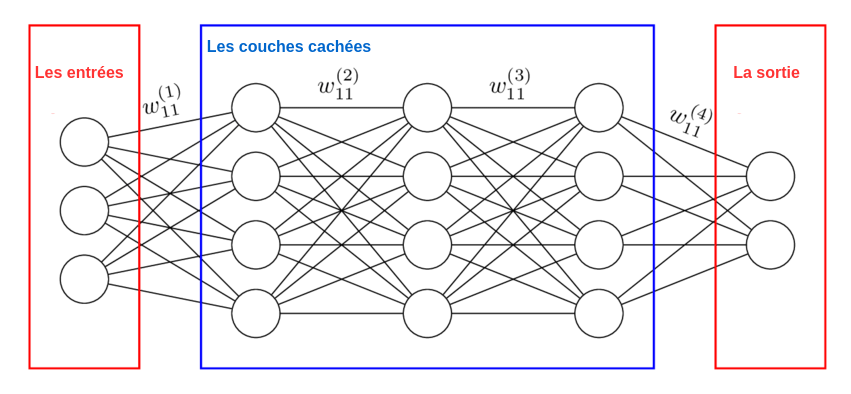}
	\end{center}
	\caption{Les couches principales d'un réseau de neurones.} 
	\label{fig:AnnexC_coucheseNeurones}
\end{figure}

Les réseaux de neurones multicouches sont composés de trois parties principales (voir la figure \ref{fig:AnnexC_coucheseNeurones}) :
\begin{itemize}
    \item \textbf{Couche d'entrée (\textit{input layer})} : elle forme la première couche du modèle. Elle est constituée d'un ensemble de neurones qui portent les entrées du modèle. Tous les neurones de cette couche d'entrée sont reliés aux neurones de la couche suivante.
    \item \textbf{Couches cachées (\textit{hidden layers})} : les couches postérieures à la couche d'entrée sont appelées couches cachées, car elles ne sont pas directement exposées à l'entrée. Ce sont un ensemble de couches où chacune est formée d'un ensemble de neurones. Le réseau de neurones peut avoir plusieurs couches cachées. Lorsque le nombre de couches cachées est supérieur à deux le réseau est appelé un réseau de neurones profond. Le choix du bon nombre de neurones par couche et le bon nombre de couches est très difficile. Généralement, deux couches suffisent pour la plupart des problèmes. Aller au-delà de 6 à 10 couches cause très souvent des problèmes de surapprentissage\footnote{Le problème de surapprentissage apparaît lorsque le réseau  a tellement appris qu’il ne peut plus généraliser.} (overfitting). En pratique, on a souvent au moins autant de neurones par couche que ce qu'on avait d'entrées.
     \item \textbf{Couche de sortie (\textit{output layer})} : la couche finale est appelée la couche de sortie. Elle représente le résultat final (la prédiction) de notre réseau. 
\end{itemize}
\section{Entraînement du Réseau de neurones}
Une fois configuré, le réseau de neurones doit être entrainé sur un ensemble de données appelé \textit{dataset} d'entainement. Ce processus est effectué sur plusieurs étapes: 
\subsection{Préparation de données}
\par D'abord, il faut préparer les données pour former le réseau de neurones. Les données fournies aux réseaux de neurones doivent être numériques. Si le \textit{dataset} contient des valeurs qui ne sont pas numériques il faut les transformer en valeurs numériques. L'une des techniques de transformation qui peut être utilisé pour ça est appelée le codage one-hot\footnote{Le hot encoding est l’une des techniques utilisée pour transformer les données non numériques en données numériques. Si nous disposons de données catégorielles, telles qu’un attribut de sexe avec les valeurs «masculin» et «féminin», nous pouvons utiliser le codage one-hot pour les représenter en valeur réelle. C'est ici qu'une nouvelle colonne est ajoutée pour chaque valeur de classe (deux colonnes dans le cas du sexe masculin et féminin) et un 0 ou 1 est ajouté pour chaque ligne en fonction de la valeur de classe pour cette ligne.}. Ce principe est aussi appliqué sur la variable de sortie dans les problèmes de classification multiple. Ce qui permet de facilité la classification. Les réseaux de neurones nécessitent que l'entrée soit mise à l'échelle d'une manière cohérente. Elles peuvent être redimensionnées dans la plage comprise entre 0 et 1. Nous appelons ça la normalisation. Une autre technique répandue consiste à la standardiser de telle sorte que chaque colonne soit centrée réduite.
\subsection{Propagation de l’information et rétropropagation du gradient}
\par Lorsque la donnée est fournie au modèle, les sorties de tous les neurones la première couche peuvent être calculées en appliquant la formule précédente (somme pondérée plus biais puis fonction d'activation). Avec la sortie de la première couche nous pouvons calculer la sortie de la deuxième couche et ainsi de suite jusqu'à la sortie. Donc l'information est propagée dans l'ensemble du réseau. Nous appelons ça la propagation de l'information. C'est le même processus qui est utilisé pour prédire une nouvelle entrée après que le réseau est entrainé.

\par La sortie prédite par le modèle est comparée avec la sortie attendue puis l'erreur commise est calculé. Cette erreur est propagée en arrière (\textit{propagated back}) couche par couche, et les poids sont mis à jour en fonction du degré de leurs contributions à l'erreur. Nous appelons ça la rétropropagation du gradient.

\par Le processus est répété pour tous les exemples (\textit{samples}) du \textit{dataset} d'entraînement. L'une des passes pour entraîner le réseau de neurones avec tous l'ensemble de données est appelé nombre d'itération (\textit{epochs}). Le réseau de neurones pout être entrainé pour dizaines, centaines, ou même des milliers d'itérations afin d'améliorer la précision du modèle à chaque fois. Le nombre d'itération est l'un des paramètres qui doit être bien choisie pour avoir un bon modèle.

\subsection{Mise à jour du poids (\textit{updating weights})}
\par Les poids dans les réseaux de neurones peuvent être mis à jour à partir des erreurs calculées pour chaque exemple des données de traitements. Nous appelons ça l'apprentissage en ligne. Cala permet de  faire des mises à jour rapides mais également chaotiques au réseau.
\par Une autre alternative consiste à enregistrer l'erreur au niveau de tous les exemples du \textit{dataset} d'entraînement, et puis les mises à jour sont effectuées vers la fin. Cela s'appelle l'apprentissage par lots (\textit{batch learning}) qui est souvent plus stable. Vu que le nombre d'exemples dans le \textit{dataset} est très grand, la taille du lot est réduite au petit nombre d'exemple tel que cent exemples afin d'augmenter l'efficacité de calcul.
\par La valeur dont les poids sont mis à jour est contrôlée par un paramètre appelé le taux d'apprentissage (\textit{learning rate}). La modification apportée au poids du réseau pour une erreur donnée est souvent de petite taille telle que $0.1$ ou $0.01$ ou plus petit.
\subsection{Prédiction}
Une fois le réseau de neurones est formé, il peut être utilisé pour prédire de nouvelles données. Nous pouvons effectuer des prédictions sur les données de test ou de validation afin d'évaluer la précision du modèle pour des nouvelles données non déjà vues. Le modèle peut être également déployé pour être utilisé pour la prédiction d'une manière continue. La topologie du modèle et les dernières valeurs du poids et de biais est tout ce qu'il doit être sauvegardé du modèle. La prédiction est établie en fournissant une entrée au modèle, puis ce dernier effectue la propagation ce qui lui permettra de générer la sortie qui peut être utilisée après comme une prédiction.
\section{Classes principales de réseau de neurones}
\par De nouveaux modèles sont publiés chaque jours. Comme il n'est pas facile de distinguer de ce qui fonctionne bien les praticiens recommande d'attendre qu'un modèle sera stable avant de l'adopter.
\par Il existe trois classes de réseaux de neurones artificiels les plus recommandées :
\begin{itemize}
    \item Les réseaux de neurones multicouches (MLPs) 
    \item Les réseaux de neurones convolutifs (CNNs) 
    \item Les réseaux de neurones récursifs (RNNs) 
\end{itemize}
\par Ces trois types sont plus flexibles et prouvent leurs utilités dans la résolution de grands nombres de problèmes. Ils ont également de nombreux sous-types spécialiser dans la résolution de certains problèmes spécifiques.

\par Les MLPs sont le type classique de réseaux de neurones. Ils sont composés d'une ou plusieurs couches. L'entrée est connectée avec la sortie à travers les couches intermédiaires ou appeler couches cachées. Ils conviennent au problème de prédiction de classification\footnote{Problème de classification est un type de problèmes dans lequel une classe ou une étiquette est attribuée aux entrées.} et aussi de régression\footnote{Problème de régression est un type de problèmes dans lequel une valeur réelle est prédite à partir d’un ensemble d’entrées.} là où les données sont souvent fournisses sous forme de tableau, comme dans un fichier CSV\footnote{Format de fichier utilisé pour stocker les données.} ou une feuille de calcul. Ils sont très flexibles et peuvent généralement être utilisés pour apprendre à mapper des entrées en sorties.\\

\par Les \textit{CNNs} sont un type de réseaux de neurones désigné pour mapper les données sous formes d’images à une variable de sortie. Ils se sont avérés si efficaces qu’ils sont le premier choix pour tout type de problème de prédiction impliquant les images en entrée. Les \textit{CNNs} fonctionnent bien avec des données ayant une relation spatiale\footnote{La relation spatiale est une propriété des données et pour laquelle nous utilisons les \textit{CNNs}. Cette propriété définit que les points d'une unité de données sont liés les unes aux autres de telle sorte qu'elles ne peuvent pas être séparées. Et par conséquent, si nous faisant ça, ou si nous les modifiant indépendamment l'unité de donnée sera corrompue. Les images et les données audio sont de type de données qui représente une relation spatiale.}.\\
\par Les \textit{RNNs} sont créés pour traiter les problèmes de prédiction de séquence de données comme les textes. Les \textit{RNNs} ont eu le plus de succès lorsqu'ils sont utilisés avec des séquences de mots et des paragraphes, généralement appelées traitement de langage naturel (\textit{NLP}). Ils sont également utilisés comme modèles génératifs nécessitant une sortie séquentielle non seulement avec du texte, mais également dans des applications telles que la génération d'écriture manuscrite. Les réseaux de neurones récurrents ne sont pas appropriés dans le cas de données tabulaires, comme dans un fichier CSV ou un tableur. Ils ne sont également pas appropriés pour les entrées de types images.

\newpage
\chapter{ Amélioration des performances du réseau de neurones }
\label{optim_model}
\par Il existe plusieurs techniques qui visent à améliorer les performances des réseaux de neurones, car initialement, ces performances peuvent souvent ne pas être satisfaisantes.  Les principaux facteurs conceptuels causant la dégradation de performances sont le problème du sur apprentissage et le mauvais choix des hyperparamètres. Nous avons sélectionné certaines techniques d’amélioration de performance pour les tester sur notre modèle.
\section{Diagnostiquer le modèle pour vérifier surapprentissage} 
\begin{figure}[H]
	\begin{center}
		\includegraphics[scale=0.4,frame]{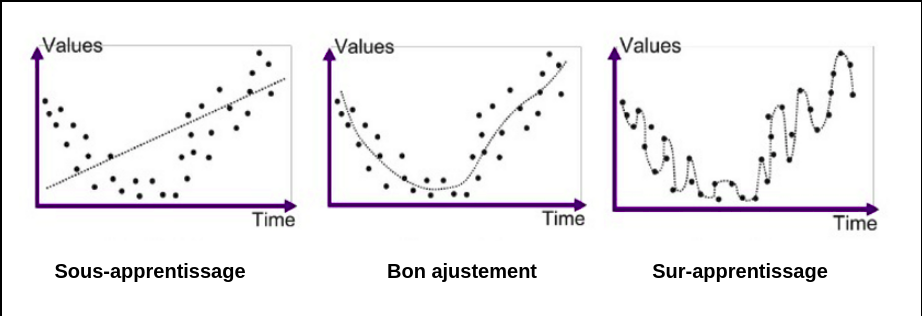}
	\end{center}
	\caption{Dispersion des points dans les différents cas de l'apprentissage (sous-apprentissage, modèle robuste et surapprentissage).} 
	\label{fig:surApprentissage}
\end{figure}

Le surapprentissage\footnote{Appelé aussi sur-ajustement.} survient lorsque le modèle commence à mémoriser les valeurs de données d’apprentissage au lieu d’apprendre avec. Par conséquent lorsque le modèle rencontre de nouvelles données qu’il n’a jamais vues auparavant, il n’est pas en mesure de prédire correctement. Nous pouvons vérifier ce phénomène en comparant la précision du modèle pour  le \textit{dataset} du test par celle du \textit{dataset} d’entraînement. Si la précision de l'entraînement est bien supérieure à celle du test, il est fort possible que le modèle a été sur-ajusté. Nous pouvons également vérifier le surapprentissage en visualisant les points prédits par le modèle sur un graphe et étudier leur dispersion (voir figure \ref{fig:surApprentissage}).
Les techniques suivantes peuvent être utilisées pour remédier au problème de surapprentissage : 
\section{Régularisation de données} 
La régularisation est une technique qui apporte de légères modifications à l’algorithme d’apprentissage de telle sorte que le modèle généralise mieux en pénalisant les poids des neurones. Cela améliore également les performances du modèle sur les données non déjà vue. Les régularisations $L1$ et $L2$ représentent les types de régularisation les plus répondus. Elles permettent de mettent à jour la fonction de coût générale en ajoutant un autre terme appelé terme de régularisation : fonction du coût devient : $Loss_{Cross\_entropy} = Loss_{Cross\_entropy} + terme\_de\_régularisation$
\subsection{Technique du \textit{Dropouts}} 
La technique du \textit{Dropouts} permet d’ignorer certains neurones choisis aléatoirement pendant la phase d’entraînement. Les neurones développent une co-dépendance pendant l'entraînement, ce qui limite la puissance individuelle de chaque neurone menant ainsi le modèle au surapprentissage. L’utilisation de cette technique permet d’abandonner aléatoirement des connexions entre les neurones et oblige le réseau à trouver de nouveaux chemins.
\subsection{Technique de \textit{early stopping}}
Appelée aussi l’arrêt précoce, cette technique permet  d’interrompre l’entraînement avant que les poids ne convergent. Souvent, nous nous arrêtons lorsque la précision cesse d’améliorer pour l’ensemble de validation.

\section{Choix de bons hyperparamètres}
Les hyperparamètres sont des valeurs d’initialisation données au réseau. Elles ne peuvent pas être apprises par le réseau pendant l’entraînement. Certains hyperparamètres sont de type architecturals tels que le nombre de couches de réseau et  le nombre de neurones dans chaque couche cachée, d'autres sont comportementals, à savoir la fonction d’activation, la fonction de perte (\textit{loss function}), l’optimiseur, la taille du lots (\textit{batch size}) et le nombre d’itérations (\textit{epochs}). Il n'existe pas une méthode directe qui permet d’identifier les meilleurs hyperparamètres, ils sont principalement obtenus par tests. Or, il existe certaines bonnes pratiques que nous avons prises en considération pour choisir certains hyperparamètres. 

\subsection{Fonction d’activation} 
Le choix de la bonne fonction d'activation permet au modèle de mieux apprendre. Certain fonctions d'activations comme le \textit{Sigmoid} et le \textit{Tanh} sont incapables de résoudre le problème de la disparition des gradients\footnote{Les valeurs des gradients diminuent lorsqu'elles atteignent les couches initiales.} (\textit{vanishing gradients}) contrairement à la fonction \textit{ReLU}\footnote{ReLu est fonction d'activation qui retourne le max entre le $0$ et la valeur d’entrée $x$ ($ReLu = max (0, x)$).} qui est devenue  la fonction  la plus utilisée. Cette dernière résout ce problème, elle a permis aux architectures de réseaux de neurones d’être plus profondes et d’voir des tailles plus grandes. 
\subsection{Taux d’apprentissage (learning rate)}
Le choix du bon taux d’apprentissage est important car il détermine si le modèle converge ou non vers les minima globaux. Un taux élevé ne mène presque jamais au minimum global. Un très petit taux peut prend énormément de temps pour converger vers un minima global. Il peut également être piéger dans un minimum local et donc, le réseau risque de converger vers un minimum local et ne pourra pas en sortir à cause du faible taux d'apprentissage. Généralement le taux d'apprentissage est une puissance de $10$ et varie entre $10 ^{-1}$ et $10 ^{-4}$).

\subsection{Taille de lot et nombre d'itérations}
Il n'y a pas de valeurs standards pour la taille du lot et le nombre d'itérations qui fonctionnent pour tous les types de problèmes. Il faut tester les différentes valeurs. Les puissances de deux $(8, 16, 32, etc.)$ sont en général les tailles de lot les plus utilisées. Le nombre d'itérations nécessaire est généralement défini par les tests ou par la technique de \textit{Early stopping} motionné ci-dessus.

\subsection{Initialisation du poids}
Lors de l’entraînement des réseaux de neurones, les poids initiaux sont attribués de manière aléatoire. Pour des architectures multicouches, les poids initiaux aléatoires ne fonctionnent pas bien. Il est possible de fournir des poids initiaux optimaux. Plusieurs méthodes peuvent être utilisées pour initialiser les poids, tel que l'initialisation aléatoire et l’initialisation des poids \textit{Xavier}\footnote{Une heuristique qui crée une distribution uniforme par couche du gradient.}. La méthode qui convient le mieux au problème est celle qui doit être choisie. Ceci est déterminé en effectuant des tests de comparaisons entre les méthodes.

%%%%%%%%%%%%%%%% End annexe D

\newpage
\begingroup
\renewcommand{\chapter}[1]{\textbf{\begin{center}Bibliographie\end{center}}}
\bibliographystyle{apacite}

%\bibliography{ref}
\endgroup
\end{onehalfspace}
\end{document}